\documentclass[twocolumn]{aastex701}

\usepackage[encapsulated]{CJK}
\usepackage{amsmath}
\usepackage{subcaption} 
\usepackage{hyperref}
\usepackage{url} 

\received{\today}
\revised{XXX}
\accepted{YYY}
\graphicspath{{./}{figs/}}

\begin{document}

\title{Everything Every Band All at Once II: The Relationship Between Optical Size and Stellar Mass Over Eight Billion Years of Cosmic History}

\shorttitle{The Optical Size-Mass Relationship from $0.5<z<8$}
\shortauthors{Miller et al.}

\author[0000-0001-8367-6265]{Tim B. Miller}
\affiliation{Center for Interdisciplinary Exploration and Research in Astrophysics (CIERA), Northwestern University, 1800 Sherman Ave, Evanston, IL 60201, USA}
\email[show]{timothy.miller@northwestern.edu}

\author[0000-0001-6454-1699]{Yunchong Zhang}
\affiliation{Department of Physics and Astronomy and PITT PACC, University of Pittsburgh, Pittsburgh, PA 15260, USA}
\email{yuz369@pitt.edu}

\author[0000-0002-0108-4176]{Sedona H. Price}
\affiliation{Space Telescope Science Institute, 3700 San Martin Drive, Baltimore, Maryland 21218, USA}
\email{sprice@stsci.edu}

\author[0000-0002-1714-1905]{Katherine A. Suess}
\affiliation{Department for Astrophysical \& Planetary Science, University of Colorado, Boulder, CO 80309, USA}
\email{Wren.Suess@colorado.edu}

\author[0000-0001-5063-8254]{Rachel Bezanson}
\affiliation{Department of Physics and Astronomy and PITT PACC, University of Pittsburgh, Pittsburgh, PA 15260, USA}
\email{rachel.bezanson@pitt.edu}

\author[0000-0003-4075-7393]{David J. Setton}
\thanks{Brinson Prize Fellow}
\affiliation{Department of Astrophysical Sciences, Princeton University, Princeton, NJ 08544, USA}
\email{davidsetton@princeton.edu}
\author[0000-0002-2057-5376]{Ivo Labbe}
\affiliation{Centre for Astrophysics and Supercomputing, Swinburne University of Technology, Melbourne, VIC 3122, Australia}
\email{ivolabbe@gmail.com}

\author[0000-0003-2680-005X]{Gabriel Brammer}
\affiliation{Cosmic Dawn Center (DAWN), Niels Bohr Institute, University of Copenhagen, Jagtvej 128, K{\o}benhavn N, DK-2200, Denmark}
\email{gabriel.brammer@nbi.ku.dk}

\author[0000-0002-7031-2865]{Sam E. Cutler}
\affiliation{Department of Physics and Astronomy, Tufts University, 574 Boston Ave., Medford, MA 02155, USA}
\email{secutler@umass.edu}

\author[0000-0001-6278-032X]{Lukas J. Furtak}
\affiliation{Cosmic Frontier Center, The University of Texas at Austin, Austin, TX 78712, USA}
\affiliation{Department of Astronomy, The University of Texas at Austin, Austin, TX 78712, USA}
\email{furtak@utexas.edu}

\author[0000-0001-6755-1315]{Joel Leja}
\affiliation{Department of Astronomy \& Astrophysics, The Pennsylvania State University, University Park, PA 16802, USA}
\affiliation{Institute for Computational \& Data Sciences, The Pennsylvania State University, University Park, PA 16802, USA}
\affiliation{Institute for Gravitation and the Cosmos, The Pennsylvania State University, University Park, PA 16802, USA}
\email{joel.leja@psu.edu}

\author[0000-0002-9651-5716]{Richard Pan}
\affiliation{Department of Physics and Astronomy, Tufts University, 574 Boston Ave., Medford, MA 02155, USA}
\email{ Richard.Pan@tufts.edu}

\author[0000-0001-9269-5046]{Bingjie Wang}
\thanks{NHFP Hubble Fellow}
\affiliation{Department of Astrophysical Sciences, Princeton University, Princeton, NJ 08544, USA}
\email{bjwang@princeton.edu}

\author[orcid=0000-0003-1614-196X]{John R. Weaver}
\thanks{Brinson Prize Fellow}
\affiliation{Kavli Institute for Astrophysics and Space Research, Massachusetts Institute of Technology, Cambridge, MA 02139, USA}
\email{john.weaver.astro@gmail.com} 

\author[0000-0001-7160-3632]{Katherine E. Whitaker}
\affiliation{Department of Astronomy, University of Massachusetts, Amherst, MA 01003, USA}
\affiliation{Cosmic Dawn Center (DAWN), Denmark} 
\email{kwhitaker@astro.umass.edu}

\author[0000-0001-8460-1564]{Pratika Dayal}
\affiliation{Canadian Institute for Theoretical Astrophysics, 60 St George St, University of Toronto, Toronto, ON M5S 3H8, Canada}
\affiliation{David A. Dunlap Department of Astronomy and Astrophysics, University of Toronto, 50 St George St, Toronto ON M5S 3H4, Canada}
\affiliation{Department of Physics, 60 St George St, University of Toronto, Toronto, ON M5S 3H8, Canada}
\email{pdayal@cita.utoronto.ca}

\author[0000-0002-1109-1919]{Robert Feldmann}
\affiliation{Department of Astrophysics, Universität Zürich, Winterthurerstrasse 190, CH-8044 Zurich, Switzerland}
\email{robert.feldmann@uzh.ch}

\author[0000-0001-7201-5066]{Seiji Fujimoto}
\affiliation{David A. Dunlap Department of Astronomy and Astrophysics, \\ University of Toronto, 50 St. George Street, Toronto, Ontario, M5S 3H4, Canada}
\affiliation{Dunlap Institute for Astronomy and Astrophysics, 50 St. George Street, Toronto, Ontario, M5S 3H4, Canada}
\email{seiji.fujimoto@astro.utoronto.ca}

\author[0000-0002-3254-9044]{K. Glazebrook}
\affiliation{Centre for Astrophysics and Supercomputing, Swinburne University of Technology, PO Box 218, Hawthorn, VIC 3122, Australia}
\email{kglazebrook@swin.edu.au}

\author[orcid=0000-0002-2380-9801]{Anna de Graaff}
\thanks{Clay Fellow}
\affiliation{Center for Astrophysics, Harvard \& Smithsonian, 60 Garden St, Cambridge, MA 02138, USA}
\affiliation{Max-Planck-Institut f\"ur Astronomie, K\"onigstuhl 17, D-69117 Heidelberg, Germany}
\email{anna.de_graaff@cfa.harvard.edu}

\author[0000-0002-5612-3427]{Jenny E. Greene}
\affiliation{Department of Astrophysical Sciences, Princeton University, 4 Ivy Lane, Princeton, NJ 08544, USA}
\email{jgreene@astro.princeton.edu}

\author[0000-0002-5588-9156]{Vasily Kokorev}
\affiliation{Cosmic Frontier Center, The University of Texas at Austin, Austin, TX 78712, USA}
\affiliation{Department of Astronomy, The University of Texas at Austin, Austin, TX 78712, USA}
\email{vasily.kokorev.astro@gmail.com}

\author[0000-0001-9002-3502]{Danilo Marchesini}
\affiliation{Department of Physics and Astronomy, Tufts University, 574 Boston Ave., Medford, MA 02155, USA}
\email{danilo.marchesini@tufts.edu}

\author[0000-0002-9330-9108]{Adam Muzzin}
\affiliation{Department of Physics and Astronomy, York University, 4700 Keele St., Toronto, Ontario, M3J 1P3, Canada}
\email{muzzin@yorku.ca}

\author[0000-0003-2804-0648 ]{Themiya Nanayakkara}
\affiliation{Centre for Astrophysics and Supercomputing, Swinburne University of Technology, PO Box 218, Hawthorn, VIC 3122, Australia}
\affiliation{Sydney Institute for Astronomy, School of Physics, The University of Sydney, Sydney, NSW 2006, Australia}
\email{wnanayakkara@swin.edu.au}

\author[0000-0002-7524-374X]{Erica J. Nelson}
\affiliation{Department for Astrophysical \& Planetary Science, University of Colorado, Boulder, CO 80309, USA}
\email{Erica.Nelson-1@colorado.edu}

\author[0000-0002-5027-0135]{Arjen van der Wel}
\affiliation{Sterrenkundig Observatorium, Universiteit Gent, Krijgslaan 281 S9, 9000 Gent, Belgium}
\email{arjen.vanderwel@ugent.be}

\begin{abstract}
While the size-mass relation provides insight into the structural evolution of galaxies, the data available and methods employed have hindered our ability to study a detailed and comprehensive description of this key relation across cosmic history. The first paper in this series presents a morphology catalog based on 20 band JWST data in the field of Abell 2744. In this paper we utilize this catalog to measure the size-mass relation from $0.5<z<8$ and $0.5<z<3$ for star-forming and quiescent galaxies respectively. We perform a global fit to our sample using B-splines to flexibly model the redshift evolution which enforces smooth evolution and can account for all observational uncertainties. Symbolic regression is used to derive simple and portable expressions that describe the redshift evolution of the size-mass relation. Analyzing the size evolution of star-forming galaxies in the context of previous work at $z\sim0$ and $z>10$, we discuss three distinct phases: Rapid growth at $z>5$, growth that mimics dark matter halos at $5< z <1$ and a late plateau at $0.5<z<1$. For quiescent galaxies we confirm previous findings that the size-mass relation flattens at $\log\ M_*/M_\odot < 10$, which inverts at $z>1$. Our results imply that quiescent galaxies are smaller than their star-forming counterparts only at around $\log M_*/M_\odot = 10$; the two populations have similar sizes at lower and higher masses.
\end{abstract}


\section{Introduction}

The relation between galaxy size and stellar mass is a crucial diagnostic of the structural evolution of galaxies. The distribution of populations of galaxies in this two parameter space, and their evolution with redshift provides key insights into the physical process that affect the formation of galaxies. It has lead to insight into the inside-out growth of star-forming galaxies~\citep{vanderwel2014,nelson2016b,tacchella2018}, specific quenching pathways~\citep{vandokkum2015,Matharu2020,suess2021,Cutler2024}, the continued growth of quiescent galaxies through minor mergers~\citep{bezanson2009, newman2012} and the relationship between galaxies and their dark matter halos.\citep{shen2003,kravtsov2013,somerville2018,mowla2019b,kawinwanichakij2021}

The size of galaxy is most commonly quantified using the effective radius (also known as the half-light radius which contains $50\%$ of the total light). In this work we focus on the rest-frame-optical wavelengths, which we define as $\lambda_{\rm rest} = 5000$\AA. This wavelength represents a historical benchmark and is reasonable proxy of the underlying stellar mass distribution. \citep[Although this may not always be a reasonable assumption, see][]{suess2019, miller2022,vanderwel2024,Martorano2025} It is readily observable from the ground at $z<0.5$, out to $z=2$ with HST and now to to dawn of galaxy formation at $z\sim8$ with JWST.~\citep{Gardner2023} The relationship between optical size and stellar mass has been extensively studied over many decades, covering several orders of magnitude in stellar mass, over two orders of magnitude in radii and 13 billion years of cosmic history.~\citep{shen2003, vanderwel2014,lange2015,mowla2019,kawinwanichakij2021,Cutler2022,nedkova2021,Carlsten2021,Damjanov2023,Allen2024,ward2024,George2024,Morishita2024,Miller2025,Yang2025,Asali2025}


However, our ability to self-consistently study the detailed evolution of the size-mass relationship across the entirety of cosmic time has been limited by our data and methods. First, due to the wavelength coverage from the ground and with HST we were only able to observe rest-frame optical light out to $z\sim2$, missing the growth of the first generation of galaxies. Since the launch of JWST, and in particular due to the capabilities of the NIRCam instrument~\citep{Rieke2023}, we can now study the rest-frame optical emission from galaxies across $0.5<z<8$ in a consistent and uniform manner.~\citep{ward2024, ormerod2024,Allen2024,Yang2025, Miller2025}. With this influx we should reconsider the methods typically used to measure sizes and the size-mass relation to ensure we are fully utilizing this wealth of data. 


The sizes of galaxies are measured by ``profile-fitting'' however traditional methods~\citep[e.g. \texttt{GALFIT},][]{peng2010} are inconsistent. Up to $~\sim40\%$ of fits fail~\citep{vanderwel2012}, and parameters lack meaningful uncertainties. Errors on measured sizes needed to be estimated in a post-hoc manner~\citep{vanderwel2012, Cutler2024,Allen2024}. Inference on the scatter around the size-mass relation, which is extremely sensitive to the assumed errors on the measured sizes, was therefore limited. This scatter is useful probe of galaxy formation and is thought to be sensitive to the burstiness of star-formation~\citep{elbadry2017,emami2021,McClymont2025} and the galaxy-halo connection ~\citep{shen2003,somerville2018,Rohr2022}.

Beyond the measuring sizes themselves, our method to fit the size-mass relation have been stagnant. Rhe size-mass relation is often fit in discrete redshift bins with a typical bin width of $\Delta z\approx0.5-1$. This yields a coarse-grained view of galaxy evolution where only broad trends, with widths greater than the can be studied. The evolution implied by independent fits can be uneven and un-physical. Overall, fitting this size-mass relation in separate redshift bins fails to meaningfully connect the structural evolution of galaxies across different epochs.

In this work we aim to provide a benchmark for the redshift evolution of the size-mass relationship from $0.5<z<8$ based on JWST data from the UNCOVER~\citep{Bezanson2022} and MegaScience surveys~\citep{Suess2024}. These surveys both image the lensing cluster Abell 2744. The combination covers all 20 wide and medium bands from NIRCam providing precise stellar masses, photometric redshift as well as multi-wavelength morphology. The depth, combined with the boost of gravitational lensing, make UNCOVER/MegaScience the ideal dataset to study the evolution of low mass galaxies in particular.

In the first paper in this series, \citet{zhang2026}, we present a morphology catalog for UNCOVER/MegaScience including 29,608 sources and all 20 NIRCam bands. Using the \texttt{pysersic}~\citep{pasha2023} we fit single S\'ersic profiles to all sources using Bayesian inference techniques to capture uncertainties on the morphological parameters. Through a series of quality checks and tests we show that we recover robust morphological parameters for $>90\%$ of all sources with reliable uncertainties. 

Beyond utilizing this new dataset we improve the methods which are used to measure the size-mass relationship. One of the most notable improvements is fitting a continuous evolution model connecting all of the galaxies in our sample. We employ Basis splines (B-splines) to flexibly model the redshift evolution so that the results are not biased by relying on simplistic parameterizations. A continuous model also allows us to naturally account for co-variant uncertainties in the photometric redshifts and stellar masses within our inference procedure to fully account for all sources of error. We can then robustly model the scatter around the size-mass relation.

In Section~\ref{sec:data} we introduce the data and sample of galaxies used in our analysis. Our inference procedure is introduced in Section~\ref{sec:fitting} and the results are displayed in Section~\ref{sec:res}. We use symbolic regression to distill the redshift evolution of our parameters into simple and interpretable equations in Section~\ref{sec:sr}. Section~\ref{sec:lit_comp} discusses differences between our results and previous studies in the literature and Section~\ref{sec:disc} discusses new insights into galaxy growth across cosmic time provided by this study.

Throughout this study we assume a $\Lambda$CDM cosmology following the 9-year results of the WMAP mission~\citep{Hinshaw2013}, a \citet{chabrier2003} initial mass function and all magnitudes are presented using the AB system~\citep{Oke1983}

\section{Data and Galaxy Properties}
\label{sec:data}
\begin{figure*}
    \centering
    \includegraphics[width=0.9\textwidth]{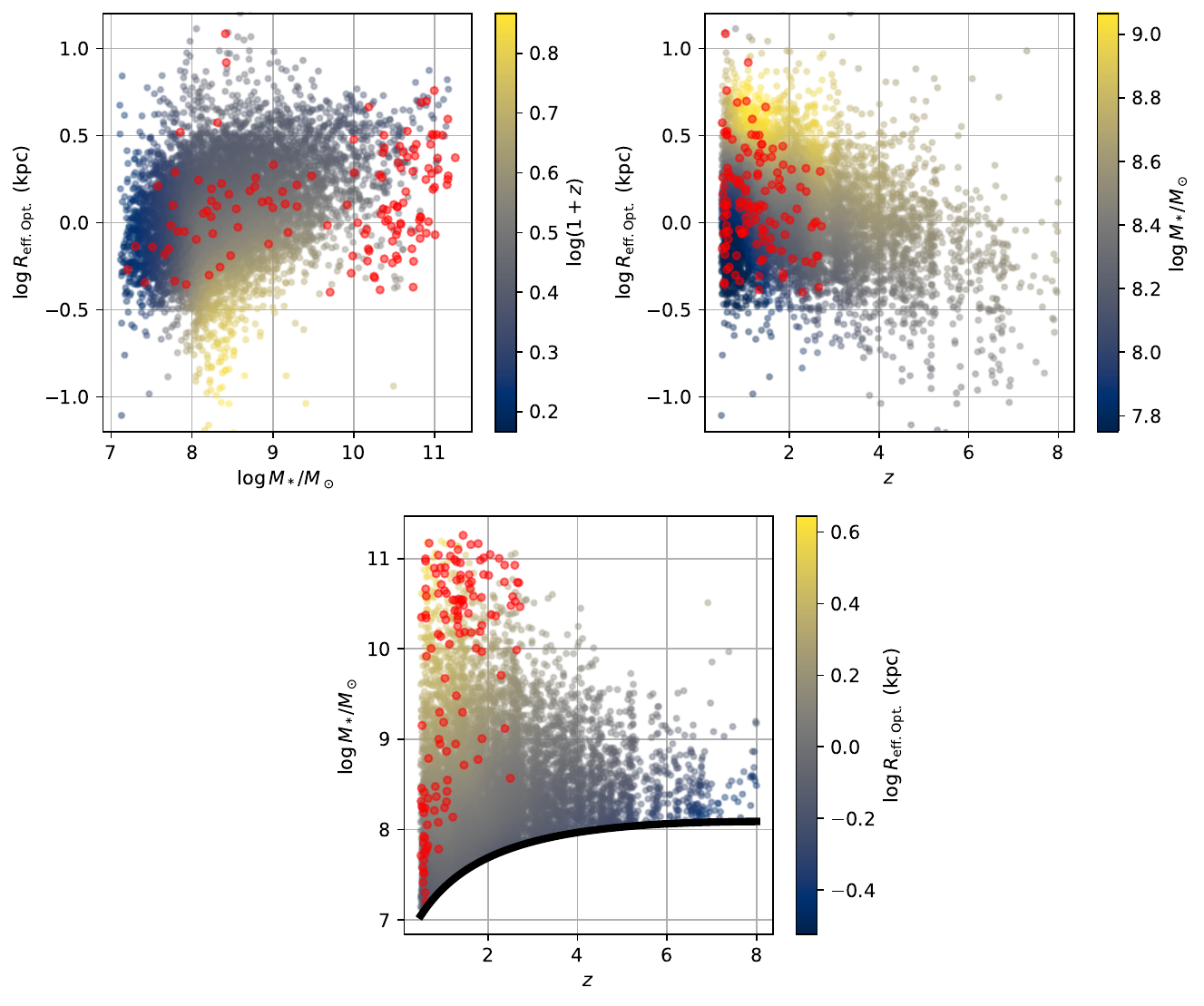}
    \caption{An overview of the sample of galaxies from UNCOVER/MegaScience used in our global size mass fit. We show three panels each with a different slice of the three dimensional relationship between stellar mass, rest-frame optical size and redshift. In each panel star forming galaxies are shown with the colored points where the color describes the median of the third parameter calculated using the \texttt{loess} algorithm.\citep{cleveland1979,cappellari2013}. Quiescent galaxies are shown as individual red points.}
    \label{fig:sample}
\end{figure*}

This study is based on data obtained from the UNCOVER~\citep{Bezanson2022} and MegaScience surveys~\citep{Suess2024} targeting the lensing cluster Abell 2744. These surveys combine to produce imaging in 20 NIRcam bands across 30 arcmin sq. This study is built off of the photometric catalog~\citep{Weaver2024,Suess2024}, lensing model~\citep{Furtak2023,Price2024} and stellar population catalog~\citep{Wang2024,Suess2024} along with the morphology catalog presented in \citet{zhang2026}

We refer readers to the relevant papers for full details on the creation of each catalog but provide a short synopsis here. The photometric catalog is created using noise-equalized images containing the F277W, F356W and F444W filters for the detection image using the \texttt{aperpy} software~\citep{Weaver2024}. The photometric redshift and stellar population catalog is created using Prospector-$\beta$, which utilizes redshift-dependent informative priors for the star-formation history and stellar mass~\citep{johnson2021,Wang2023}. We use the \texttt{v2.0} lensing model which includes spectroscopic redshifts from the UNCOVER survey\citet{Furtak2023,Price2024}. 

The foundation of this study is the morphological catalog presented in \citet{zhang2026}. We use \texttt{pysersic}~\citep{pasha2023} to fit S\'{e}rsic models to all sources with $S/N > 10$ separately in all 20 NIRCam bands. We perform a series of quality tests and find the $>90\%$ of sources are successfully fit along with injection-recovery to ensure the uncertainties measured are reliable. In practice we do not use the fits to individual bands but instead the functional form presented for the wavelength dependence of the effective radius. For each $\log R_{\textrm eff}$ is parameterized as a quadratic in$\log\lambda$. as long as two or more bands have successful S\'{e}rsic fits. We utilize the best fit parameters of this quadratic function and their covariances.


To ensure robust SED fits and the correct wavelength coverage we make a series of additional cuts:
\begin{itemize}
    \item $\chi^2_{\textrm sps} / N_{\textrm bands} < 5$
    \item $z_{84} - z_{16} < 1. $
    \item $0.5 < z_{50} < 8$
    \item $z_{16} > 0.4$ and $z_{84} < 9$
    \item $\log\lambda_{\rm success, min} - 0.1 < 5,000 \mathring{A}\,(1+z_{50}) < \log \lambda_{\rm success, max} + 0.1$
    \item $\mu(z= z_{50}) < 4$
\end{itemize}
The first two cuts ensure the galaxy photometry is well fit by the stellar population model and the photometric redshift is reasonably constrained. The redshift range approximates where the JWST NIRCam bands provide coverage of the rest-frame optical. $\lambda_{\rm success, min}$ and $\lambda_{\rm success, max}$ represent the lowest and highest wavelengths for which a successful Sersic model is fit to each galaxy. This ensures the rest-frame optical is not extrapolating far beyond where the measurements are. The final magnification cut limits our sample to moderate magnification as for high magnification the correction for lensing becomes more complicated and often bespoke solutions are required. Additionally we remove any sources that are photometrically or spectroscopically classified as `little red dots' in ~\citet{Labbe2023} or ~\citet{Greene2024}, as their inferred stellar mass is highly uncertain~\citep[e.g.][]{Wang2024b}. We split our sample into star-forming and quiescent galaxies following the prescription presented in~\citet{antwidanso2022} based on synthetic $ugi$ colors. These are calculated using the best fit Prospector$-\beta$ model for each galaxy.

The lower redshift limit of $z=0.5$ is chosen for two main reasons; First we do not wish to select galaxies in the cluster itself and second the coverage from JWST does not sample the $u$ band well making the star-forming / quiescent selection more ambiguous. For quiescent galaxies we further limit the redshift range to $0.5<z<3$. This is simply due to the fact that we have very few quiescent galaxies at higher redshifts due to the declining number density~\citep{Carnall2023,Zhang2026_q} and the limited volume probed by the UNCOVER/MegaScience surveys. The on-sky area is only 30 arcmin$^2$ and the volume is further reduced by the effect of gravitational lensing. For star-forming galaxies the upper limit of $z=8$ is chosen as the highest redshift at which the rest frame $5,000$ \AA{} is covered by NIRCam.  An overview of the population of galaxies used for analysis is shown in Figure~\ref{fig:sample}.

We calculate the stellar mass completeness limit for each group independently following the procedure of~\citet{Pozzetti2010}. Briefly, we take galaxies with the lowest 30\% (de-lensed) stellar mass and rescale their mass such that their magnitude in the detection image matches the detection limit, which we assume to be 29.5~\citet{Weaver2024}. We then take the 95\% of this rescaled distribution as the limiting stellar mass. For star-forming galaxies we calculate the limiting mass in 10 redshift bins, each with an equal number of galaxies, and fit the redshift dependence by a quadratic in $\log(1+z)$, specifically:
\begin{equation}
    \log M_{*,\textrm{lim, SF}}/M_\odot = -1.70[\log(1+z)]^2 + 3.25\log(1+z) + 6.34 
    \label{eqn:mlim_sf}
\end{equation}
This same procedure is repeated for Quiescent galaxies, except due to the smaller number only use 3 bins and fit a linear relation following:
\begin{equation}
    \log M_{*,\textrm{lim, Q}}/M_\odot =  2.45\log(1+z) + 6.6
    \label{eqn:mlim_q}
\end{equation}
After removing galaxies below this completeness limit we arrive at a sample of 137 quiescent galaxies and 8524 star-forming galaxies. An overview of this sample is presented in Figure~\ref{fig:sample}. 

This empirical procedure to measure completeness is mainly sensitive to the $S/N$ limits that are imposed during the S\'ersic fitting procedure. Variation in the fraction of galaxies that pass the other quality checks are largely not captured. As shown in the appendix of~\citet{zhang2026} this can vary as a function of mass and redshift. However above our mass completeness line there is only moderate variation, with the minimum successful fit fraction being roughly $30\%$ at $\log M_*/M_\odot \lesssim9$ and $z>7$. The key assumption we make is that this failure rate does not depend on the size of a galaxy so therefore we do not correct for it in our fitting. We believe this is a reasonable assumption because these failures can be caused by numerous reasons that are unrelated to the physical properties of the galaxy, e.g. nearby bright objects, crowded fields or data issues.

\section{Fitting the Size-Mass Relationship}
\label{sec:fitting}
\subsection{Parameterizing the Size-Mass Relationship and Its Evolution}
\label{sec:param}
\begin{figure}
    \centering
    \includegraphics[width=0.99\columnwidth]{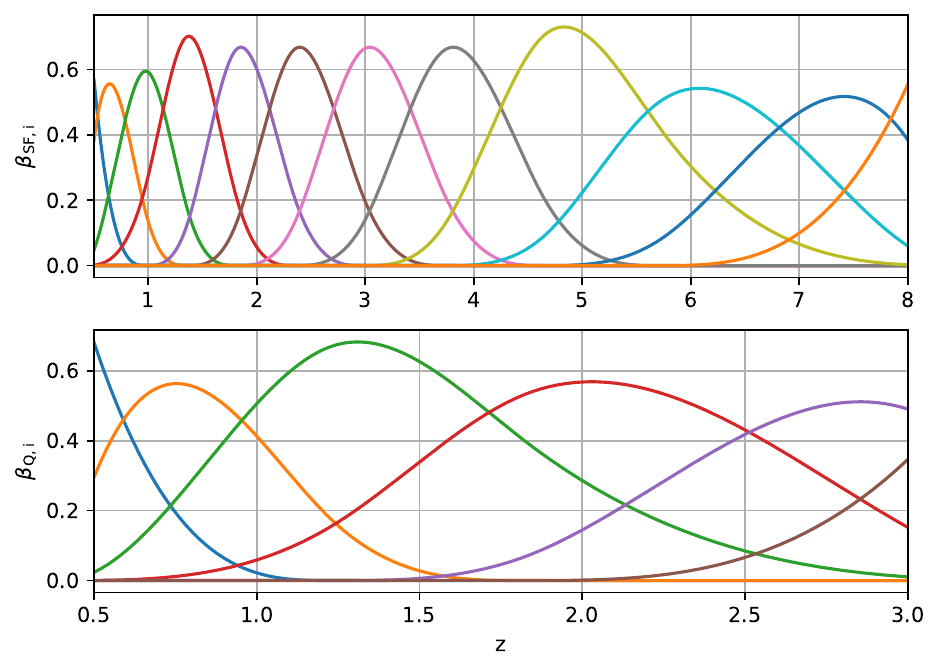}
    \caption{Basis functions using in the B-Splines to describe the redshift evolution of the size-mass parameters for star-forming galaxies (\emph{top}) and quiescent galaxies (\emph{bottom}). For star-forming galaxies we using a 3rd order spline with 12 basis function calculated using knots that are spaced equally in $\log(1+z)$ across $0.4<z<8.5$. For quiescent galaxies we also used a 3rd order spline with the same spacing but only use 6 basis function between $0.4<z<3.5$}
    \label{fig:bspl_basis}
\end{figure}

For star-forming galaxies we follow numerous previous studies~\citep[e.g.][]{shen2003,vanderwel2014,lange2015} and parameterize the size-mass relation with a simple power law, following:

\begin{equation}
    \log R_{\rm eff, SF}\,(M_*,z) = m(z)* (\log M_*/M_{\odot} - 8.5) + b(z)
    \label{eqn:sf_pl}
\end{equation}

where $m(z)$ is the logarithmic slope which is a function of redshift and $b(z)$ is the intercept, or the average logarithmic size of galaxies with $\log M_*/M_\odot = 8.5$. Additionally we model the scatter around the size-mass relation and assume it follows a log-normal distribution. We model the scatter, $\sigma_{\rm \log r}$ as a function of stellar mass and redshift, specifically modeled as following:
\begin{equation}
    X(M_*,z) = \,n(z)*(\log M_*/M_{\odot} - 8.5) + c(z)
\end{equation}
\begin{equation}
    \sigma_{\rm log\, R,SF}\,(M_*,z) =  \frac{1}{1 + e^{-X(M_*,z)}}
    \label{eqn:sf_sig}
\end{equation}

The second equation represents a Sigmoid function and bounds the results between 0 and 1. The redshift dependent slope and intercept of this relation are given by $n(z)$ and $c(z)$ respectively. Using the sigmoid function we ensure the resulting scatter will be contained to positive and realistic values. This simplifies the parameterization and increases the sampling efficiency since there is no need for complex priors on $n(z)$ and $b(z)$ to ensure $\sigma_{\rm log \ R, SF} > 0$. In these equations the parameters, $b(z), m(z), c(z)$ and $n(z)$ are the free parameters, and their evolution with redshift will be fit for.

For quiescent galaxies we parameterize the size-mass relationship as a double power law following \citet{mowla2019}, following the equation:
\begin{equation}
    R_{\rm eff., Q} = r_p(z)\left(\frac{M_*}{M_p(z)} \right)^{\alpha(z)}\left(\frac{1}{2} \left[ 1 + \left(\frac{M_*}{M_p(z)}\right)^\delta \right]\right)^{\frac{\beta(z) - \alpha(z)}{\delta}}
    \label{eqn:q_dpl}
\end{equation}
Where $\alpha,\beta, r_p$ and $M_p$ are the free parameters. $\alpha$ represents the power law slope at low mass and $\beta$ at high masses. $M_p$ is the pivot mass and represents the location of the transition between these two regimes. $r_p$ is the average value of $R_{\rm eff}$ at the pivot mass. The parameter $\delta$ controls the smoothness of the transition between the low and high mass slopes. Following previous studies~\citep{mowla2019,kawinwanichakij2021} we keep $\delta$ fixed to 6. Similar to star forming galaxies, the scatter around the size-mass relation is assumed to be Gaussian in $\log R_{\rm eff}$ with a standard deviation $\sigma_{\rm log\, R,Q}$. Unlike star-forming galaxies, and partly due to the fact that we have far fewer quiescent galaxies in our sample, we assume that the scatter depends only on redshift and not on stellar mass. As such we do not need to use a sigmoid function and simply fit for the logarithmic scatter, $\sigma_{\rm \log\, R,\, Q}(z)$ .

To parameterize the evolution of the parameters with redshift we use Basis splines. These provide a flexible representation on the redshift evolution without any bias imparted due to a specific choice of parameterization. An example of this for the parameter $m(z)$ is shown below,
\begin{equation}
    m(z) = \sum_i m_i B_i(z)
    \label{eqn:bspl_ex}
\end{equation}
Where $B_i(z)$ are the pre-defined Basis components and $m_i$ are the free parameters or coefficients. B-splines are a flexible parameterization that ensure continuity and differentiability. For star-forming galaxies in this study we use a 3rd order B-spline with 12 components equally spaced in $\log (1+z)$ ranging from z=0.4-8.5. For quiescent galaxies it is a similar setup but 6 components ranging from z=0.4-3.5. These components are displayed in Figure~\ref{fig:bspl_basis}. Spacing the components in $\log(1+z)$ provides more flexibility at $z\lesssim2$ where we have many galaxies and less so at higher redshift where we have fewer galaxies and thus expect our constraints to be weaker.

We place broad uniform priors on the spline coefficients for each variable; however when using B-splines it is common for un-realistic ``wiggles'' to appear. To enforce smoothness of the resulting redshift evolution we implement the Penalized B-spline framework first outlined in \citet{Eilers1996}. This study suggests adding an additional term to the likelihood to enforce similarity between neighboring coefficients, specifically one that penalizes second order differences, i.e. $\Delta_i^2 = 2m_{i-1} - m_i - m_{i-2}$. The strength of this smoothing is then by a smoothing parameter. We follow \citet{Lang2004} who extend this to a Bayesian context by treating the smoothing parameter as a random variable. In practice we add an additional term to the likelihood: A Gaussian in $\Delta_i^2$ for each parameter with a standard deviation $\rho$. Where $\rho$ is the smoothing parameter and is sampled from an exponential distribution following~\citet{Simpson2017} with a rate of 5. We use a separate smoothing $\rho$ for each of the size-mass parameters.

\subsection{Inference Procedure}
\label{sec:method}
The observational constraints on the redshift dependent size-mass relationship come from the prospector-$\beta$ stellar population catalog, morphological and lensing  catalogs~\citep{Furtak2023,Wang2024,zhang2026}. To properly account for uncertainties in the stellar mass, redshift and observed size the input to our modeling procedure are the posteriors produced by these fitting procedures. To account for the complex posterior on stellar mass and redshift from prospector-$\beta$ we model it as a 2D gaussian mixture model (GMM). Using \texttt{scikit-learn}~\citep{scikit-learn}, we fit a three component model for each galaxy using a diagonal covariance matrix. This allows us to capture the complex posteriors produced from SED fitting in a parameterized way.

For the size measurement we do not rely on measurements from individual bands but instead the functional form, $R_{\rm eff}(\lambda)$, described in \citet{zhang2026}. It is parameterized with $\log R_{\rm eff}$ is a quadratic in $\log \lambda$, i.e. $\log R_{\rm eff} \propto a(\log \lambda)^2 + b\log \lambda + c$. We use the covariance matrices provided to account for uncertainties and correlations between the polynomial parameters. For each galaxy, at each step, we calculate observed size at $\lambda_{\textrm rest} = 5,000$\AA using these parameters and the redshift sampled using the procedure described above. We then convert this observed size to a physical size accounting for lensing and angular diameter distance, again accounting for difference in the redshift. We `de-lens' our sizes using the simple correction of  $1/\sqrt{\mu(z)}$, where we assume a 10\% uncertainty on $\mu(z)$. 

Following~\citet{Miller2025} we allow for the possibility of outliers from the parameterized size-mass relationship for both star-forming and quiescent galaxies. We make no assumption about the cause of these outliers, only that they do not follow the typical distribution of sizes at the specific mass and redshift. We model this outlier distribution following \citet{hogg2010} by adding an additional term to the likelihood function whose relative importance is controlled by the fraction of the population that is considered an outlier, $q_{\rm ol}$. 

This distribution of the outlier population is parameterized as a log-normal distribution in $\log\, R_{\rm eff}$ defined by three parameters: $q_{\rm ol}$, the fraction of galaxies classified as outlier (which we assume is independent of galaxy properties), $\mu_{\rm ol}$, the mean of the distribution and $\sigma_O$, the scatter in logarithmic space. We set the prior on $\mu_{\rm ol}\sim \mathcal{N}(0,1)$ the standard normal and $\sigma_{\rm ol}$ with a a half-normal with a mean of 0 and standard-deviation of 1, in units of $\log\,  R_{\rm eff, opt.}/{\textrm{kpc}}$. We set a broad prior on the outlier fraction for the star-forming galaxy fit of $\mathcal{U}(0,0.25)$. For quiescent galaxies we enforce a stricter prior on the outlier fraction of $\mathcal{N}(0.25,0.1)$, truncated at 0 and 0.5. This is informed by the two most likely candidates for outliers; a $7.8\%$ photo-z failure rate reported in ~\citet{Suess2024} and roughly $20\%$ contamination fraction reported in ~\citet{antwidanso2022}.

To perform inference for the size-mass relationship we implement the parameterization for each population described above in the \texttt{numpyro}~\citep{phan2019} probabilistic programming language. We fit each population separately. We use the No U-turn (NUTS) sampler with four chains for 2,000 warm up and 3,000 sampling steps each. We ensure that the chains are converged by checking the effective sample size and $\hat{r}$ metrics~\citep{vehtari2021}. The posterior distribution for all of the size-mass parameters have effective sample size $>1,500$ and $\hat{r}$, which compares inter- to intra- chain variance, $<1.02$.

\section{Results of The Global Size-Mass Fit}
\label{sec:res}
\subsection{Star-forming Galaxies}
\label{sec:sf_res}
\begin{figure*}
    \centering
    \includegraphics[width=0.99\textwidth]{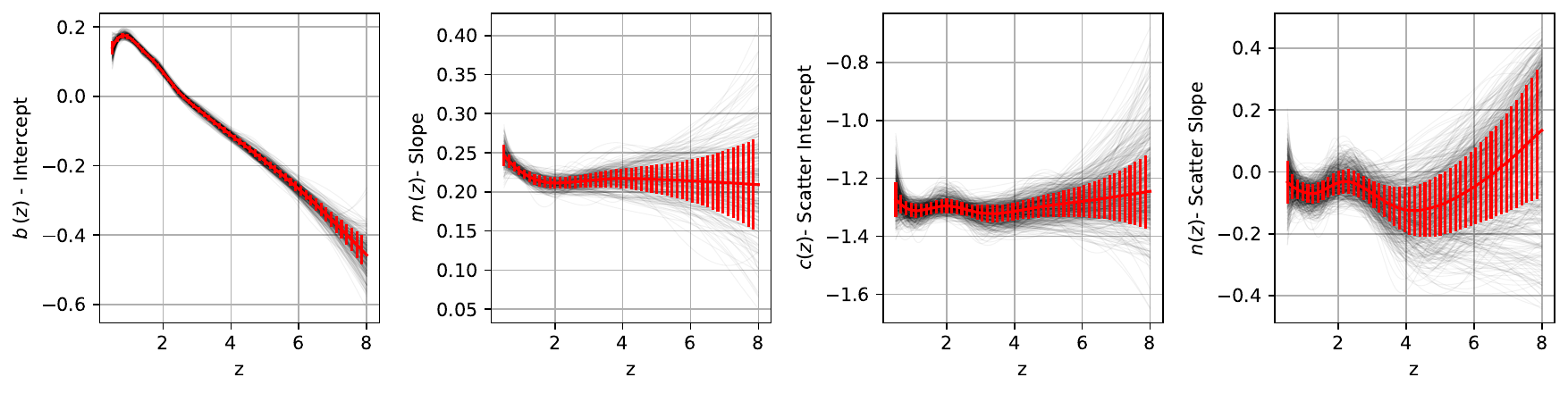}
    \caption{The measured evolution of the parameters describing the size-mass relationship for star-forming galaxies (defined in Eqn.~\ref{eqn:sf_pl} and Eqn.~\ref{eqn:sf_sig}. The red line showcases the median of the posterior with the error bars denoting the 16th - 84th percentile interval. 250 individual draws from the posterior are shown as the thin black lines. Our method utilizing B-splines to model the redshift evolution allows for complex and non-monotonic evolution while maintaining continuity and smoothness. Qualitatively we find that that the average size of galaxy, as indicated by the intercept $b$, decrease as a function of redshift, except for a plateau at $z<1$, while the slope of the size-mass relation, $m$, stays decreases from $z=0.5$ to $z=1$ but stays constant afterwards.}
    \label{fig:sf_params}
\end{figure*}

\begin{figure}
    \centering
    \includegraphics[width=0.99\columnwidth]{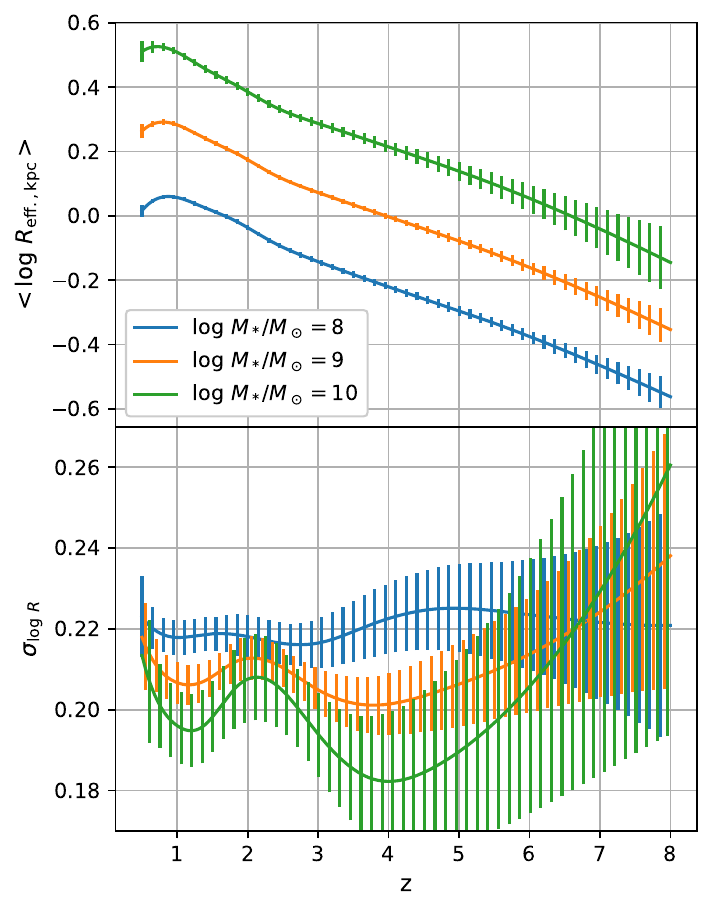}
    \caption{The average size and intrinsic scatter as a function of mass and redshift as measured in this study. The redshift evolution of the average size as function of redshift for three different masses are shown in the top panel. The error bars here denote the measurement uncertainty, not the scatter within the population. At $z>1.5$ the rate of growth is similar across all masses as the slope of the size-mass relation is constant, however at $z<1$ we find more of a turnover in low mass galaxies compared to higher masses at $\log M_*/M_\odot > 9$. The measured scatter around the size-mass relation is shown as a function of mass and redshift in the bottom panel. We find a consistent negative mass dependence of the scatter, galaxies at $\log M_*/M_\odot = 10$ show a scatter of around 0.2 dex whereas for low mass galaxies at $\log M_*/M_\odot = 8$ the scatter is measured to be 0.22 dex. However, our constraints on the scatter are weak at $z>4$ due to the limited sample size of galaxies at this redshift.}
    \label{fig:sf_res_overview}
\end{figure}

The inferred redshift evolution of the parameters describing the size mass relation are displayed in Figure~\ref{fig:sf_params}. These are: $b$, the intercept or average size at $\log M_*/M_\odot =8.5$, $m$, the slope of the size-mass relation, $c$, the normalization of the scatter and $n$, which describes the mass dependence of the scatter. We find non-monotonic but smooth evolution with redshift suggesting our penalized B-spline framework allows the flexibility necessary while maintaining realistic evolution. The intercept, $b(z)$, and slope, $m(z)$, of the size mass relation are well constrained with typical uncertainties $\lesssim 0.02$, especially at $z<6$. The parameters $c(z)$ and $n(z)$, which control the scatter show increased uncertainty at $z\gtrsim4$. The scatter is a more subtle measurement of the population and therefore $c$ and $n$ have larger uncertainties on average. This is compounded at $z>4$ where the number of galaxies, and the dynamic range in mass, decease drastically. 

The redshift evolution of these parameters, translated into physical quantities is showcased in Figure~\ref{fig:sf_res_overview}. The average sizes of galaxies generally decreases at higher redshift, except for a plateau at $z<1$. The behavior at these low redshifts depends on mass and is caused by the evolution of the slope of the size-mass relation, $m$ in Figure~\ref{fig:sf_params}. At $\log M_*/M_\odot=8$, there is an actual turnover, with the size decreasing by about 10\% between the maximum at $z=0.9$ and $z=0.5$. At higher masses there is no evidence for a decrease, only a flattening. At $z>1$, the rate of size growth is largely consistent across masses as the slope of the size-mass relation is largely constant.

The mass dependent scatter around the size-mass relationship measured in this study is displayed in the bottom panel of Figure~\ref{fig:sf_res_overview} a. We find the overall normalization of of the scatter does not evolve significantly with redshift, as shown with the evolution of the parameter $c$ in Fig~\ref{fig:sf_params}, which remains constant at a a value of $-1.3$ corresponding to an observed scatter of 0.21. The slope of the scatter, indicated $n(z)$, is also constant, within uncertainties at a value of $-0.06$. This leads to a mild, negative dependence of the scatter with stellar mass at all redshifts, however the uncertainties are large at $z>3$.  Due to the nature of the UNCOVER/MegaScience survey our sample of galaxies in this mass and redshift range is quite limited, as can be seen in Fig.~\ref{fig:sample}. It will be worth re-visiting this finding including wider field survey data, such as from the MINERVA Survey~\citep{Muzzin2025}, which will greatly increase the number of massive galaxies observed at $z>3$ with full medium band coverage.

In our fit for star-forming galaxies we measure an outlier fraction of $4.1 \pm 0.9\%$. To better understand the causes of these outliers we calculate the outlier probability of each individual galaxy by comparing the relative probability of the outlier distribution to the size-mass distribution at its specific size, mass and redshift. For galaxies with a high outlier probability, $>50\%$, we find they typically fall into two distinct regions on the size-mass plane. They are either high mass ($\log\, M_*/M_\odot \sim 10.5$) and below the measured size-mass or low mass ($\log\, M_*/M_\odot \sim 8.5$) and above the measured relationship. 

The former is consistent with quiescent galaxies that are mis-identified as star-forming galaxies. At this mass quiescent galaxies are significantly smaller and so would appear as outliers if mislabeled as star-forming. We find that about 2/3 of these galaxies lie near the quiescent galaxy selection in $ugi$ color space, supporting this hypothesis. The latter, lower mass galaxies with large radii, arise from one of two scenarios. The first possibility is a systematic issue where they are photo-z failures, i.e in reality they are actually more massive galaxies at a higher redshift. This is the most common failure mode~\citep[See Fig 5. in][]{Suess2024}, but the rate of photo-z failures is measured to be $7.2\%$, higher than our measured mass-size outlier fraction. Additionally they could be a real population of galaxies that are outliers in the size-mass plane; possibly progenitors of diffuse dwarf galaxies than have been studied in the local Universe.~\citep[e.g.][]{Danieli2019,Li2023} The location of these outliers on the size-mass plane, $\log M_*/M_\odot\sim 8.5$ and $R_{\textrm eff,opt} \sim 3$ kpc are roughly consistent with those studied in the local Universe but more work is needed to establish a potential evolutionary link.

\subsection{Quiescent Galaxies}
\label{sec:q_res}
\begin{figure}
    \centering
    \includegraphics[width=0.99\columnwidth]{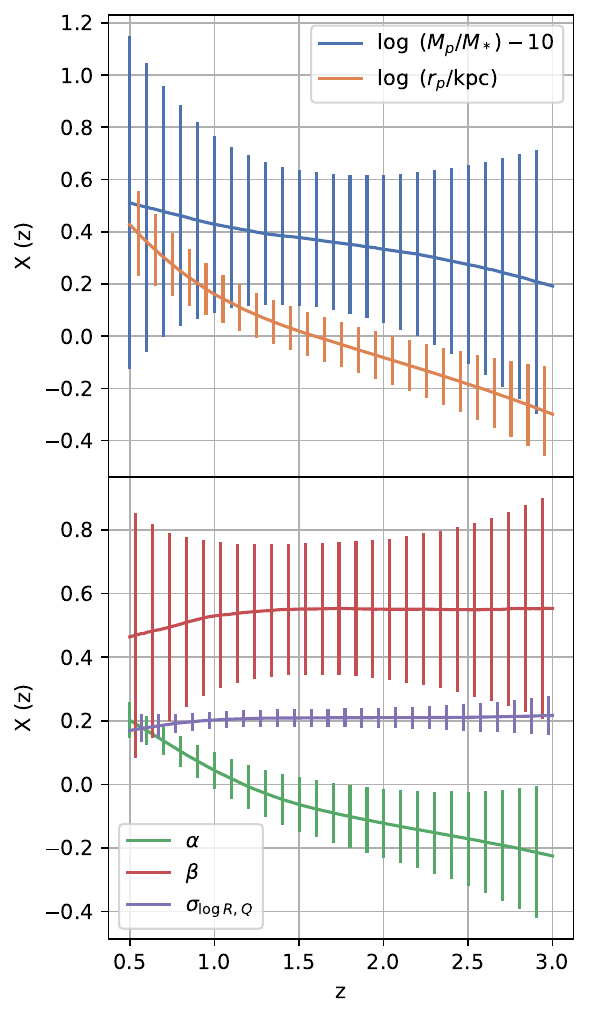}
\caption{The measured redshift evolution of the parameters defining the double-power law model for the size-mass relationship of quiescent galaxies (Eqn.~\ref{eqn:q_dpl} . In the top panel we display the evolution of the normalizations, the pivot radius, $r_p$ and mass, $M_p$. The bottom panel displays the low mass slope, $\alpha$, high mass slope, $\beta$ and the log-normal scatter. We find the pivot radius and low mass slope both decrease towards high redshift while the other three parameters show no evolution.}
    \label{fig:q_param}
\end{figure}

Figure~\ref{fig:q_param} displays the redshift evolution of the five parameters which define the quiescent galaxy size-mass relation. The top panel displays the evolution of the pivot mass and radius which define the normalization of the relationship. We find the pivot mass, $M_p$, to be constant within uncertainties across the redshift range probed, whereas the pivot radius sharply declines with redshift, mirroring the known rapid evolution in the sizes of quiescent galaxies~\citep{trujillo2006,vandokkum2008, mowla2019}. The bottom panel displays the evolution of the two slopes which define the broken power-law. The high mass slope, $\beta$, remains constant at a value of 0.53, whereas the low mass slope is positive at $z<1$ but decreases to become inverted with a negative slope of $-0.2$ at $z>1.5$. The measured scatter remains constant across $0.5<z<3$ at a value of $0.20$ 

\section{Symbolic Regression to Distill the Redshift Evolution of The Size-Mass Parameters}

\label{sec:sr}
The B-spline parameterization used in this study for the redshift evolution of the size-mass parameters offers flexibility but lacks interpretability. The main benefit is that we do not have to pre-suppose any parameterization for how the size-mass parameters vary with redshift that can bias the results. However, once we have performed the measurements using this flexible model it can be useful to distill the results down to a simple equation. A simple parameterization provides future studies an easy to implement version of the results as a benchmark or comparison. In the future we hope to use a similar framework to pursue physical interpretation, i.e. compare the redshift evolution to models for how dark matter halos grow and evolve over time. 

To this end we aim to derive a parameterization based on the results of our inference procedure using symbolic regression. Symbolic regression is a broad term for a collection of methods that aim to search for mathematical expressions that best explain the data~\citep[See a recent review here;][]{Makke2024}. Simple expressions, like addition, subtraction and exponentiation, are combined to form more complex equations. The goal of symbolic regression is often two fold: accuracy and simplicity. Which is to say less complex --- alternatively more interpretable --- equations are preferred. For this study we will employ the package \texttt{pysr} to perform symbolic regression on the redshift evolution of the size-mass parameters~\citep{Cranmer2023}. This package uses a genetic algorithm to add, remove or mutate mathematical expressions to evolve a series of equations across a range of complexities. Complexity in \texttt{pysr} is measured as the sum total of the number of operators, constants and variables included in the equation. The result of this procedure is a list of ``best fit'' equations at varying complexities.

In this work we apply symbolic regression to model the redshift evolution of all parameters that define the size-mass relationship for star-forming galaxies, $m,b$. We also apply it to two of the five parameters for quiescent galaxies, $\log\, r_p$ and $\alpha$. The other parameters, $n$ and $c$ for star-forming galaxies and $\log\, M_p, \beta$ and $\sigma_{\log R, Q}$ for quiescent galaxies, remain constant across the redshift range probed, within uncertainties. In our initial testing including these other parameters, we found the $\chi^2$ increased by less than a factor of 2 between constant values and the most complex equations. For star-forming galaxies we include an additional data-point at $z=0$ using the results for \citet{Asali2025} for their SDSS/NSA isolated sample. Without this we found the resulting equations often diverged as $z\rightarrow0$ and adding this data-point, even with inflated error bars, ensured realistic extrapolations.

Using \texttt{pysr}~\citep{Cranmer2023} we fit each of the 6 parameters considered at a series of redshift points using the median of the posterior with an L2 loss that is weighted by the inverse variance, essentially using a $\chi^2$ loss function. For each data point the variance is calculated as the square of half the 16th-84th percentile range. We run each fit for 50,000 iterations including the following binary operators: $+,-,*,/$, along with the following unary operators:$\log_{10}, \textrm{square}, \sqrt{}$ and exponential. For all the parameters except $b$ we set a maximum complexity of 15. For $b$, since it is the most well constrained and shows complex behavior we increase the maximum complexity to 18. In \texttt{pysr}, complexity is defined as the number of components of the equation, so operators, variables and numeric constants all contribute 1 complexity. We set the parsimony value to $10^{-6}$ for $b$ and $10^{-5}$ for all other variables and the adaptive parsimony scaling to 5000, to encourage greater exploration of low complexity equations.

We then must decide on a final equation for each parameter. To begin we order the list of each parameter by the ``score'' value in \texttt{pysr}, defined as $\Delta[\log( \textrm{loss})]/\Delta[\log\,( \textrm{complexity})]$. This measures the increase in accuracy per unit complexity and is used as an initial metric for finding the best equation. We consider the top three ``scored'' equations for each parameter and plot them compared to the input data. Based on how well it matches the data and its complexity we make a decision on a final equation for each parameter. We discuss this process more in Appendix~\ref{app:sr} and show the other equations found.

Once we have decided on a final equation we perform MCMC sampling on the constants within each equation to further optimize and estimate the uncertainties. This is performed again using numpyro and a NUTS sampler. We use a full covariance matrix for the loss function estimated directly from the posterior using the Ledoit-Wolf estimator~\citep{ledoit2004}. We sample from the posterior distribution using 4 chains with 1,000 warm up and 2,000 sampling steps. The final equations and optimized constants with uncertainties which describe the redshift evolution of the slope and intercept of the star-forming galaxy size-mass relationship are shown below:

\begin{equation}
    \label{eqn:b}
    \begin{aligned}
    b(z) =&\ \left( b_0 + \exp[b_1z] \right) \left(z^2 + b_2\right)\\
    b_0 =&\ (-7.4 \pm 0.3)\times10^{-3}\\
    b_1 =&\ -1.88 \pm 0.02\\
    b_2 =&\ 0.16 \pm 0.03\\
\end{aligned}
\end{equation}

\begin{equation}
    \label{eqn:m}
    \begin{aligned}
    m(z) =&\ \left[ m_0\,z^{3/2}\exp(-z) +m_1\, z^{1/2} + m_2 \right]^{1/2}\\
    m_0 =&\ (-4.2 \pm 2.6 )\times 10^{-2}\\
    m_1 =&\ (-1.3 \pm 0.5 )\times 10^{-2}\\
    m_2 =&\ (7.9 \pm 1.3 )\times 10^{-2}\\
\end{aligned}
\end{equation}

For the parameters describing the scatter around the size-mass relationship we find the best fitting equations provide no major improvement over a constant value, therefore we adopt a constant value with no redshift evolution following:

\begin{equation}
    \label{eqn:c}
    \begin{aligned}
    c(z) = -1.30 \pm 0.02
\end{aligned}
\end{equation}

\begin{equation}
    \label{eqn:n}
    \begin{aligned}
    n(z) = -0.06 \pm 0.02
\end{aligned}
\end{equation}

For quiescent galaxies we find the redshift evolution of the low-mass slope and pivot radius are well described by:

\begin{equation}
    \label{eqn:alpha}
    \begin{aligned}
    \alpha(z) =&\ \alpha_0 + \alpha_1\exp[-z] \\
    \alpha_0 =&\ -0.20\pm 0.13\\
    \alpha_1 =&\ -0.67\pm 0.27\\
\end{aligned}
\end{equation}

\begin{equation}
    \label{eqn:lrp}
    \begin{aligned}
    \log_{10}\, r_p(z)/\textrm{kpc} =&\ z*\left(r_0\exp (-z) + r_1 \right) + r_2\\
    r_0 =&\ -0.95 \pm 0.77\\
    r_1 =&\ -0.32 \pm 0.14\\
    r_2 =&\ 0.82 \pm 0.43\\
\end{aligned}
\end{equation}
The other three parameters, again, do not show significant improvement over a constant redshift evolution over $0.5<z<3$, therefore we assume they do not evolve with redshift. The values are given by are given by:

\begin{equation}
    \label{eqn:Mp}
    \log_{10}(M_p/M_\odot) = 10.36 \pm 0.26
\end{equation}

\begin{equation}
    \label{eqn:ls_q}
    \sigma_{\log R, Q} = 0.20 \pm 0.02
\end{equation}

\begin{equation}
    \label{eqn:beta}
    \beta = 0.53 \pm 0.20
\end{equation}

The uncertainties on the constants here do not account for co-variances in the posterior distributions. For the parameters with multiple constants, taking these at face value will likely result in uncertainties that are too large. We report the uncertainties here\footnote{\url{https://github.com/tbmiller-astro/Miller26_size_mass}}.

\section{Comparison with Previous Studies}
\label{sec:lit_comp}

\subsection{Differences in Methodology}
\label{sec:meth_comp}

The first and most obvious improvement our method makes is using a continuity model or global fit to the evolving size-mass relation. Instead of relying on independent bins of redshift we parameterize the redshift evolution using a smooth function to maximize constraints on the inferred parameters. The global fit also allows us to fully account for all of the uncertainties involved when fitting the size mass. It is common to include uncertainties in both stellar mass and size \citep[e.g.][]{ward2024,Allen2024} however we take it a step further by modeling the joint uncertainties between $z_{\rm phot}$ and $\log M_*$.

There have been a few examples of this style of modeling for the size-mass relationship~\citep[e.g.][]{Morishita2024,Miller2025,Song2025}. Although we are the first to allow freedom for all the size-mass parameters while studying the entire redshift frame where JWST covers the rest-frame optical. To maintain flexibility in the redshift evolution we employ B-splines. This flexibility ensures we are not biased by any a-priori parameterization yet it retains smooth and continuous behavior.

Our parameterization allows the intrinsic scatter around the size-mass relation to vary with stellar mass. This is not a novel idea, \citet{shen2003} study the mass and magnitude dependence of scatter around the size-mass relation in SDSS galaxies. In more recent works, especially those focused at higher redshifts, $z>1$, it is common to  assume that the scatter around the size-mass relation does not depend on mass. ~\citep[e.g.][]{vanderwel2014,mowla2019,kawinwanichakij2021,Allen2024,Yang2025}. 

In our model we include the possibility of outliers which do not fall on the typical size-mass distribution. Our procedure does not discriminate between the specific causes of these outliers but likely candidates are; photometric redshift failures, mis-identified star-forming vs. quiescent galaxies or perhaps physically distinct populations. In any case we do not wish for these to bias our inference of the general population. Past studies such as ~\citet{vanderwel2014} model the size-mass relation of star-forming and quiescent galaxies simultaneously and account for mis-identification of galaxies during the fitting. This accounts for one type of outlier whereas our approach is agnostic to any specific cause.

\subsection{Star-forming Galaxies}

\begin{figure*}
    \centering
    \includegraphics[width=0.99\textwidth]{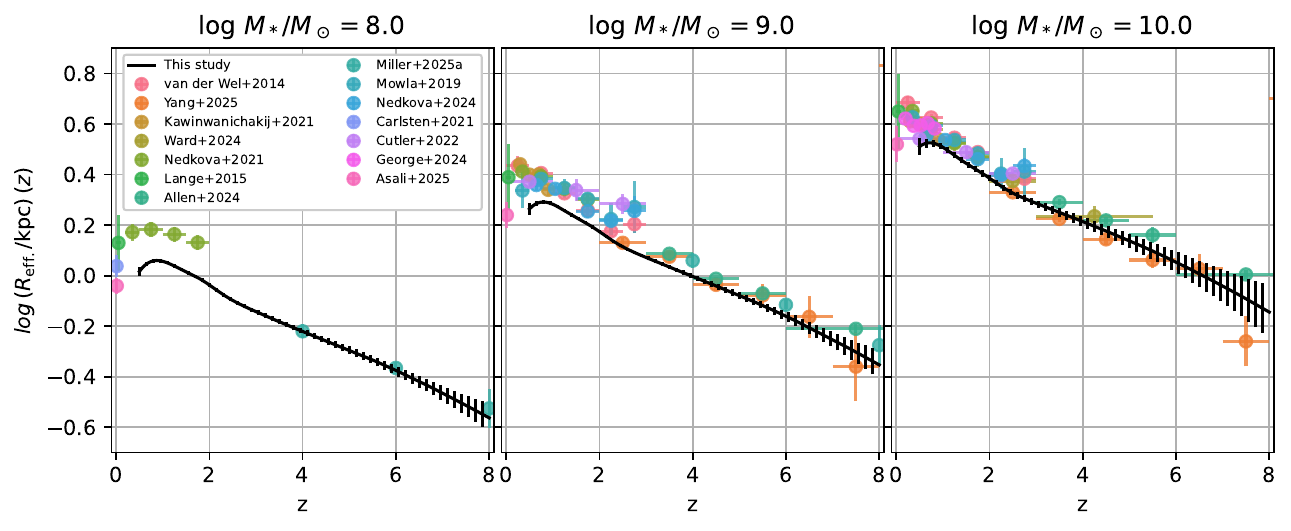}
    \caption{The evolution of galaxy sizes at fixed stellar mass from this study, shown in the black line, compared to numerous results from the literature. These are represented as colored points and include studies at $z\lesssim1$ using data from the ground~\citep{lange2015,Carlsten2021,kawinwanichakij2021,George2024,Asali2025}, out to cosmic noon $z\sim 2$ with HST~\citep{vanderwel2014,mowla2019,nedkova2021,Cutler2022,Nedkova2024} and with JWST out to $z\sim8$~\citep{ward2024,Allen2024,Yang2025, Miller2025}. Given the incredible variation in type of data and methods used among these many studies  the level of general agreement is heartening. Our results agree well with the synthesis of the literature at $z\gtrsim3$ and/or $\log M_*/M_\odot\gtrsim9$ but at low masses and redshifts we tend to find systematically lower sizes at a fixed mass.}
    \label{fig:sf_size_lit}
\end{figure*}

To place our results in context of past work, Figure~\ref{fig:sf_size_lit} compares our results on the redshift evolution of the average size of galaxies at three different stellar masses. It is worth noting that the studies included span an impressive range of methods and data used, so the level of general agreement is heartening. We find the average size of galaxies at all masses decreases with increasing redshift. However, there is a notable plateau at $z\lesssim1$ especially at low masses, $\log M_*/M_\odot \lesssim 9$. This flattening in the size growth of low mass galaxies is not often discussed in the literature but can be observed in the comparison of results using HST and JWST at $z>0.5$ to local studies using ground based imaging. The finding of this plateau is reinforced by our study using a consistent dataset and framework to model the redshift evolution.

When comparing in detail, we find that our analysis produces smaller average sizes at a fixed mass for galaxies at $z<2$. This difference is more pronounced at lower masses, with the difference between our study and \citet{nedkova2021} reaching 0.15 dex at $\log\ M_*/M_\odot = 8$. At higher masses this difference shrinks with only a minor difference, $<0.05$ dex, between our results and the literature consensus at $\log\ M_*/M_\odot = 10$. This difference also disappears at higher redshifts, at $z>2.5$ our analysis agrees, within uncertainties, with recent studies including \citet{ward2024,Allen2024,Miller2025,Yang2025}. We discuss possible causes of this offset, and other differences, below.

\begin{figure}
    \centering
    \includegraphics[width=0.99\columnwidth]{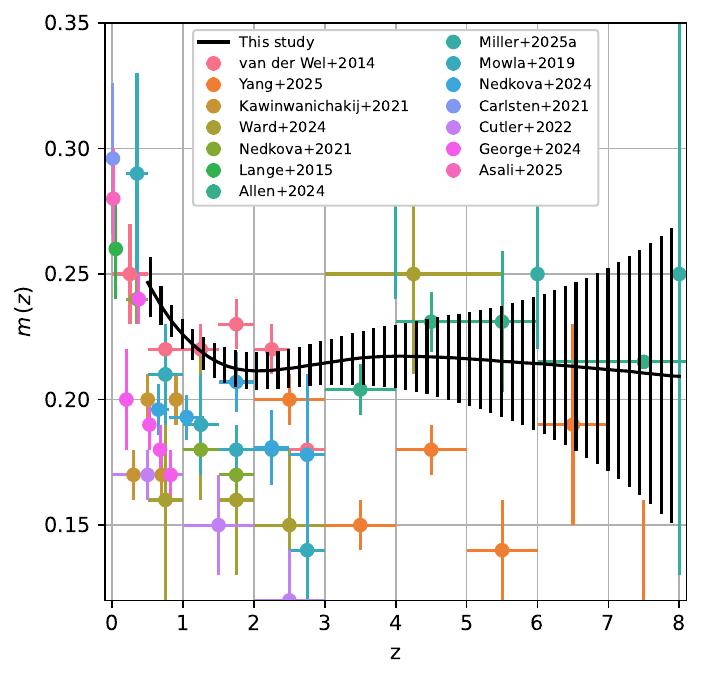}
    \caption{The power-law slope of the size-mass relationship for star-forming galaxies across $0.5<z<8$ measured in our study along with our literature comparison sample. Our results indicate mild evolution; the slope is measured to be $0.25$ at $z=0.5$, decreases to a minimum of $0.21$ at $z\sim 2$ and stays relatively constant between $2<z<8$. This is qualitatively consistent with the consensus of the literature however we find a slightly steeper slope around cosmic noon ($1<z<3$). }
    \label{fig:sf_slope_lit}
\end{figure}

The slope of the size-mass relationship measured in this study shows only minor variations across cosmic history. Displayed in Figure~\ref{fig:sf_slope_lit}, the slope at $z=0.5$ is approximately 0.25, decreases to a minimum of 0.21 at $z=2$ and remains relatively flat to higher redshift. This behavior qualitatively matches the consensus of the literature and is in good quantitative agreement at both low and high redshifts. In many studies in the local Universe, the slope is measured to be in the range of $0.25-0.3$~\citep{lange2015, Carlsten2021, Asali2025} consistent with the extrapolation of our results to $z\sim 0$. At $z>3$ our results are in quantitative agreement with previous work, except for \citet{Yang2025} who find a shallower slope. It's worth noting that the \cite{Yang2025} study is based on COSMOS-Web imaging, which is much shallower than UNCOVER/MegaScience but over a much larger area, so the mass ranges of the galaxies studies do not overlap. 

At cosmic noon, $1<z<3$ we find a steeper slope of the size-mass relation compared to the bulk of the previously published values. Over this redshift range the slope we measure is roughly constant 0.21. This is consistent, within uncertainties with a few studies, including \citet{vanderwel2014}, but the consensus of studies in the literature is a shallower slope with an average value of $\sim 0.17$. This difference in slope is intertwined with the mass-dependent average size offset that we observe in Fig.~\ref{fig:sf_size_lit}. There are many differences between these studies, including the type of data and methods used. Below we discuss likely causes of these discrepancies.

\begin{figure}
    \centering
    \includegraphics[width=0.99\columnwidth]{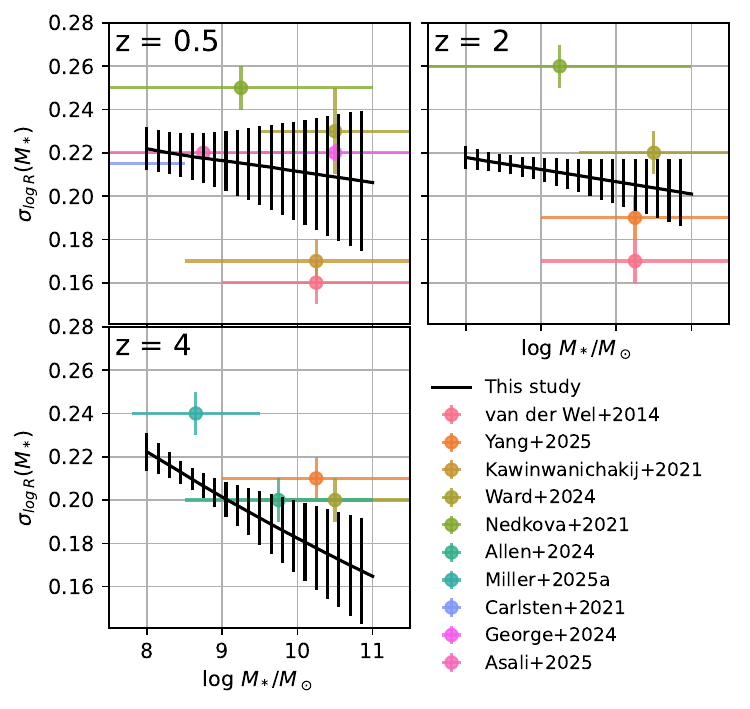}
    \caption{Showcasing the evolution of the stellar-mass dependent scatter around the size-mass relationship measured in this paper. Three redshifts are displayed across the three panels with our results (black lines) compared to previous studies in the literature ~\citep{vanderwel2014,Carlsten2021,kawinwanichakij2021,nedkova2021,Allen2024,ward2024,Yang2025,Miller2025,Asali2025}. At $z<1$ we find a flat relationship between scatter and stellar mass that steepens at higher redshift, controlled by the evolution of the parameter $n(z)$ seen in Fig.~\ref{fig:sf_params}. For the literature studies we also show the approximate mass ranges probed in each study. We find broad agreement between our study and the literature.}
    \label{fig:sf_slope_sig}
\end{figure}

We find largely consistent results when comparing the scatter around the size-mass relationship to the previous literature in Fig.~\ref{fig:sf_slope_sig}. It is worth emphasizing that the measurement of the scatter is dependent on accurate uncertainties on the measured masses and sizes of the galaxy sample. Thus it is prone to unknown or unaccounted for systematic uncertainties that can vary from study-to-study based on the data and methods used. All of the studies compared in Fig.~\ref{fig:sf_slope_sig} generally agree, within about 0.02 dex, but it is  difficult to proceed with more detailed comparisons without a much deeper investigation of the systematic uncertainties affecting each study.

To attempt to explain the differences between our study and previous works we consider differences in the measurement of the size and stellar mass of galaxies along with our inference method. The first possibility we consider is in the measured sizes. This would likely cause the biggest differences between our work and those using non-JWST based sizes given the difference in resolution and depth. However we don't think that this is likely the cause of the disagreement. First the difference is a strong function of mass and redshift, whereas for a difference in the measured sizes one would expect a consistent systematic offset. Studies such as~\citet{Martorano2023} which have directly compared the results of S\'{e}rsic fitting applied to HST and JWST do not find systematic differences. Additionally studies such as~\citet{ward2024} find consistent results at $z<2$ with HST based studies~\citep{vanderwel2014,mowla2019} in the size evolution of star-forming galaxies. Another distinction is we are in a lensing field, whereas many other studies are performed in non-lensing legacy fields. This could cause systematic differences in the lensing correction, for example. However other studies in lensing fields, such as~\citep{nedkova2021}, do not find similar offsets.

The more likely culprit, we think, is differences in stellar mass measurements. Since we are comparing the evolution of sizes at a fixed stellar mass, and there is a consistent relationship between size and mass, systematic differences in stellar mass measurements can lead to differences in the implied size evolution. We use the \texttt{prospector-}$\beta$ model to measure stellar masses which utilizes a non-parameteric star-formation history (SFH)~\citep{leja2017} along with an evolving redshift-dependent prior~\citep{Wang2023}. Compared to the traditional parametric star-formation histories, like the delayed-tau model, the non-parametric SFH is more extended~\citep{leja2017}. This leads to more mass being put into older stellar populations, with a high M/L ratio, leading to a higher inferred stellar mass.\citep{leja2019}. Crucially the magnitude of this difference depends strongly on the shape of the SFH, therefore varies as a function of mass and redshift. In particular, \citet{leja2019} show that this effect is stronger at lower masses and lower redshifts, which matches the trends in difference between size in our study compared to previous works which employ delayed-tau models to measure stellar masses. \citep[e.g.][]{vanderwel2014,kawinwanichakij2021,nedkova2021,ward2024}

To test if this difference in stellar mass measurements can explain the offset in size-mass parameters we create a fake population of galaxies, apply a systematic offset, and re-measure the size-mass parameters. We test this at two redshifts; $z=0.75$ and $z=2$. We begin with the distribution of stellar mass in our sample with $\Delta z = 0.25$ of each redshift and assign sizes following a perfect power-law using the parameters found in Sec.~\ref{sec:sf_res}. We then apply the median offset to the stellar masses measured using prospector, used in our study, and FAST, similar to previous literature, found in~\citet{leja2019} (Figure 3 in their paper) as a function of mass and redshift along with an additional scatter of 0.2 dex. We then re-measure the size-mass relation using a simple least squares approach. 

Our tests show that systematic differences in stellar mass measurements shift the studies closer together but cannot fully explain the observed differences in the sizes of galaxies at a fixed mass. At $z=0.75$ we find that the intercept of the size-mass relationship increases from 0.17 to 0.23, while the slope stays roughly the same leading to a consistent offset of 0.06 dex. This eases a lot of the tension at $\log M_*/M_\odot > 9$ but does not fully explain the difference at lower masses. At $z=2$, where the offset in stellar mass measurements is lower on average, we find that the intercept increases from 0.07 to 0.1. However the effect is a stronger function of mass so the slope decreases from 0.21 to 0.19, more in line with the bulk of the literature, however the systematic offset in sizes still persists.

\begin{figure}
    \centering
    \includegraphics[width=0.99\columnwidth]{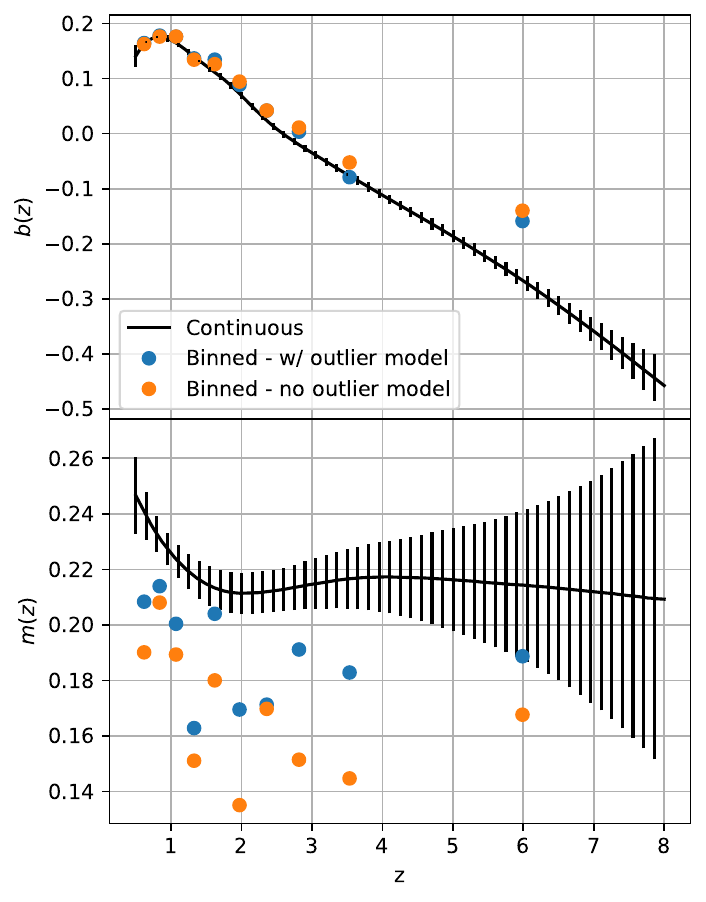}
    \caption{Showcasing variations on our method to infer the size-mass relationship and the impact on the measured parameters. In particular we focus on fitting the relationship in separate redshift bins, rather than our continuous evolution model, along with excluding our treatment of outliers. These two changes, which would better match methods more commonly used in the literature, could help explain the differences in slope we have observed in our study.}
    \label{fig:binned_comp}
\end{figure}

Our inference procedure for measuring the parameters of the size-mass relationship is distinct enough that it could affect the measurements. The two key differences we identify are the continuous model, rather than considering galaxies in bins of redshift, and the modeling of outliers. To test differences in modeling choices we fit the size-mass relationship to galaxies binned by redshift. We account for uncertainties in the size and mass in the loss but assume that they are simply Gaussian. We test two methods, one including a model for outliers, similar to what is implemented in Sec.~\ref{sec:method}. We use 10 bins, space in redshift such that each contains approximately 900 galaxies. These results are shown in Figure~\ref{fig:binned_comp}. Fitting galaxies in discrete bins has little systematic effect on the intercept leads to a lower slope, which is further decreased if an outlier model is not included. This binned redshift slope is more consistent with the consensus of the literature, as seen Fig.~\ref{fig:sf_slope_lit}.

One explanation for this offset in slope is asymmetric scattering of galaxies from lower redshift bins due to photo-z uncertainties; since the number density of galaxies decreases with redshift more scatter in from below than above. Additionally size generally decreases with redshift and these are most likely to be low-mass galaxies, given the mass distribution of our sample and that they tend to have larger photo-z uncertainties caused by lower S/N photometry. This leads to an excess of larger galaxies at low masses, biasing the slope. Additionally given the outliers in our inference tend to be large low mass galaxies or small high-mass galaxies, not accounting for these will also decrease the measured slope.

Overall we find that the differences in our inferred size-mass parameters to the previous literature can largely, but not fully, be explained by these two differences. The difference in slope between our value and the literature can be well explained by the change from galaxies binned by redshift to a continuous model. The difference in stellar mass measurements, leading to a systemic shift in the observed sizes of galaxies at a fixed stellar mass, is an appealing explanation for the offset because it is mass and redshift dependent, similar to the difference in size at a fixed mass. However, our tests indicate this may only explain roughly half of the difference, meaning there is likely another unknown systematic difference causing the observed difference.

\subsection{Quiescent Galaxies}
\begin{figure*}
    \centering
    \includegraphics[width=0.99\textwidth]{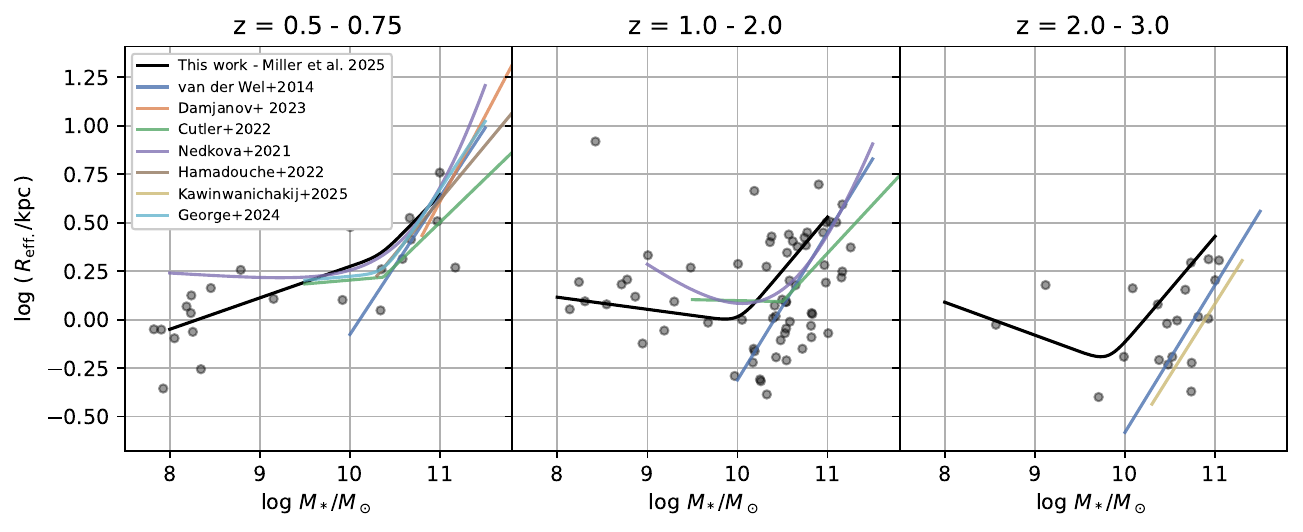}
    \caption{The evolution of the size-mass relationship for quiescent galaxies displayed across three panels. Individual galaxies from our sample are shown as points with our measured relationship as a black line. We include size-mass relationships measured from previous studies.~\citep{vanderwel2012, Hamadouche2022,nedkova2021,Cutler2022,Damjanov2023,Kawinwanichakij2026} We find good agreement with these previous studies including a flattening of the size-mass relationship for quiescent galaxies at $\log\, M_*/M_\odot < 10$, which inverts at $z>2$.}
    \label{fig:q_lit}
\end{figure*}

Our measured size-mass relation for quiescent galaxies is compared to previous measurements of in Figure~\ref{fig:q_lit}. The evolution of the low-mass slope, $\alpha$, we find is consistent with studies in the local universe that find a slope of $\sim0.3$~\citep{Carlsten2021,Asali2025} and studies at higher redshift which find a flat or inverted slope at $z\gtrsim1$.~\citep{nedkova2021,Cutler2022,Cutler2024}. Our measured high mass slope is shallower than studies that fit a single power law and focus exclusively on massive quiescent galaxies~\citep{vanderwel2014,mowla2019,Hamadouche2022,Damjanov2023,Kawinwanichakij2026}. This difference can likely be attributed to the different mass ranges probed and parameterizations. As can be seen in Fig~\ref{fig:q_lit} these published relationships match the size-mass distribution of high-mass quiescent galaxies in our sample well. 

In this study we use a broken power-law to describe the size-mass relationship for quiescent galaxies which is continuous across the entire mass range. This is inspired by previous studies~\citep{mowla2019b,kawinwanichakij2021, Cutler2022} and appears to be an adequate description of the data at intermediate redshifts, $z\sim 0.5-1$ but may break down at higher redshifts. At $2<z<3$ the low mass slope becomes inverted to match the galaxies with $\log M_*/M_\odot\sim 9$ and may be less accurate for galaxies near the pivot mass. However, there are only a handful of galaxies in our sample at this redshift, especially at low masses, which limits the constraining power and possible complexity we can add to the parameterization. It is possible that two separate, independent, relationships for low and high masses would provide a better description. Physically this would be realistic if, as suggested in \citet{Cutler2024} and \citet{Hamadouche2025}, there are two distinct classes of quiescent galaxies possibly with separate formation mechanisms. In future studies, including a larger sample of quiescent galaxies from the on-going MINERVA survey~\citep{Muzzin2025}, we hope to explore alternative parameterizations.

\section{The Evolution of the Size-Mass Distribution Across Cosmic History}
\label{sec:disc}

\subsection{The Differences in Sizes Between Star-forming and Quiescent Galaxies}

\begin{figure*}
    \centering
    \includegraphics[width=0.99\textwidth]{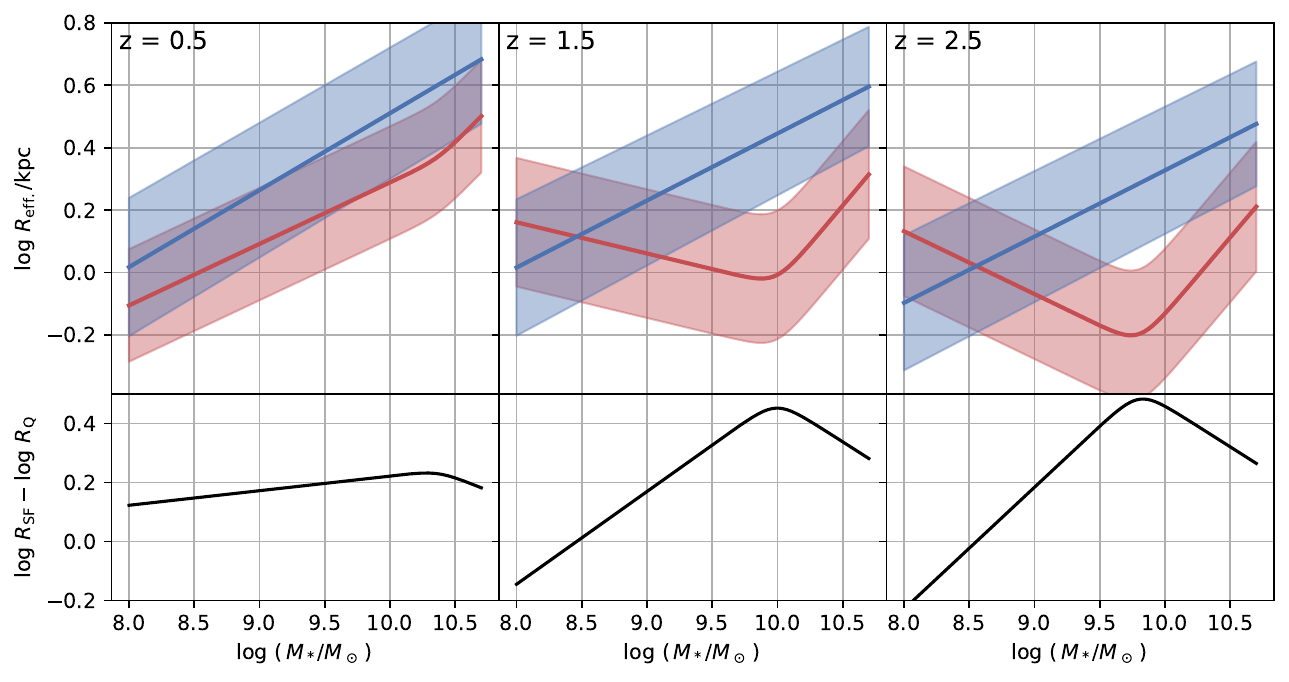}
    \caption{Comparing the size-mass relationship of star-forming and quiescent galaxies at three different redshifts. We use the median value of the posterior for each parameter and show star-forming galaxies in blue and quiescent in red. The shaded regions show the 1-$\sigma$ intrinsic scatter around each relation. The bottom panels display the difference between the two relations as a function of stellar mass.}
    \label{fig:sf_q_comp}
\end{figure*}

When studying the size-mass relationship it is common practice to separate star-forming and quiescent galaxies as they appear to follow two separate relationships. This is interpreted as two distinct populations being affected by separate physical mechanisms~\citep{vandokkum2015, suess2021} with quiescent galaxies being consistently smaller on average. This has led to empirical relationships between central density and quenching,~\citep{Franx2008,whitaker2017}. This difference can be seen in Figure~\ref{fig:sf_q_comp}, where the difference in size is largest around the pivot mass $\log M_*/M_\odot\sim10$ with a difference of around 0.4 dex which increases at higher redshift. At lower masses we re-affirm the results of previous studies~\citep{nedkova2021,Cutler2022,Cutler2024} that the sizes of star-forming and quiescent galaxies at $z\gtrsim1$ are similar. When combined with the observations of local dwarfs which find similar sizes between star-forming and quiescent galaxies \citep[e.g.][]{Carlsten2021,Asali2025}, the implication is that at all redshifts there is little size difference at lower masses ($\log M_*/M_\odot \lesssim9$). Additionally Wide area surveys have found that the sizes converge at high masses with $\log M_*/M_\odot \gtrsim 11$ ~\citep{kawinwanichakij2021,mowla2019}. Taken together with this work it suggests that quiescent galaxies are only smaller than star-forming galaxies at $\log M_*/M_\odot \sim 10-10.5$.


\subsection{Stellar-Mass Dependent Scatter}

In this study we find that the scatter around the size-mass relationship depends on stellar mass, opening new avenues to study the physics of galaxy formation. At all redshifts we find a lower scatter at higher masses. This is qualitatively consistent with what ~\citet{shen2003} observed in the local Universe. In detail they parameterize the dependence of scatter on stellar mass as a pseudo-step function whereas we model the dependence as linear. 

The dependence of the intrinsic scatter on stellar mass can be interpreted through the lens of the galaxy-halo connection or baryonic physics. \citet{shen2003} interpret their results using a model for the galaxy-halo connection and disk formation following~\citet{mo1998}. In order to account for the mass-dependent scatter the authors need to vary the bulge-to-disk ratio with galaxy spin. Alternatively if we assume that the ratio between virial radius and half-light radius is constant this could be a reflection of an increase in scatter of virial radii (and by extension halo mass) at a fixed stellar mass at lower masses. There is some observational ~\citep{Smercina2018,Danieli2023} and theoretical evidence~\citep{Matthee2017, pillepich2018,Munshi2021} of a mass dependent scatter although the results are mixed and there is not a consensus~\citep[see Fig, 8 in][]{Wechsler2018}. Additionally it is often discussed as scatter in stellar mass at fixed halo mass, whereas scatter in the size-mass relation may be probing scatter in halo mass at fixed stellar mass, in the orthogonal direction, further complicating comparisons. For now this is mostly speculative but in the future we hope to combine our results with galaxy-halo connection modeling \citep[e.g.][]{kravtsov2013,somerville2018,Behroozi2022} to further constrain the relationship between stellar radii and halo properties.

Alternatively, baryonic processes could be responsible for the increase in scatter at low masses. Bursty star formation histories, with large variations in the instantaneous SFR, are thought to be able to drive time-varying gas inflows and outflows that can dramatically affect the observed size of a galaxy. \citet{elbadry2016} find that in the FIRE simulations, short term variations in the sizes of galaxies can account for nearly all of the scatter around the size-mass relation This effect is strongest in galaxies with lower masses, therefore smaller potential well where the time varying potential of the gas, can have a larger impact. However~\citet{Geda2025} find much more gradual size evolution. \citet{emami2021} finds evidence for a correlation between bursts in star formation and increased radii in local dwarf galaxies. The difficulty in correlating these observables is that they may not be contemporaneous; outflows may not be perfectly correlated with peaks of star-formation and the time-varying potential needs time to cause migration of stars to larger orbits. Using spectroscopic and photometric constraints from JWST we can now begin to constrain burstiness across a wider redshift and mass range~\citep[e.g.][]{Clarke2025,Mintz2025,Simmonds2025}. In future studies we plan to use these methods combined with the morphological catalog to directly test the connection between star-formation histories and morphology.

\subsection{Galaxy Size Growth Spanning 13 Gyr}

\begin{figure*}
    \centering
    \includegraphics[width=0.99\textwidth]{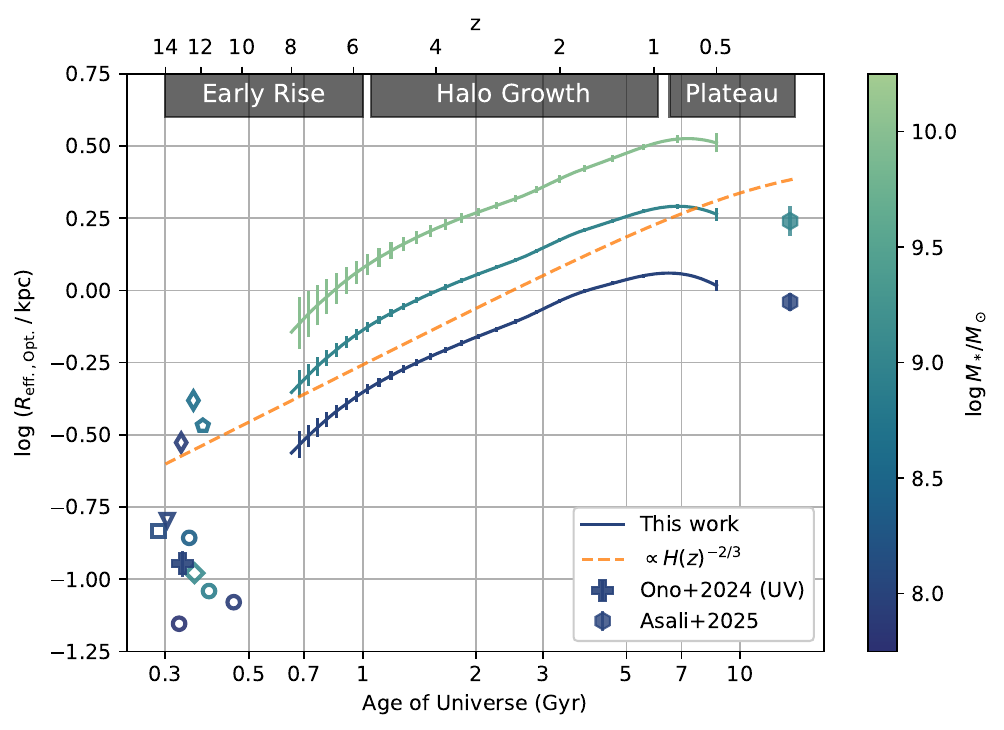}
    \caption{The evolution of the sizes of star-forming galaxies at a fixed mass across cosmic history. Results from this study are shown as lines with the color corresponding to the stellar mass. The error bars show uncertainty on the measurement, rather than the scatter within the population. We supplement this work with measurements of low-redshift galaxies from~\citet{Asali2025} as the filled hexagons and a compilation of galaxies at $z>10$ including individually reported galaxies as open points: \citet{Naidu2025} as a square, \citet{finkelstein2022} as a pentagon, \citet{Carniani2024} as trianlges, \citet{Wang2023b} as thin diamonds, \citet{Castellano2024} as a diamond, \citet{Robertson2023} as circles. A population study by~\citet{Ono2025} is displayed with the filled $+$. Note that these measurements are made in at rest-frame UV wavelengths however at this epoch there is little evidence for significant variation in size as a function of wavelength.~\citep{Treu2023,Miller2025} The thin orange line displays the relative evolution of $H(z)^{-2/3}$ which is related to the growth of the virial radius of halos and has been linked to the growth of galaxies sizes~\citep{vanderwel2014}. The figure displays a complex and non-monotonic evolution in the sizes of galaxies. We identify three distinct phases: Rapid growth at $z>5$, gradual growth, likely related to the halo at $5>z>1$, and a late plateau at $z<1$.}
    \label{fig:z_evo_SF}
\end{figure*}

In this study we have used the global morphology catalog of Abell2744 from \citet{zhang2026} and modeled the size-mass relationship of star-forming galaxies from $0.5<z<8$, 8 billion years of cosmic history. In Figure~\ref{fig:z_evo_SF} we plot the evolution of the size of galaxies measured in our study at three different stellar masses values, $\log\, M_*/M_\odot = 8,9,10$, as lines denoted with different colors. The uncertainties shown are measurement uncertainties on the size evolution, not the scatter intrinsic to the population. 

To connect our study to the beginnings and end of galaxy formation, spanning the entire $>13$ billion year history of the Universe, we have added additional data-points at $z>10$ and $z\sim 0$. For the local populations we show measurements of the size-mass relationship from~\citet{Asali2025} using data from the SAGA survey with filled hexagons. At $z>10$ we show several individual galaxies with data taken from each discovery paper as open data points, along with a recent comparison from~\citet{Ono2025} as the filled ``+'' sign. It's worth noting that due to the wavelength limits of NIRCam on JWST these sizes are measured at UV wavelengths, $\lambda_{\rm rest}\lesssim 0.2\, \mu m$, compared to optical wavelengths for this study. Observationally there has not been evidence for a systematic difference between UV and Optical sizes in the early Universe~\citep{Treu2023,Miller2025} but there are some theoretical predictions that size should vary with wavelength at these masses and epochs.~\citep{marshall2022}. For the comparison samples, similar to the lines denoting the results from this study, the error bars denote measurement uncertainty, not scatter within the population. It's worth noting that while the combination of these studies with our work seems to provide a consistent picture there still may be unknown systematic differences due to the differences in data available and methods used.

Figure~\ref{fig:z_evo_SF} reveals a complex growth history of galaxy sizes throughout cosmic history. When comparing our results out to $z\sim8$ to studies at $z>10$ we find a rapid growth in the sizes of galaxies in the first 1 Gyr of the Universe. As discussed in ~\citet{Miller2025} this is suggestive of a different mode of galaxy growth at these early epochs. There is growing evidence from strongly lensed galaxies that -- at least some fraction of -- galaxies at this epoch are dominated by a several to a few dozen isolated clusters or clumps~\citep{Adamo2024,Mowla2024, Bradac2025, Claeyssens2025, Fujimoto2025, Nakane2025}. Recent theoretical work on the formation of galaxies at $z>8$, such as feedback-free bursts~\citep{Dekel2023,Li2024} or super-Eddington outflows~\citep{Ferrara2023,Ferrara2024} predicts formation in dense clusters, with sizes similar to what is observed. This is in contrast to latter epochs where star-formation is observed to be more spatially homogeneous. If this style of highly clustered star-formation is indeed common among all galaxies at this epoch then the physics behind their growth, e.g. merging and spatial distribution of these clumps within galaxies, likely plays an important role in the evolution of the overall sizes.

At $5>z>1$ we find the growth rate of galaxy sizes roughly matches that of dark-matter halos in the $\Lambda$CDM framework. This is displayed as the orange line which shows the relative evolution of $H(z)^{-2/3}$, which describes the evolution of the virial radius of a halo. We will be directly comparing the virial radius to the projected radius at optical wavelength however it is likely more appropriate to relate it to the 3-D mass-weighted radii. The conversion from projected to 3-D radii is generally assumed to be a simple scalar conversion~\citep{vandeven2021, degraaff2022}, but the conversion from light- to mass-  weighted radii is more complicated, and may evolve with redshift~\citep[see][for further discussion]{suess2019,miller2022,vanderwel2024}. 

The equivalence in the evolution of virial radius and galaxy radius was observed in~\citet{vanderwel2014} for galaxies at $\log\,M_*/M_\odot = 10.3$ at $z<3$. We find this behavior extends down to $\log\,M_*/M_\odot = 8$, but it does not continue at redshifts below 1. The connection between galaxy and halo sizes has been popular since the initial theories of disk formation~\citep{Fall1980,mo1998}. In detail this model fails to explain observations of the connection between halo and galaxy radii~\citep{jiang2018,Zanisi2020,Behroozi2022}. However there are numerous observational ~\citep{shen2003, kravtsov2013,somerville2018,mowla2019b,kawinwanichakij2021} and theoretical~\citep{Rohr2022} studies which show an intimate connection between halo and galaxy radii. In the context of our study which provides a consistent measurement of the size-mass relation across an unprecedented range of mass and redshift it will be interesting to revisit the connection.

At $z<1$ our results indicate a flattening in the evolution of galaxy sizes at a fixed mass. This plateau is corroborated by the recent results of~\citet{Asali2025}, shown as hexagons, which are consistent with the extrapolation of results and are lower than the peak value at measure in our study at $z\sim 0.75$. This lack of evolution in the sizes of low-mass star-forming galaxies was previously noted in~\citet{nedkova2021}. This result stands in contrast to the common wisdom that average sizes of galaxies at a fixed mass decreases monotonically as redshift increases. In our study we find that this flattening is present up to at least $\log M_*/M_\odot = 9$. We find the slope of the size-mass relation increases across the same redshift range so this behavior is likely not for higher mass galaxies, $\log M_*/M_\odot \gtrsim 10$. 

Since these galaxies are still actively star-forming, whatever physical mechanism that drives this stagnation likely shifts the typical inside-out formation to a self-similar or outside-in spatial configuration. \citet{Nelson2021} find that galaxies at this mass and redshift range ($\log\,M_*/M_\odot \sim 9$ and $z\sim1$) show flat specific-SFR profile in both observations and simulations, consistent with our finding of a lack of size growth. Recently~\citet{Matharu2025} find rising profiles in EW(Pa$\alpha$) and EW(H$\alpha$) profiles in galaxies at similar mass and redshifts, indicative of continued inside-out growth and somewhat at odds with previous results. This shift in radial distribution of stellar populations can also lead to color gradients that can cause evolution of the observed sizes of galaxies without altering the underlying mass distribution~\citep{suess2019,Miller2022b,Martorano2025}. It's worth noting that these two studies study stacks of galaxies which can wash out a lot of scatter in the spatial distribution of star-formation in individual galaxies~\citep{Orr2017}. The cause of this shift in the spatial distribution of star-formation is still to be investigated. 

Theoretically there have been different explanations for the change in the nature of star-formation and galaxy growth at these epochs related to the change in the balance of gas-inflows and stellar feedback.~\citet{Stern2021} examine the properties of the circumgalactic-medium (CGM) in the FIRE simulations and find a drop in temperature in the inner CGM promotes stable disk formation stopping the bursty star-formation phase by allowing smooth accretion and confining outflows caused by feedback. \citet{Hopkins2023} perform a set of idealized simulations and suggest that when the gravitational potential of the galaxy and halo becomes large and centrally concentrated enough it stabilizes a disk that is more resilient to stellar feedback. This causes a shift away from a chaotic and bursty mode of galaxy growth to be more stable and ordered. Crucially either of these scenarios leads to ordered disks which are efficient at funneling gas to the center of the galaxy leading to a stagnation or decrease in the observed radii of galaxies. Further understanding the onset and mass-dependence of this plateau in the growth of sizes will help us learn about how feedback affects galaxy formation at intermediate redshifts.

Environmental effects are another possible physical mechanism affecting the size evolution of low-mass galaxies. It has been known for decades that local environment can affect a galaxy's morphology~\citep[e.g.][]{Dressler1980,Postman1984} and lower mass galaxies at lower redshift are more likely to be satellites, however we disfavor this interpretation for the cause of the plateau.  First the direction of the effect may be wrong. Recent observational~\citep{Ghosh2024} and theoretical~\citep{Mercado2025} studies suggest that galaxies in denser environments are likely larger. Additionally, according to the stellar-to-halo mass modeling performed by~\citet{Shuntov2022}, only about 30\% of galaxies at this epoch are satellites and it is unclear if this is enough to meaningfully alter the observed relations. Additionally one might expect an increase in scatter caused by a distribution of local environments at later times as the Universe becomes more clustered, however this is not observed in our modeling. 

We discuss three phases in the growth of galaxy sizes: An early rapid rise at $z>5$, sustained growth at $5<z<1$, that we theorize is related to the growth of dark matter halos, and a plateau at $z<1$. This is a complex and non-monotonic picture for the evolution of galaxies that is not well captured by simple parameterizations. In particular it is common to parameterize the size evolution as a power law in $1+z$, or the inverse scale factor, or $H(z)$, the evolving Hubble constant. The benefit of these parameterizations is they are directly related to physical quantities and so can be easily compared to models, e.g. for the growth of dark matter halos. However they are not complex enough to accurately model the size evolution across 13 Billion years of cosmic history. It may be that for certain masses and redshifts the size evolution can be well described by power laws of these quantities but as a whole they are not sufficient.

\section{Conclusions}
In this study we utilize the catalog of morphology in the field of Abell 2744 presented in \citet{zhang2026} to measure the evolving relationship between optical size and stellar mass. Our aim is to provide an updated benchmark for the size-mass relationship using the 20 bands of NIRCam coverage provided by the combination of the UNCOVER and MegaScience surveys that provide exquisite stellar masses, photometric redshifts and multi-band morphology~\citep{Bezanson2022,Suess2024}. We model the size-mass relationship of 8524 star-forming galaxies with a power-law that evolves across $0.5<z<8$, eight billion years of cosmic history, with a mass range of $\log\, M_*/M_\odot = 7.5-11$. For quiescent galaxies we employ a broken power-law to model the size-mass relationship using 137 galaxies over  $0.5<z<3$. 

We implement a number of improvements in measuring the size mass relationship. The first is allowing for additional flexibility in modeling the intrinsic scatter of the relation by adding in a mass-dependence. While modeling the relation we forgo the traditional method of assigning galaxies into discrete bins of redshift and consider all galaxies simultaneously using a ``continuity'' model. For both populations the redshift evolution is described using B-splines that are flexible and free of any a-priori parameterization. We additionally account for all related uncertainties, including in the photometric redshifts, and model outliers to be robust against anomalous populations (whether physically meaningful or not). The B-splines provide a flexible method of inferring the redshift evolution but they lack in interpretability and portability. To provide simple representations we use \texttt{pysr}~\citep{Cranmer2023} to perform symbolic regression on all of the parameters of the star-forming and quiescent size-mass relationships. These results are presented in Equations~\ref{eqn:b}-\ref{eqn:beta}. Our results for the B-spline continuity model, the parameterizations a along with example scripts are uploaded github\footnote{\url{https://github.com/tbmiller-astro/Miller26_size_mass}}.

While our results qualitatively agree with the census of the literature for star-forming galaxies there are notable systematic differences in detail. First we find a lower overall normalization of the size-mass relationship, with the difference being larger at lower masses and redshifts. Second we find a slightly steeper slope to the size mass relation at $1<z<3$, 0.05 larger than the literature average. We discuss these differences and find that they can mostly be explained by the combination of two differences: Our use of non-parametric star-formation histories when performing SED fitting and using a continuous evolution model instead of redshift bins. For quiescent galaxies we find largely consistent results with previous studies although differing parameterizations and the low numbers in this study prevent more detailed comparisons.

The breadth of our galaxy sample and quality of data used, along with the methodological improvements lead to key insights into the evolution of galaxies across cosmic history. Given our data come from deep observations in a lensing field we are sensitive to low mass galaxies ($\log\, M_*/M_\odot \sim 8$) across cosmic history. Using this large statistical sample of low mass galaxies we investigate the stellar mass dependence of the intrinsic scatter around the size-mass relation. Our results suggest that there is lower scatter a higher masses, mirroring results in the local Universe~\citep{shen2003}. One potential explanation is that this difference is due to different star-formation histories with low-mass galaxies displaying burstier histories on short timescales. This opens up a new avenue to studying the connection between star-formation and morphology in the early Universe.

We analyze the difference between star-forming and quiescent galaxies as a function of mass and redshift. When synthesized with previous results we find that quiescent galaxies only appear smaller than their star-forming galaxies at around $\log\, M_*/M_\odot = 10$. At both higher and lower masses the sizes converge.

Finally the wavelength coverage of NIRCam allows us to study rest-frame optical morphology across $0.5<z<8$, eight billion years of cosmic history, using the same dataset. When combined with results in the local Universe~\citep[e.g.][]{Asali2025} and UV morphology at higher redshifts~\citep[e.g.][]{Ono2025} we can construct the size growth of galaxies spanning the entire history of the Universe. This combination presents a complex growth history that is not well described by simple parameterizations. We suggest that there are three distinct phases:
\begin{enumerate}
    \item $14>z>5$ - Early rapid growth in the sizes of galaxies during the first Billion years of the Universe. While the physics behind this growth remain uncertain, the growing number of strongly lensed galaxies at this epoch which show star-formation which is concentrated in clumps or clusters may provide clues.
    \item $5 > z > 1$ - Continued, sustained growth which matches the evolution of $H(z)^{-2/3}$, describing the evolution halo's virial radii. This correlation between the growth of galaxy and halo radii is well studied~\citep{kravtsov2013,vanderwel2014,somerville2018, mowla2019b,kawinwanichakij2021} and suggests an intimate connection. 
    \item $1>z>0.5$ - A late plateau, and possible contraction, in the average sizes of galaxies at a fixed stellar mass. This is a qualitative shift in the evolution of galaxy sizes. We discuss this observation in the context of theoretical work~\citep{Stern2021,Hopkins2023} that examines the shift from a chaotic and bursty mode of galaxy formation to a stable, ordered one.
\end{enumerate}

These results provide a new benchmark for the relation between optical size and stellar mass to build on in future studies. The exquisite SED coverage provided by UNCOVER/MegaScience combined with methodological improvements in measuring sizes~\citep{zhang2026} and this paper combine to provide the unprecedented constraints on the evolution of the size-mass relationship across 8 billion years of cosmic history. Due to the relatively small volume probed by UNCOVER/MegaScience we do not have significant statistics to study rare populations in detail such as massive and/or quiescent galaxies. Our work will be complimented by those studying wide-area surveys such as the widest surveys, like COSMOS-Web ~\citep{Casey2023,Yang2025} or the on-going MINERVA survey~\citep{Muzzin2025} which provides similar level of NIRCam medium band coverage across a significantly larger area.

In the era of JWST it is possible to study the morphological evolution of galaxies across their entire life-cycle. Our work represents a step in this process using an established standard of optical morphology. Further work building off of this census, such as studying wavelength dependent morphology \citep{suess2022} and color gradients~\citep{Miller2022b,Martorano2025}, will provide further insight into the structural evolution of galaxies and the physical processes which shape them.

\begin{acknowledgments}
This work is based in part on observations made with the NASA/ESA/CSA James Webb Space Telescope. The data were obtained from the Mikulski Archive for Space Telescopes at the Space Telescope Science Institute, which is operated by the Association of Universities for Research in Astronomy, Inc. (AURA), under NASA contract NAS 5-03127 for JWST. These observations are associated with JWST-GO-2561, JWST-GO-4111, JWST-GO-2883, JWST-GO-3516, JWST-3538, JWST-ERS-1324, and JWST-DD-2756. Support for program JWST-GO-2561 was provided by NASA through a grant from the Space Telescope Science Institute under NASA contract NAS 526555. The specific original observations used to produce our UNCOVER catalogs can be accessed via \dataset[doi: 10.17909/7yvw-xn77]{http://dx.doi.org/10.17909/7yvw-xn77}.

Some of the data products presented herein were retrieved from the Dawn JWST Archive (DJA). DJA is an initiative of the Cosmic Dawn Center (DAWN), which is funded by the Danish National Research Foundation under grant DNRF140.

The Cosmic Dawn Center is funded by the Danish National Research Foundation (DNRF) under grant \#140. 

RB gratefully acknowledges support from the Research Corporation for Scientific Advancement (RCSA) Cottrell Scholar Award ID No: 27587.

P. Dayal warmly acknowledges support from an NSERC discovery grant (RGPIN-2025-06182).

This research was supported in part by the University of Pittsburgh Center for Research Computing and Data, RRID:SCR\_022735, through the resources provided. Specifically, this work used the H2P cluster, which is supported by NSF award number OAC-2117681. 

DJS and JRW acknowledge that support for this work was provided by The Brinson Foundation through a Brinson Prize Fellowship grant. 

TBM was supported by a CIERA Fellowship. This work used computing resources provided by Northwestern University and the Center for Interdisciplinary Exploration and Research in Astrophysics (CIERA). This research was supported in part through the computational resources and staff contributions provided for the Quest high performance computing facility at Northwestern University, which is jointly supported by the Office of the Provost, the Office for Research, and Northwestern University Information Technology.
\end{acknowledgments}

\facilities{ 
JWST (NIRCam)
}
\bibliography{all}{}

\begin{thebibliography}{}
\expandafter\ifx\csname natexlab\endcsname\relax\def\natexlab#1{#1}\fi
\providecommand{\url}[1]{\href{#1}{#1}}
\providecommand{\dodoi}[1]{doi:~\href{http://doi.org/#1}{\nolinkurl{#1}}}
\providecommand{\doeprint}[1]{\href{http://ascl.net/#1}{\nolinkurl{http://ascl.net/#1}}}
\providecommand{\doarXiv}[1]{\href{https://arxiv.org/abs/#1}{\nolinkurl{https://arxiv.org/abs/#1}}}

\bibitem[{A. {Adamo} {et~al.}(2024){Adamo}, {Bradley}, {Vanzella}, {Claeyssens}, {Welch}, {Diego}, {Mahler}, {Oguri}, {Sharon}, {Abdurro'uf}, {Hsiao}, {Xu}, {Messa}, {Lassen}, {Zackrisson}, {Brammer}, {Coe}, {Kokorev}, {Ricotti}, {Zitrin}, {Fujimoto}, {Inoue}, {Resseguier}, {Rigby}, {Jim{\'e}nez-Teja}, {Windhorst}, {Hashimoto}, \& {Tamura}}]{Adamo2024}
{Adamo}, A., {Bradley}, L.~D., {Vanzella}, E., {et~al.} 2024, \bibinfo{title}{{Bound star clusters observed in a lensed galaxy 460 Myr after the Big Bang},} \nat, 632, 513, \dodoi{10.1038/s41586-024-07703-7}

\bibitem[{N. {Allen} {et~al.}(2024){Allen}, {Oesch}, {Toft}, {Matharu}, {McPartland}, {Weibel}, {Brammer}, {Bowler}, {Ito}, {Gottumukkala}, {Rizzo}, {Valentino}, {Varadaraj}, {Weaver}, \& {Whitaker}}]{Allen2024}
{Allen}, N., {Oesch}, P.~A., {Toft}, S., {et~al.} 2024, \bibinfo{title}{{Galaxy Size and Mass Build-up in the First 2 Gyrs of Cosmic History from Multi-Wavelength JWST NIRCam Imaging},} arXiv e-prints, arXiv:2410.16354, \dodoi{10.48550/arXiv.2410.16354}

\bibitem[{J. {Antwi-Danso} {et~al.}(2022){Antwi-Danso}, {Papovich}, {Leja}, {Marchesini}, {Marsan}, {Martis}, {Labb{\'e}}, {Muzzin}, {Glazebrook}, {Straatman}, \& {Tran}}]{antwidanso2022}
{Antwi-Danso}, J., {Papovich}, C., {Leja}, J., {et~al.} 2022, \bibinfo{title}{{Beyond UVJ: Color Selection of Galaxies in the JWST Era},} arXiv e-prints, arXiv:2207.07170.
\newblock \doarXiv{2207.07170}

\bibitem[{Y. {Asali} {et~al.}(2025){Asali}, {Geha}, {Kado-Fong}, {Mao}, {Wechsler}, {de los Reyes}, {Pasha}, {Kallivayalil}, {Nadler}, {Tollerud}, {Wang}, {Weiner}, \& {Wu}}]{Asali2025}
{Asali}, Y., {Geha}, M., {Kado-Fong}, E., {et~al.} 2025, \bibinfo{title}{{The SAGA Survey. VI. The Size-Mass Relation for Low-Mass Galaxies Across Environments},} arXiv e-prints, arXiv:2509.25335, \dodoi{10.48550/arXiv.2509.25335}

\bibitem[{P. {Behroozi} {et~al.}(2022){Behroozi}, {Hearin}, \& {Moster}}]{Behroozi2022}
{Behroozi}, P., {Hearin}, A., \& {Moster}, B.~P. 2022, \bibinfo{title}{{Observational measures of halo properties beyond mass},} \mnras, 509, 2800, \dodoi{10.1093/mnras/stab3193}

\bibitem[{R. {Bezanson} {et~al.}(2009){Bezanson}, {van Dokkum}, {Tal}, {Marchesini}, {Kriek}, {Franx}, \& {Coppi}}]{bezanson2009}
{Bezanson}, R., {van Dokkum}, P.~G., {Tal}, T., {et~al.} 2009, \bibinfo{title}{{The Relation Between Compact, Quiescent High-redshift Galaxies and Massive Nearby Elliptical Galaxies: Evidence for Hierarchical, Inside-Out Growth},} \apj, 697, 1290, \dodoi{10.1088/0004-637X/697/2/1290}

\bibitem[{R. {Bezanson} {et~al.}(2022){Bezanson}, {Labbe}, {Whitaker}, {Leja}, {Price}, {Franx}, {Brammer}, {Marchesini}, {Zitrin}, {Wang}, {Weaver}, {Furtak}, {Atek}, {Coe}, {Cutler}, {Dayal}, {van Dokkum}, {Feldmann}, {Forster Schreiber}, {Fujimoto}, {Geha}, {Glazebrook}, {de Graaff}, {Greene}, {Juneau}, {Kassin}, {Kriek}, {Khullar}, {Maseda}, {Mowla}, {Muzzin}, {Nanayakkara}, {Nelson}, {Oesch}, {Pacifici}, {Pan}, {Papovich}, {Setton}, {Shapley}, {Smit}, {Stefanon}, {Taylor}, \& {Williams}}]{Bezanson2022}
{Bezanson}, R., {Labbe}, I., {Whitaker}, K.~E., {et~al.} 2022, \bibinfo{title}{{The JWST UNCOVER Treasury survey: Ultradeep NIRSpec and NIRCam ObserVations before the Epoch of Reionization},} arXiv e-prints, arXiv:2212.04026, \dodoi{10.48550/arXiv.2212.04026}

\bibitem[{M. {Brada{\v{c}}} {et~al.}(2025){Brada{\v{c}}}, {Jude{\v{z}}}, {Willott}, {Rihtar{\v{s}}i{\v{c}}}, {Martis}, {Harshan}, {Felicioni}, {Asada}, {Desprez}, {Clowe}, {Gonzalez}, {Jones}, {Lemaux}, {Markov}, {Mowla}, {Noirot}, {Peter}, {Robertson}, {Sarrouh}, {Sawicki}, {Schrabback}, \& {Tripodi}}]{Bradac2025}
{Brada{\v{c}}}, M., {Jude{\v{z}}}, J., {Willott}, C., {et~al.} 2025, \bibinfo{title}{{Star Formation under a Cosmic Microscope: Highly magnified z = 11 galaxy behind the Bullet Cluster},} arXiv e-prints, arXiv:2509.20446, \dodoi{10.48550/arXiv.2509.20446}

\bibitem[{M. {Cappellari} {et~al.}(2013){Cappellari}, {McDermid}, {Alatalo}, {Blitz}, {Bois}, {Bournaud}, {Bureau}, {Crocker}, {Davies}, {Davis}, {de Zeeuw}, {Duc}, {Emsellem}, {Khochfar}, {Krajnovi{\'c}}, {Kuntschner}, {Morganti}, {Naab}, {Oosterloo}, {Sarzi}, {Scott}, {Serra}, {Weijmans}, \& {Young}}]{cappellari2013}
{Cappellari}, M., {McDermid}, R.~M., {Alatalo}, K., {et~al.} 2013, \bibinfo{title}{{The ATLAS$^{3D}$ project - XX. Mass-size and mass-{\ensuremath{\sigma}} distributions of early-type galaxies: bulge fraction drives kinematics, mass-to-light ratio, molecular gas fraction and stellar initial mass function},} \mnras, 432, 1862, \dodoi{10.1093/mnras/stt644}

\bibitem[{S.~G. {Carlsten} {et~al.}(2021){Carlsten}, {Greene}, {Greco}, {Beaton}, \& {Kado-Fong}}]{Carlsten2021}
{Carlsten}, S.~G., {Greene}, J.~E., {Greco}, J.~P., {Beaton}, R.~L., \& {Kado-Fong}, E. 2021, \bibinfo{title}{{Structures of Dwarf Satellites of Milky Way-like Galaxies: Morphology, Scaling Relations, and Intrinsic Shapes},} \apj, 922, 267, \dodoi{10.3847/1538-4357/ac2581}

\bibitem[{A.~C. {Carnall} {et~al.}(2023){Carnall}, {McLeod}, {McLure}, {Dunlop}, {Begley}, {Cullen}, {Donnan}, {Hamadouche}, {Jewell}, {Jones}, {Pollock}, \& {Wild}}]{Carnall2023}
{Carnall}, A.~C., {McLeod}, D.~J., {McLure}, R.~J., {et~al.} 2023, \bibinfo{title}{{A surprising abundance of massive quiescent galaxies at 3 < z < 5 in the first data from JWST CEERS},} \mnras, 520, 3974, \dodoi{10.1093/mnras/stad369}

\bibitem[{S. {Carniani} {et~al.}(2024){Carniani}, {Hainline}, {D'Eugenio}, {Eisenstein}, {Jakobsen}, {Witstok}, {Johnson}, {Chevallard}, {Maiolino}, {Helton}, {Willott}, {Robertson}, {Alberts}, {Arribas}, {Baker}, {Bhatawdekar}, {Boyett}, {Bunker}, {Cameron}, {Cargile}, {Charlot}, {Curti}, {Curtis-Lake}, {Egami}, {Giardino}, {Isaak}, {Ji}, {Jones}, {Kumari}, {Maseda}, {Parlanti}, {P{\'e}rez-Gonz{\'a}lez}, {Rawle}, {Rieke}, {Rieke}, {Del Pino}, {Saxena}, {Scholtz}, {Smit}, {Sun}, {Tacchella}, {{\"U}bler}, {Venturi}, {Williams}, \& {Willmer}}]{Carniani2024}
{Carniani}, S., {Hainline}, K., {D'Eugenio}, F., {et~al.} 2024, \bibinfo{title}{{Spectroscopic confirmation of two luminous galaxies at a redshift of 14},} \nat, 633, 318, \dodoi{10.1038/s41586-024-07860-9}

\bibitem[{C.~M. {Casey} {et~al.}(2023){Casey}, {Kartaltepe}, {Drakos}, {Franco}, {Harish}, {Paquereau}, {Ilbert}, {Rose}, {Cox}, {Nightingale}, {Robertson}, {Silverman}, {Koekemoer}, {Massey}, {McCracken}, {Rhodes}, {Akins}, {Allen}, {Amvrosiadis}, {Arango-Toro}, {Bagley}, {Bongiorno}, {Capak}, {Champagne}, {Chartab}, {Ch{\'a}vez Ortiz}, {Chworowsky}, {Cooke}, {Cooper}, {Darvish}, {Ding}, {Faisst}, {Finkelstein}, {Fujimoto}, {Gentile}, {Gillman}, {Gould}, {Gozaliasl}, {Hayward}, {He}, {Hemmati}, {Hirschmann}, {Jahnke}, {Jin}, {Khostovan}, {Kokorev}, {Lambrides}, {Laigle}, {Larson}, {Leung}, {Liu}, {Liaudat}, {Long}, {Magdis}, {Mahler}, {Mainieri}, {Manning}, {Maraston}, {Martin}, {McCleary}, {McKinney}, {McPartland}, {Mobasher}, {Pattnaik}, {Renzini}, {Rich}, {Sanders}, {Sattari}, {Scognamiglio}, {Scoville}, {Sheth}, {Shuntov}, {Sparre}, {Suzuki}, {Talia}, {Toft}, {Trakhtenbrot}, {Urry}, {Valentino}, {Vanderhoof}, {Vardoulaki}, {Weaver}, {Whitaker}, {Wilkins}, {Yang}, \& {Zavala}}]{Casey2023}
{Casey}, C.~M., {Kartaltepe}, J.~S., {Drakos}, N.~E., {et~al.} 2023, \bibinfo{title}{{COSMOS-Web: An Overview of the JWST Cosmic Origins Survey},} \apj, 954, 31, \dodoi{10.3847/1538-4357/acc2bc}

\bibitem[{M. {Castellano} {et~al.}(2024){Castellano}, {Napolitano}, {Fontana}, {Roberts-Borsani}, {Treu}, {Vanzella}, {Zavala}, {Arrabal Haro}, {Calabr{\`o}}, {Llerena}, {Mascia}, {Merlin}, {Paris}, {Pentericci}, {Santini}, {Bakx}, {Bergamini}, {Cupani}, {Dickinson}, {Filippenko}, {Glazebrook}, {Grillo}, {Kelly}, {Malkan}, {Mason}, {Morishita}, {Nanayakkara}, {Rosati}, {Sani}, {Wang}, \& {Yoon}}]{Castellano2024}
{Castellano}, M., {Napolitano}, L., {Fontana}, A., {et~al.} 2024, \bibinfo{title}{{JWST NIRSpec Spectroscopy of the Remarkable Bright Galaxy GHZ2/GLASS-z12 at Redshift 12.34},} \apj, 972, 143, \dodoi{10.3847/1538-4357/ad5f88}

\bibitem[{G. Chabrier(2003)Chabrier}]{chabrier2003}
Chabrier, G. 2003, \bibinfo{title}{{The Galactic Disk Mass Function: Reconciliation of the Hubble Space Telescope and Nearby Determinations},} The Astrophysical Journal, 586, L133, \dodoi{10.1086/374879}

\bibitem[{A. {Claeyssens} {et~al.}(2025){Claeyssens}, {Adamo}, {Messa}, {Dessauges-Zavadsky}, {Richard}, {Kramarenko}, {Matthee}, \& {Naidu}}]{Claeyssens2025}
{Claeyssens}, A., {Adamo}, A., {Messa}, M., {et~al.} 2025, \bibinfo{title}{{Tracing star formation across cosmic time at tens of parsec-scales in the lensing cluster field Abell 2744},} \mnras, 537, 2535, \dodoi{10.1093/mnras/staf058}

\bibitem[{L. {Clarke} {et~al.}(2025){Clarke}, {Shapley}, {Lam}, {Topping}, {Brammer}, {Sanders}, {Reddy}, \& {Karthikeyan}}]{Clarke2025}
{Clarke}, L., {Shapley}, A.~E., {Lam}, N., {et~al.} 2025, \bibinfo{title}{{The Star-forming Main Sequence and Bursty Star-formation Histories at $z>1.4$ in JADES and AURORA},} arXiv e-prints, arXiv:2510.06681, \dodoi{10.48550/arXiv.2510.06681}

\bibitem[{W.~S. Cleveland(1979)Cleveland}]{cleveland1979}
Cleveland, W.~S. 1979, \bibinfo{title}{Robust locally weighted regression and smoothing scatterplots,} Journal of the American statistical association, 74, 829

\bibitem[{M. {Cranmer}(2023){Cranmer}}]{Cranmer2023}
{Cranmer}, M. 2023, \bibinfo{title}{{Interpretable Machine Learning for Science with PySR and SymbolicRegression.jl},} arXiv e-prints, arXiv:2305.01582, \dodoi{10.48550/arXiv.2305.01582}

\bibitem[{S.~E. {Cutler} {et~al.}(2022){Cutler}, {Whitaker}, {Mowla}, {Brammer}, {van der Wel}, {Marchesini}, {van Dokkum}, {Momcheva}, {Song}, {Akhshik}, {Nelson}, {Bezanson}, {Franx}, {Kriek}, {Lange-Vagle}, {Leja}, {MacKenty}, {Muzzin}, \& {Shipley}}]{Cutler2022}
{Cutler}, S.~E., {Whitaker}, K.~E., {Mowla}, L.~A., {et~al.} 2022, \bibinfo{title}{{Diagnosing DASH: A Catalog of Structural Properties for the COSMOS-DASH Survey},} \apj, 925, 34, \dodoi{10.3847/1538-4357/ac341c}

\bibitem[{S.~E. {Cutler} {et~al.}(2024){Cutler}, {Whitaker}, {Weaver}, {Wang}, {Pan}, {Bezanson}, {Furtak}, {Labbe}, {Leja}, {Price}, {Cheng}, {Clausen}, {Cullen}, {Dayal}, {de Graaff}, {Dickinson}, {Dunlop}, {Feldmann}, {Franx}, {Giavalisco}, {Glazebrook}, {Greene}, {Grogin}, {Illingworth}, {Koekemoer}, {Kokorev}, {Marchesini}, {Maseda}, {Miller}, {Nanayakkara}, {Nelson}, {Setton}, {Shipley}, \& {Suess}}]{Cutler2024}
{Cutler}, S.~E., {Whitaker}, K.~E., {Weaver}, J.~R., {et~al.} 2024, \bibinfo{title}{{Two Distinct Classes of Quiescent Galaxies at Cosmic Noon Revealed by JWST PRIMER and UNCOVER},} \apjl, 967, L23, \dodoi{10.3847/2041-8213/ad464c}

\bibitem[{I. {Damjanov} {et~al.}(2023){Damjanov}, {Sohn}, {Geller}, {Utsumi}, \& {Dell'Antonio}}]{Damjanov2023}
{Damjanov}, I., {Sohn}, J., {Geller}, M.~J., {Utsumi}, Y., \& {Dell'Antonio}, I. 2023, \bibinfo{title}{{Size and Spectroscopic Evolution of HectoMAP Quiescent Galaxies},} \apj, 943, 149, \dodoi{10.3847/1538-4357/aca88f}

\bibitem[{S. {Danieli} {et~al.}(2023){Danieli}, {Greene}, {Carlsten}, {Jiang}, {Beaton}, \& {Goulding}}]{Danieli2023}
{Danieli}, S., {Greene}, J.~E., {Carlsten}, S., {et~al.} 2023, \bibinfo{title}{{ELVES. IV. The Satellite Stellar-to-halo Mass Relation Beyond the Milky Way},} \apj, 956, 6, \dodoi{10.3847/1538-4357/acefbd}

\bibitem[{S. {Danieli} \& P. {van Dokkum}(2019){Danieli} \& {van Dokkum}}]{Danieli2019}
{Danieli}, S., \& {van Dokkum}, P. 2019, \bibinfo{title}{{Revisiting the Size-Luminosity Relation in the Era of Ultra Diffuse Galaxies},} \apj, 875, 155, \dodoi{10.3847/1538-4357/ab14f3}

\bibitem[{A. {de Graaff} {et~al.}(2022){de Graaff}, {Trayford}, {Franx}, {Schaller}, {Schaye}, \& {van der Wel}}]{degraaff2022}
{de Graaff}, A., {Trayford}, J., {Franx}, M., {et~al.} 2022, \bibinfo{title}{{Observed structural parameters of EAGLE galaxies: reconciling the mass-size relation in simulations with local observations},} \mnras, 511, 2544, \dodoi{10.1093/mnras/stab3510}

\bibitem[{A. {Dekel} {et~al.}(2023){Dekel}, {Sarkar}, {Birnboim}, {Mandelker}, \& {Li}}]{Dekel2023}
{Dekel}, A., {Sarkar}, K.~C., {Birnboim}, Y., {Mandelker}, N., \& {Li}, Z. 2023, \bibinfo{title}{{Efficient formation of massive galaxies at cosmic dawn by feedback-free starbursts},} \mnras, 523, 3201, \dodoi{10.1093/mnras/stad1557}

\bibitem[{A. {Dressler}(1980){Dressler}}]{Dressler1980}
{Dressler}, A. 1980, \bibinfo{title}{{Galaxy morphology in rich clusters: implications for the formation and evolution of galaxies.},} \apj, 236, 351, \dodoi{10.1086/157753}

\bibitem[{P.~H. Eilers \& B.~D. Marx(1996)Eilers \& Marx}]{Eilers1996}
Eilers, P.~H., \& Marx, B.~D. 1996, \bibinfo{title}{Flexible smoothing with B-splines and penalties,} Statistical science, 11, 89

\bibitem[{K. {El-Badry} {et~al.}(2016){El-Badry}, {Wetzel}, {Geha}, {Hopkins}, {Kere{\v{s}}}, {Chan}, \& {Faucher-Gigu{\`e}re}}]{elbadry2016}
{El-Badry}, K., {Wetzel}, A., {Geha}, M., {et~al.} 2016, \bibinfo{title}{{Breathing FIRE: How Stellar Feedback Drives Radial Migration, Rapid Size Fluctuations, and Population Gradients in Low-mass Galaxies},} \apj, 820, 131, \dodoi{10.3847/0004-637X/820/2/131}

\bibitem[{K. {El-Badry} {et~al.}(2017){El-Badry}, {Wetzel}, {Geha}, {Quataert}, {Hopkins}, {Kere{\v{s}}}, {Chan}, \& {Faucher-Gigu{\`e}re}}]{elbadry2017}
{El-Badry}, K., {Wetzel}, A.~R., {Geha}, M., {et~al.} 2017, \bibinfo{title}{{When the Jeans Do Not Fit: How Stellar Feedback Drives Stellar Kinematics and Complicates Dynamical Modeling in Low-mass Galaxies},} \apj, 835, 193, \dodoi{10.3847/1538-4357/835/2/193}

\bibitem[{N. {Emami} {et~al.}(2021){Emami}, {Siana}, {El-Badry}, {Cook}, {Ma}, {Weisz}, {Gharibshah}, {Alaee}, {Scarlata}, \& {Skillman}}]{emami2021}
{Emami}, N., {Siana}, B., {El-Badry}, K., {et~al.} 2021, \bibinfo{title}{{Testing the Relationship between Bursty Star Formation and Size Fluctuations of Local Dwarf Galaxies},} \apj, 922, 217, \dodoi{10.3847/1538-4357/ac1f8d}

\bibitem[{S.~M. {Fall} \& G. {Efstathiou}(1980){Fall} \& {Efstathiou}}]{Fall1980}
{Fall}, S.~M., \& {Efstathiou}, G. 1980, \bibinfo{title}{{Formation and rotation of disc galaxies with haloes.},} \mnras, 193, 189, \dodoi{10.1093/mnras/193.2.189}

\bibitem[{A. {Ferrara}(2024){Ferrara}}]{Ferrara2024}
{Ferrara}, A. 2024, \bibinfo{title}{{Super-early JWST galaxies, outflows, and Ly{\ensuremath{\alpha}} visibility in the Epoch of Reionization},} \aap, 684, A207, \dodoi{10.1051/0004-6361/202348321}

\bibitem[{A. {Ferrara} {et~al.}(2023){Ferrara}, {Pallottini}, \& {Dayal}}]{Ferrara2023}
{Ferrara}, A., {Pallottini}, A., \& {Dayal}, P. 2023, \bibinfo{title}{{On the stunning abundance of super-early, luminous galaxies revealed by JWST},} \mnras, 522, 3986, \dodoi{10.1093/mnras/stad1095}

\bibitem[{S.~L. {Finkelstein} {et~al.}(2022){Finkelstein}, {Bagley}, {Arrabal Haro}, {Dickinson}, {Ferguson}, {Kartaltepe}, {Papovich}, {Burgarella}, {Kocevski}, {Huertas-Company}, {Iyer}, {Larson}, {P{\'e}rez-Gonz{\'a}lez}, {Rose}, {Tacchella}, {Wilkins}, {Chworowsky}, {Medrano}, {Morales}, {Somerville}, {Yung}, {Fontana}, {Giavalisco}, {Grazian}, {Grogin}, {Kewley}, {Koekemoer}, {Kirkpatrick}, {Kurczynski}, {Lotz}, {Pentericci}, {Pirzkal}, {Ravindranath}, {Ryan}, {Trump}, {Yang}, {Almaini}, {Amor{\'\i}n}, {Annunziatella}, {Backhaus}, {Barro}, {Behroozi}, {Bell}, {Bhatawdekar}, {Bisigello}, {Bromm}, {Buat}, {Buitrago}, {Calabr{\'o}}, {Casey}, {Castellano}, {Ch{\'a}vez Ortiz}, {Ciesla}, {Cleri}, {Cohen}, {Cole}, {Cooke}, {Cooper}, {Cooray}, {Costantin}, {Cox}, {Croton}, {Daddi}, {Dav{\'e}}, {de la Vega}, {Dekel}, {Elbaz}, {Estrada-Carpenter}, {Faber}, {Fern{\'a}ndez}, {Finkelstein}, {Freundlich}, {Fujimoto}, {Garc{\'\i}a-Argum{\'a}nez}, {Gardner}, {Gawiser}, {G{\'o}mez-Guijarro}, {Guo}, {Hamilton}, {Hathi},
  {Holwerda}, {Hirschmann}, {Hutchison}, {Jha}, {Jogee}, {Juneau}, {Jung}, {Kassin}, {Le Bail}, {Leung}, {Lucas}, {Magnelli}, {Mantha}, {Matharu}, {McGrath}, {McIntosh}, {Merlin}, {Mobasher}, {Newman}, {Nicholls}, {Pandya}, {Rafelski}, {Ronayne}, {Santini}, {Seill{\'e}}, {Shah}, {Shen}, {Simons}, {Snyder}, {Stanway}, {Straughn}, {Teplitz}, {Vanderhoof}, {Vega-Ferrero}, {Wang}, {Weiner}, {Willmer}, {Wuyts}, \& {Zavala}}]{finkelstein2022}
{Finkelstein}, S.~L., {Bagley}, M.~B., {Arrabal Haro}, P., {et~al.} 2022, \bibinfo{title}{{A Long Time Ago in a Galaxy Far, Far Away: A Candidate z \raisebox{-0.5ex}\textasciitilde 14 Galaxy in Early JWST CEERS Imaging},} arXiv e-prints, arXiv:2207.12474.
\newblock \doarXiv{2207.12474}

\bibitem[{M. {Franx} {et~al.}(2008){Franx}, {van Dokkum}, {F{\"o}rster Schreiber}, {Wuyts}, {Labb{\'e}}, \& {Toft}}]{Franx2008}
{Franx}, M., {van Dokkum}, P.~G., {F{\"o}rster Schreiber}, N.~M., {et~al.} 2008, \bibinfo{title}{{Structure and Star Formation in Galaxies out to z = 3: Evidence for Surface Density Dependent Evolution and Upsizing},} \apj, 688, 770, \dodoi{10.1086/592431}

\bibitem[{S. {Fujimoto} {et~al.}(2025){Fujimoto}, {Ouchi}, {Kohno}, {Valentino}, {Gim{\'e}nez-Arteaga}, {Brammer}, {Furtak}, {Kohandel}, {Oguri}, {Pallottini}, {Richard}, {Zitrin}, {Bauer}, {Boylan-Kolchin}, {Dessauges-Zavadsky}, {Egami}, {Finkelstein}, {Ma}, {Smail}, {Watson}, {Hutchison}, {Rigby}, {Welch}, {Ao}, {Bradley}, {Caminha}, {Caputi}, {Espada}, {Endsley}, {Fudamoto}, {Gonz{\'a}lez-L{\'o}pez}, {Hatsukade}, {Koekemoer}, {Kokorev}, {Laporte}, {Lee}, {Magdis}, {Ono}, {Rizzo}, {Shibuya}, {Shimasaku}, {Sun}, {Toft}, {Umehata}, {Wang}, \& {Yajima}}]{Fujimoto2025}
{Fujimoto}, S., {Ouchi}, M., {Kohno}, K., {et~al.} 2025, \bibinfo{title}{{Primordial rotating disk composed of at least 15 dense star-forming clumps at cosmic dawn},} Nature Astronomy, 9, 1553, \dodoi{10.1038/s41550-025-02592-w}

\bibitem[{L.~J. {Furtak} {et~al.}(2023){Furtak}, {Zitrin}, {Weaver}, {Atek}, {Bezanson}, {Labb{\'e}}, {Whitaker}, {Leja}, {Price}, {Brammer}, {Wang}, {Marchesini}, {Pan}, {Dayal}, {van Dokkum}, {Feldmann}, {Fujimoto}, {Franx}, {Khullar}, {Nelson}, \& {Mowla}}]{Furtak2023}
{Furtak}, L.~J., {Zitrin}, A., {Weaver}, J.~R., {et~al.} 2023, \bibinfo{title}{{UNCOVERing the extended strong lensing structures of Abell 2744 with the deepest JWST imaging},} \mnras, 523, 4568, \dodoi{10.1093/mnras/stad1627}

\bibitem[{J.~P. {Gardner} {et~al.}(2023){Gardner}, {Mather}, {Abbott}, {Abell}, {Abernathy}, {Abney}, {Abraham}, {Abraham}, {Abul-Huda}, {Acton}, {Adams}, {Adams}, {Adler}, {Adriaensen}, {Aguilar}, {Ahmed}, {Ahmed}, {Ahmed}, {Albat}, {Albert}, {Alberts}, {Aldridge}, {Allen}, {Allen}, {Altenburg}, {Altunc}, {Alvarez}, {{\'A}lvarez-M{\'a}rquez}, {Alves de Oliveira}, {Ambrose}, {Anandakrishnan}, {Andersen}, {Anderson}, {Anderson}, {Anderson}, {Anderson}, {Aprea}, {Archer}, {Arenberg}, {Argyriou}, {Arribas}, {Artigau}, {Arvai}, {Atcheson}, {Atkinson}, {Averbukh}, {Aymergen}, {Bacinski}, {Baggett}, {Bagnasco}, {Baker}, {Balzano}, {Banks}, {Baran}, {Barker}, {Barrett}, {Barringer}, {Barto}, {Bast}, {Baudoz}, {Baum}, {Beatty}, {Beaulieu}, {Bechtold}, {Beck}, {Beddard}, {Beichman}, {Bellagama}, {Bely}, {Berger}, {Bergeron}, {Bernier}, {Bertch}, {Beskow}, {Betz}, {Biagetti}, {Birkmann}, {Bjorklund}, {Blackwood}, {Blazek}, {Blossfeld}, {Bluth}, {Boccaletti}, {Boegner}, {Bohlin}, {Boia}, {B{\"o}ker}, {Bonaventura},
  {Bond}, {Bosley}, {Boucarut}, {Bouchet}, {Bouwman}, {Bower}, {Bowers}, {Bowers}, {Boyce}, {Boyer}, {Boyer}, {Boyer}, {Boyer}, {Bradley}, {Brady}, {Brandl}, {Brannen}, {Breda}, {Bremmer}, {Brennan}, {Bresnahan}, {Bright}, {Broiles}, {Bromenschenkel}, {Brooks}, {Brooks}, {Brown}, {Brown}, {Brown}, {Bruce}, {Bryson}, {Bujanda}, {Bullock}, {Bunker}, {Bureo}, {Burt}, {Bush}, {Bushouse}, {Bussman}, {Cabaud}, {Cale}, {Calhoon}, {Calvani}, {Canipe}, {Caputo}, {Cara}, {Carey}, {Case}, {Cesari}, {Cetorelli}, {Chance}, {Chandler}, {Chaney}, {Chapman}, {Charlot}, {Chayer}, {Cheezum}, {Chen}, {Chen}, {Cherinka}, {Chichester}, {Chilton}, {Chittiraibalan}, {Clampin}, {Clark}, {Clark}, {Clark}, {Claybrooks}, {Cleveland}, {Cohen}, {Cohen}, {Col{\'o}n}, {Coleman}, {Colina}, {Comber}, {Comeau}, {Comer}, {Conde Reis}, {Connolly}, {Conroy}, {Contos}, {Contreras}, {Cook}, {Cooper}, {Cooper}, {Correia}, {Correnti}, {Cossou}, {Costanza}, {Coulais}, {Cox}, {Coyle}, {Cracraft}, {Crew}, {Curtis}, {Cusveller}, {Da Costa Maciel},
  {Dailey}, {Daugeron}, {Davidson}, {Davies}, {Davis}, {Davis}, {Day}, {de Chambure}, {de Jong}, {De Marchi}, {Dean}, {Decker}, {Delisa}, {Dell}, \& {Dellagatta}}]{Gardner2023}
{Gardner}, J.~P., {Mather}, J.~C., {Abbott}, R., {et~al.} 2023, \bibinfo{title}{{The James Webb Space Telescope Mission},} \pasp, 135, 068001, \dodoi{10.1088/1538-3873/acd1b5}

\bibitem[{R. {Geda} {et~al.}(2025){Geda}, {Cruz}, {Wright}, {Greene}, {Brooks}, {Quinn}, {Wadsley}, \& {Keller}}]{Geda2025}
{Geda}, R., {Cruz}, A., {Wright}, A.~C., {et~al.} 2025, \bibinfo{title}{{Disk Formation and the Size-sSFR Relation of Dwarf Galaxies},} arXiv e-prints, arXiv:2510.26875, \dodoi{10.48550/arXiv.2510.26875}

\bibitem[{A. {George} {et~al.}(2024){George}, {Damjanov}, {Sawicki}, {Arnouts}, {Desprez}, {Gwyn}, {Picouet}, {Birrer}, \& {Silverman}}]{George2024}
{George}, A., {Damjanov}, I., {Sawicki}, M., {et~al.} 2024, \bibinfo{title}{{Two rest-frame wavelength measurements of galaxy sizes at z < 1: the evolutionary effects of emerging bulges and quenched newcomers},} \mnras, 528, 4797, \dodoi{10.1093/mnras/stae154}

\bibitem[{A. {Ghosh} {et~al.}(2024){Ghosh}, {Urry}, {Powell}, {Shimakawa}, {van den Bosch}, {Nagai}, {Mitra}, \& {Connolly}}]{Ghosh2024}
{Ghosh}, A., {Urry}, C.~M., {Powell}, M.~C., {et~al.} 2024, \bibinfo{title}{{Denser Environments Cultivate Larger Galaxies: A Comprehensive Study beyond the Local Universe with 3 Million Hyper Suprime-Cam Galaxies},} \apj, 971, 142, \dodoi{10.3847/1538-4357/ad596f}

\bibitem[{J.~E. {Greene} {et~al.}(2024){Greene}, {Labbe}, {Goulding}, {Furtak}, {Chemerynska}, {Kokorev}, {Dayal}, {Volonteri}, {Williams}, {Wang}, {Setton}, {Burgasser}, {Bezanson}, {Atek}, {Brammer}, {Cutler}, {Feldmann}, {Fujimoto}, {Glazebrook}, {de Graaff}, {Khullar}, {Leja}, {Marchesini}, {Maseda}, {Matthee}, {Miller}, {Naidu}, {Nanayakkara}, {Oesch}, {Pan}, {Papovich}, {Price}, {van Dokkum}, {Weaver}, {Whitaker}, \& {Zitrin}}]{Greene2024}
{Greene}, J.~E., {Labbe}, I., {Goulding}, A.~D., {et~al.} 2024, \bibinfo{title}{{UNCOVER Spectroscopy Confirms the Surprising Ubiquity of Active Galactic Nuclei in Red Sources at z > 5},} \apj, 964, 39, \dodoi{10.3847/1538-4357/ad1e5f}

\bibitem[{M.~L. {Hamadouche} {et~al.}(2022){Hamadouche}, {Carnall}, {McLure}, {Dunlop}, {McLeod}, {Cullen}, {Begley}, {Bolzonella}, {Buitrago}, {Castellano}, {Cucciati}, {Fontana}, {Gargiulo}, {Moresco}, {Pozzetti}, \& {Zamorani}}]{Hamadouche2022}
{Hamadouche}, M.~L., {Carnall}, A.~C., {McLure}, R.~J., {et~al.} 2022, \bibinfo{title}{{A combined VANDELS and LEGA-C study: the evolution of quiescent galaxy size, stellar mass, an@ARTICLE{2022MNRAS.512.1262H, author = {{Hamadouche}, M.~L. and {Carnall}, A.~C. and {McLure}, R.~J. and {Dunlop}, J.~S. and {McLeod}, D.~J. and {Cullen}, F. and {Begley}, R. and {Bolzonella}, M. and {Buitrago}, F. and {Castellano}, M. and {Cucciati}, O. and {Fontana}, A. and {Gargiulo}, A. and {Moresco}, M. and {Pozzetti}, L. and {Zamorani}, G.}, title = "{A combined VANDELS and LEGA-C study: the evolution of quiescent galaxy size, stellar mass, and age from z = 0.6 to z = 1.3}", journal = {\mnras}, keywords = {galaxies: evolution, galaxies: high-redshift, galaxies: star formation, Astrophysics - Astrophysics of Galaxies}, year = 2022, month = may, volume = {512}, number = {1}, pages = {1262-1274}, doi = {10.1093/mnras/stac535}, archivePrefix = {arXiv}, eprint = {2201.10576}, primaryClass = {astro-ph.GA}, adsurl =
  {https://ui.adsabs.harvard.edu/abs/2022MNRAS.512.1262H}, adsnote = {Provided by the SAO/NASA Astrophysics Data System} } Md age from z = 0.6 to z = 1.3},} \mnras, 512, 1262, \dodoi{10.1093/mnras/stac535}

\bibitem[{M.~L. {Hamadouche} {et~al.}(2025){Hamadouche}, {McLure}, {Carnall}, {McLeod}, {Dunlop}, {Whitaker}, {Donnan}, {Begley}, {Stanton}, {Almaini}, {Aird}, {Cullen}, {Cutler}, {Grogin}, \& {Koekemoer}}]{Hamadouche2025}
{Hamadouche}, M.~L., {McLure}, R.~J., {Carnall}, A.~C., {et~al.} 2025, \bibinfo{title}{{JWST PRIMER: strong evidence for the environmental quenching of low-mass galaxies out to z≃ 2},} \mnras, 541, 463, \dodoi{10.1093/mnras/staf971}

\bibitem[{G. {Hinshaw} {et~al.}(2013){Hinshaw}, {Larson}, {Komatsu}, {Spergel}, {Bennett}, {Dunkley}, {Nolta}, {Halpern}, {Hill}, {Odegard}, {Page}, {Smith}, {Weiland}, {Gold}, {Jarosik}, {Kogut}, {Limon}, {Meyer}, {Tucker}, {Wollack}, \& {Wright}}]{Hinshaw2013}
{Hinshaw}, G., {Larson}, D., {Komatsu}, E., {et~al.} 2013, \bibinfo{title}{{Nine-year Wilkinson Microwave Anisotropy Probe (WMAP) Observations: Cosmological Parameter Results},} \apjs, 208, 19, \dodoi{10.1088/0067-0049/208/2/19}

\bibitem[{D.~W. {Hogg} {et~al.}(2010){Hogg}, {Bovy}, \& {Lang}}]{hogg2010}
{Hogg}, D.~W., {Bovy}, J., \& {Lang}, D. 2010, \bibinfo{title}{{Data analysis recipes: Fitting a model to data},} arXiv e-prints, arXiv:1008.4686.
\newblock \doarXiv{1008.4686}

\bibitem[{P.~F. {Hopkins} {et~al.}(2023){Hopkins}, {Gurvich}, {Shen}, {Hafen}, {Grudi{\'c}}, {Kurinchi-Vendhan}, {Hayward}, {Jiang}, {Orr}, {Wetzel}, {Kere{\v{s}}}, {Stern}, {Faucher-Gigu{\`e}re}, {Bullock}, {Wheeler}, {El-Badry}, {Loebman}, {Moreno}, {Boylan-Kolchin}, \& {Quataert}}]{Hopkins2023}
{Hopkins}, P.~F., {Gurvich}, A.~B., {Shen}, X., {et~al.} 2023, \bibinfo{title}{{What causes the formation of discs and end of bursty star formation?},} \mnras, 525, 2241, \dodoi{10.1093/mnras/stad1902}

\bibitem[{F. Jiang {et~al.}(2018)Jiang, Dekel, Kneller, Lapiner, Ceverino, Primack, Faber, Macci{\`{o}}, Dutton, Genel, \& Somerville}]{jiang2018}
Jiang, F., Dekel, A., Kneller, O., {et~al.} 2018, \bibinfo{title}{{Is the dark-matter halo spin a predictor of galaxy spin and size?},} eprint arXiv:1804.07306

\bibitem[{B.~D. {Johnson} {et~al.}(2021){Johnson}, {Leja}, {Conroy}, \& {Speagle}}]{johnson2021}
{Johnson}, B.~D., {Leja}, J., {Conroy}, C., \& {Speagle}, J.~S. 2021, \bibinfo{title}{{Stellar Population Inference with Prospector},} \apjs, 254, 22, \dodoi{10.3847/1538-4365/abef67}

\bibitem[{L. {Kawinwanichakij} {et~al.}(2021){Kawinwanichakij}, {Silverman}, {Ding}, {George}, {Damjanov}, {Sawicki}, {Tanaka}, {Taranu}, {Birrer}, {Huang}, {Li}, {Onodera}, {Shibuya}, \& {Yasuda}}]{kawinwanichakij2021}
{Kawinwanichakij}, L., {Silverman}, J.~D., {Ding}, X., {et~al.} 2021, \bibinfo{title}{{Hyper Suprime-Cam Subaru Strategic Program: A Mass-dependent Slope of the Galaxy Size-Mass Relation at z < 1},} \apj, 921, 38, \dodoi{10.3847/1538-4357/ac1f21}

\bibitem[{L. {Kawinwanichakij} {et~al.}(2026){Kawinwanichakij}, {Glazebrook}, {Nanayakkara}, {Kacprzak}, {Chittenden}, {Jacobs}, {Chandro-G{\'o}mez}, {Lagos}, {Marchesini}, {Mart{\`\i}nez-Mar{\`\i}n}, {Oesch}, \& {Remus}}]{Kawinwanichakij2026}
{Kawinwanichakij}, L., {Glazebrook}, K., {Nanayakkara}, T., {et~al.} 2026, \bibinfo{title}{{Connecting Environment, Star Formation History, and Morphology of Massive Quiescent Galaxies at 3 < z < 4 with JWST},} \apj, 997, 29, \dodoi{10.3847/1538-4357/ae0a18}

\bibitem[{A.~V. Kravtsov(2013)Kravtsov}]{kravtsov2013}
Kravtsov, A.~V. 2013, \bibinfo{title}{{The size-virial radius relation of galaxies},} Astrophysical Journal Letters, 764

\bibitem[{I. {Labbe} {et~al.}(2023){Labbe}, {Greene}, {Bezanson}, {Fujimoto}, {Furtak}, {Goulding}, {Matthee}, {Naidu}, {Oesch}, {Atek}, {Brammer}, {Chemerynska}, {Coe}, {Cutler}, {Dayal}, {Feldmann}, {Franx}, {Glazebrook}, {Leja}, {Marchesini}, {Maseda}, {Nanayakkara}, {Nelson}, {Pan}, {Papovich}, {Price}, {Suess}, {Wang}, {Whitaker}, {Williams}, \& {Zitrin}}]{Labbe2023}
{Labbe}, I., {Greene}, J.~E., {Bezanson}, R., {et~al.} 2023, \bibinfo{title}{{UNCOVER: Candidate Red Active Galactic Nuclei at 3<z<7 with JWST and ALMA},} arXiv e-prints, arXiv:2306.07320, \dodoi{10.48550/arXiv.2306.07320}

\bibitem[{S. Lang \& A. Brezger(2004)Lang \& Brezger}]{Lang2004}
Lang, S., \& Brezger, A. 2004, \bibinfo{title}{Bayesian P-splines,} Journal of computational and graphical statistics, 13, 183

\bibitem[{R. Lange {et~al.}(2015)Lange, Driver, Robotham, Kelvin, Graham, Alpaslan, Andrews, Baldry, Bamford, Bland-Hawthorn, Brough, Cluver, Conselice, Davies, Haeussler, Konstantopoulos, Loveday, Moffett, Norberg, Phillipps, Taylor, L{\'{o}}pez-S{\'{a}}nchez, \& Wilkins}]{lange2015}
Lange, R., Driver, S.~P., Robotham, A.~S., {et~al.} 2015, \bibinfo{title}{{Galaxy and mass assembly (GAMA): Mass-size relations of z < 0.1 galaxies subdivided by S{\'{e}}rsic index, colour and morphology},} Monthly Notices of the Royal Astronomical Society, 447, 2603

\bibitem[{O. Ledoit \& M. Wolf(2004)Ledoit \& Wolf}]{ledoit2004}
Ledoit, O., \& Wolf, M. 2004, \bibinfo{title}{A well-conditioned estimator for large-dimensional covariance matrices,} Journal of Multivariate Analysis, 88, 365

\bibitem[{J. {Leja} {et~al.}(2017){Leja}, {Johnson}, {Conroy}, {van Dokkum}, \& {Byler}}]{leja2017}
{Leja}, J., {Johnson}, B.~D., {Conroy}, C., {van Dokkum}, P.~G., \& {Byler}, N. 2017, \bibinfo{title}{{Deriving Physical Properties from Broadband Photometry with Prospector: Description of the Model and a Demonstration of its Accuracy Using 129 Galaxies in the Local Universe},} \apj, 837, 170, \dodoi{10.3847/1538-4357/aa5ffe}

\bibitem[{J. {Leja} {et~al.}(2019){Leja}, {Tacchella}, \& {Conroy}}]{leja2019}
{Leja}, J., {Tacchella}, S., \& {Conroy}, C. 2019, \bibinfo{title}{{Beyond UVJ: More Efficient Selection of Quiescent Galaxies with Ultraviolet/Mid-infrared Fluxes},} \apjl, 880, L9, \dodoi{10.3847/2041-8213/ab2f8c}

\bibitem[{J. {Li} {et~al.}(2023){Li}, {Greene}, {Greco}, {Huang}, {Melchior}, {Beaton}, {Casey}, {Danieli}, {Goulding}, {Joseph}, {Kado-Fong}, {Kim}, \& {MacArthur}}]{Li2023}
{Li}, J., {Greene}, J.~E., {Greco}, J.~P., {et~al.} 2023, \bibinfo{title}{{Beyond Ultra-diffuse Galaxies. I. Mass-Size Outliers among the Satellites of Milky Way Analogs},} \apj, 955, 1, \dodoi{10.3847/1538-4357/ace829}

\bibitem[{Z. {Li} {et~al.}(2024){Li}, {Dekel}, {Sarkar}, {Aung}, {Giavalisco}, {Mandelker}, \& {Tacchella}}]{Li2024}
{Li}, Z., {Dekel}, A., {Sarkar}, K.~C., {et~al.} 2024, \bibinfo{title}{{Feedback-free starbursts at cosmic dawn: Observable predictions for JWST},} \aap, 690, A108, \dodoi{10.1051/0004-6361/202348727}

\bibitem[{N. Makke \& S. Chawla(2024)Makke \& Chawla}]{Makke2024}
Makke, N., \& Chawla, S. 2024, \bibinfo{title}{Interpretable scientific discovery with symbolic regression: a review,} Artificial Intelligence Review, 57, 2

\bibitem[{M.~A. Marshall {et~al.}(2022)Marshall, Wilkins, Di~Matteo, Roper, Vijayan, Ni, Feng, \& Croft}]{marshall2022}
Marshall, M.~A., Wilkins, S., Di~Matteo, T., {et~al.} 2022, \bibinfo{title}{The impact of dust on the sizes of galaxies in the Epoch of Reionization,} Monthly Notices of the Royal Astronomical Society, 511, 5475

\bibitem[{M. {Martorano} {et~al.}(2023){Martorano}, {van der Wel}, {Bell}, {Franx}, {Whitaker}, {Nersesian}, {Price}, {Baes}, {Suess}, {Nelson}, {Miller}, {Bezanson}, \& {Brammer}}]{Martorano2023}
{Martorano}, M., {van der Wel}, A., {Bell}, E.~F., {et~al.} 2023, \bibinfo{title}{{Rest-frame Near-infrared Radial Light Profiles up to z = 3 from JWST/NIRCam: Wavelength Dependence of the S{\'e}rsic Index},} \apj, 957, 46, \dodoi{10.3847/1538-4357/acf716}

\bibitem[{M. {Martorano} {et~al.}(2026){Martorano}, {van der Wel}, {Gebek}, {Baes}, {Bell}, {Brammer}, {Meidt}, {Nersesian}, {Whitaker}, \& {Wuyts}}]{Martorano2025}
{Martorano}, M., {van der Wel}, A., {Gebek}, A., {et~al.} 2026, \bibinfo{title}{{Evolution and mass dependence of UV-to-near-IR color gradients up to z = 2.5 from the Hubble Space Telescope and the James Webb Space Telescope},} \aap, 705, A236, \dodoi{10.1051/0004-6361/202555974}

\bibitem[{J. {Matharu} {et~al.}(2020){Matharu}, {Muzzin}, {Brammer}, {van der Burg}, {Auger}, {Hewett}, {Chan}, {Demarco}, {van Dokkum}, {Marchesini}, {Nelson}, {Noble}, \& {Wilson}}]{Matharu2020}
{Matharu}, J., {Muzzin}, A., {Brammer}, G.~B., {et~al.} 2020, \bibinfo{title}{{HST/WFC3 grism observations of z {\ensuremath{\sim}} 1 clusters: evidence for evolution in the mass-size relation of quiescent galaxies from post-starburst galaxies},} \mnras, 493, 6011, \dodoi{10.1093/mnras/staa610}

\bibitem[{J. {Matharu} {et~al.}(2025){Matharu}, {Shen}, {Shivaei}, {Oesch}, {Papovich}, {Brammer}, {Reddy}, {Cheng}, {van Dokkum}, {Finkelstein}, {Hathi}, {Kartaltepe}, {Koekemoer}, {Matthee}, {Pirzkal}, {Wilkins}, {Wozniak}, \& {Xiao}}]{Matharu2025}
{Matharu}, J., {Shen}, L., {Shivaei}, I., {et~al.} 2025, \bibinfo{title}{{A first look at a complete view of spatially resolved star formation at 1<z<1.8 with JWST NGDEEP+FRESCO slitless spectroscopy},} arXiv e-prints, arXiv:2511.15792, \dodoi{10.48550/arXiv.2511.15792}

\bibitem[{J. {Matthee} {et~al.}(2017){Matthee}, {Schaye}, {Crain}, {Schaller}, {Bower}, \& {Theuns}}]{Matthee2017}
{Matthee}, J., {Schaye}, J., {Crain}, R.~A., {et~al.} 2017, \bibinfo{title}{{The origin of scatter in the stellar mass-halo mass relation of central galaxies in the EAGLE simulation},} \mnras, 465, 2381, \dodoi{10.1093/mnras/stw2884}

\bibitem[{W. {McClymont} {et~al.}(2025){McClymont}, {Tacchella}, {Smith}, {Kannan}, {Puchwein}, {Borrow}, {Garaldi}, {Keating}, {Vogelsberger}, {Zier}, {Shen}, \& {Popovic}}]{McClymont2025}
{McClymont}, W., {Tacchella}, S., {Smith}, A., {et~al.} 2025, \bibinfo{title}{{The THESAN-ZOOM project: central starbursts and inside-out quenching govern galaxy sizes in the early Universe},} \mnras, 544, 1732, \dodoi{10.1093/mnras/staf1861}

\bibitem[{F.~J. {Mercado} {et~al.}(2025){Mercado}, {Moreno}, {Feldmann}, {Zeender}, {Benavides}, {Piotrowska}, {Klein}, {Wheeler}, {Necib}, {Bullock}, \& {Hopkins}}]{Mercado2025}
{Mercado}, F.~J., {Moreno}, J., {Feldmann}, R., {et~al.} 2025, \bibinfo{title}{{Effects of Galactic Environment on Size and Dark Matter Content in Low-mass Galaxies},} \apj, 983, 93, \dodoi{10.3847/1538-4357/adbf07}

\bibitem[{T.~B. {Miller} {et~al.}(2022{\natexlab{a}}){Miller}, {van Dokkum}, \& {Mowla}}]{miller2022}
{Miller}, T.~B., {van Dokkum}, P., \& {Mowla}, L. 2022{\natexlab{a}}, \bibinfo{title}{{Color gradients and half-mass radii of galaxies out to $z=2$ in the CANDELS/3D-HST fields: further evidence for important differences in the evolution of mass-weighted and light-weighted sizes},} arXiv e-prints, arXiv:2207.05895.
\newblock \doarXiv{2207.05895}

\bibitem[{T.~B. {Miller} {et~al.}(2022{\natexlab{b}}){Miller}, {Whitaker}, {Nelson}, {van Dokkum}, {Bezanson}, {Brammer}, {Heintz}, {Leja}, {Suess}, \& {Weaver}}]{Miller2022b}
{Miller}, T.~B., {Whitaker}, K.~E., {Nelson}, E.~J., {et~al.} 2022{\natexlab{b}}, \bibinfo{title}{{Early JWST Imaging Reveals Strong Optical and NIR Color Gradients in Galaxies at z 2 Driven Mostly by Dust},} \apjl, 941, L37, \dodoi{10.3847/2041-8213/aca675}

\bibitem[{T.~B. {Miller} {et~al.}(2025){Miller}, {Suess}, {Setton}, {Price}, {Labbe}, {Bezanson}, {Brammer}, {Cutler}, {Furtak}, {Leja}, {Pan}, {Wang}, {Weaver}, {Whitaker}, {Dayal}, {de Graaff}, {Feldmann}, {Greene}, {Fujimoto}, {Maseda}, {Nanayakkara}, {Nelson}, {van Dokkum}, \& {Zitrin}}]{Miller2025}
{Miller}, T.~B., {Suess}, K.~A., {Setton}, D.~J., {et~al.} 2025, \bibinfo{title}{{JWST UNCOVERs the Optical Size{\textendash}Stellar Mass Relation at 4 < z < 8: Rapid Growth in the Sizes of Low-mass Galaxies in the First Billion Years of the Universe},} \apj, 988, 196, \dodoi{10.3847/1538-4357/ade438}

\bibitem[{A. {Mintz} {et~al.}(2025){Mintz}, {Setton}, {Greene}, {Leja}, {Wang}, {Burnham}, {Suess}, {Atek}, {Bezanson}, {Brammer}, {Cutler}, {Dayal}, {Feldmann}, {Furtak}, {Glazebrook}, {Khullar}, {Kokorev}, {Labb{\'e}}, {Maseda}, {Miller}, {Mitsuhashi}, {Nanayakkara}, {Pan}, {Price}, {Weaver}, \& {Whitaker}}]{Mintz2025}
{Mintz}, A., {Setton}, D.~J., {Greene}, J.~E., {et~al.} 2025, \bibinfo{title}{{Taking a Break at Cosmic Noon: Continuum-selected Low-mass Galaxies Require Long Burst Cycles},} arXiv e-prints, arXiv:2506.16510, \dodoi{10.48550/arXiv.2506.16510}

\bibitem[{H.~J. Mo {et~al.}(1998)Mo, Mao, \& White}]{mo1998}
Mo, H.~J., Mao, S., \& White, S.~D. 1998, \bibinfo{title}{{The formation of galactic discs},} Monthly Notices of the Royal Astronomical Society, 295, 319

\bibitem[{T. {Morishita} {et~al.}(2024){Morishita}, {Stiavelli}, {Chary}, {Trenti}, {Bergamini}, {Chiaberge}, {Leethochawalit}, {Roberts-Borsani}, {Shen}, \& {Treu}}]{Morishita2024}
{Morishita}, T., {Stiavelli}, M., {Chary}, R.-R., {et~al.} 2024, \bibinfo{title}{{Enhanced Subkiloparsec-scale Star Formation: Results from a JWST Size Analysis of 341 Galaxies at 5 < z < 14},} \apj, 963, 9, \dodoi{10.3847/1538-4357/ad1404}

\bibitem[{L. {Mowla} {et~al.}(2019){Mowla}, {van der Wel}, {van Dokkum}, \& {Miller}}]{mowla2019b}
{Mowla}, L., {van der Wel}, A., {van Dokkum}, P., \& {Miller}, T.~B. 2019, \bibinfo{title}{{A Mass-dependent Slope of the Galaxy Size-Mass Relation out to z {\ensuremath{\sim}} 3: Further Evidence for a Direct Relation between Median Galaxy Size and Median Halo Mass},} \apjl, 872, L13, \dodoi{10.3847/2041-8213/ab0379}

\bibitem[{L. {Mowla} {et~al.}(2024){Mowla}, {Iyer}, {Asada}, {Desprez}, {Tan}, {Martis}, {Sarrouh}, {Strait}, {Abraham}, {Brada{\v{c}}}, {Brammer}, {Muzzin}, {Pacifici}, {Ravindranath}, {Sawicki}, {Willott}, {Estrada-Carpenter}, {Jahan}, {Noirot}, {Matharu}, {Rihtar{\v{s}}i{\v{c}}}, \& {Zabl}}]{Mowla2024}
{Mowla}, L., {Iyer}, K., {Asada}, Y., {et~al.} 2024, \bibinfo{title}{{Formation of a low-mass galaxy from star clusters in a 600-million-year-old Universe},} \nat, 636, 332, \dodoi{10.1038/s41586-024-08293-0}

\bibitem[{L.~A. {Mowla} {et~al.}(2019){Mowla}, {van Dokkum}, {Brammer}, {Momcheva}, {van der Wel}, {Whitaker}, {Nelson}, {Bezanson}, {Muzzin}, {Franx}, {MacKenty}, {Leja}, {Kriek}, \& {Marchesini}}]{mowla2019}
{Mowla}, L.~A., {van Dokkum}, P., {Brammer}, G.~B., {et~al.} 2019, \bibinfo{title}{{COSMOS-DASH: The Evolution of the Galaxy Size-Mass Relation since z {\textasciitilde} 3 from New Wide-field WFC3 Imaging Combined with CANDELS/3D-HST},} \apj, 880, 57, \dodoi{10.3847/1538-4357/ab290a}

\bibitem[{F. {Munshi} {et~al.}(2021){Munshi}, {Brooks}, {Applebaum}, {Christensen}, {Quinn}, \& {Sligh}}]{Munshi2021}
{Munshi}, F., {Brooks}, A.~M., {Applebaum}, E., {et~al.} 2021, \bibinfo{title}{{Quantifying Scatter in Galaxy Formation at the Lowest Masses},} \apj, 923, 35, \dodoi{10.3847/1538-4357/ac0db6}

\bibitem[{A. {Muzzin} {et~al.}(2025){Muzzin}, {Suess}, {Marchesini}, {Robbins}, {Willott}, {Alberts}, {Antwi-Danso}, {Asada}, {Brammer}, {Cutler}, {Iyer}, {Labbe}, {Martis}, {Miller}, {Mitsuhashi}, {Pope}, {Sajina}, {Sarrouh}, {Sharma}, {Stefanon}, {Whitaker}, {Abraham}, {Atek}, {Bradac}, {Berek}, {Bezanson}, {Brown}, {Burgasser}, {Chicoine}, {Cloonan}, {Cooper}, {Dayal}, {de Graaff}, {Desprez}, {Feldmann}, {Forrest}, {Franx}, {Fudamoto}, {Fujimoto}, {Furtak}, {Glazebrook}, {Goovaerts}, {Greene}, {Jagga}, {Jarvis}, {Kriek}, {Khullar}, {La Torre}, {Leja}, {Lin}, {Lorenz}, {Lyon}, {Markov}, {Maseda}, {McConachie}, {Merchant}, {Merida}, {Mowla}, {Myers}, {Naidu}, {Nanayakkara}, {Nelson}, {Noirot}, {Oesch}, {Omori}, {Pan}, {Porraz Barrera}, {Price}, {Ravindranath}, {Sawicki}, {Setton}, {Smit}, {Sok}, {Speagle}, {Taylor}, {Tan}, {Tripodi}, {van der Wel}, {Perez Vidal}, {Wang}, {Weaver}, {Williams}, {Withers}, \& {Zaidi}}]{Muzzin2025}
{Muzzin}, A., {Suess}, K.~A., {Marchesini}, D., {et~al.} 2025, \bibinfo{title}{{MINERVA: A NIRCam Medium Band and MIRI Imaging Survey to Unlock the Hidden Gems of the Distant Universe},} arXiv e-prints, arXiv:2507.19706, \dodoi{10.48550/arXiv.2507.19706}

\bibitem[{R.~P. {Naidu} {et~al.}(2025){Naidu}, {Oesch}, {Brammer}, {Weibel}, {Li}, {Matthee}, {Chisholm}, {Pollock}, {Heintz}, {Johnson}, {Shen}, {Hviding}, {Leja}, {Tacchella}, {Ganguly}, {Witten}, {Atek}, {Belli}, {Bose}, {Bouwens}, {Dayal}, {Decarli}, {de Graaff}, {Fudamoto}, {Giovinazzo}, {Greene}, {Illingworth}, {Inoue}, {Kane}, {Labbe}, {Leonova}, {Marques-Chaves}, {Meyer}, {Nelson}, {Roberts-Borsani}, {Schaerer}, {Simcoe}, {Stefanon}, {Sugahara}, {Toft}, {van der Wel}, {van Dokkum}, {Walter}, {Watson}, {Weaver}, \& {Whitaker}}]{Naidu2025}
{Naidu}, R.~P., {Oesch}, P.~A., {Brammer}, G., {et~al.} 2025, \bibinfo{title}{{A Cosmic Miracle: A Remarkably Luminous Galaxy at $z_{\rm{spec}}=14.44$ Confirmed with JWST},} arXiv e-prints, arXiv:2505.11263, \dodoi{10.48550/arXiv.2505.11263}

\bibitem[{M. {Nakane} {et~al.}(2025){Nakane}, {Kokorev}, {Fujimoto}, {Ouchi}, {McLeod}, {Golubchik}, {Oguri}, {Zitrin}, {Bondestam}, {Donnan}, {Brammer}, {Finkelstein}, {Willott}, {Adamo}, {Vanzella}, {Brada{\v{c}}}, {Messa}, {Yanagisawa}, {Sun}, {Ferguson}, {Lucas}, {Coe}, {Richard}, {Abdurro'uf}, {Akins}, {Allingham}, {Amor{\'\i}n}, {Asada}, {Atek}, {Bezanson}, {Bradley}, {Chisholm}, {Conselice}, {Dayal}, {Dessauges-Zavadsky}, {Diego}, {Faisst}, {Fei}, {Frye}, {Fudamoto}, {Furtak}, {Harikane}, {Hsiao}, {Jim{\'e}nez-Teja}, {Kartaltepe}, {Kiyota}, {Koekemoer}, {Lagos}, {Magdis}, {Meena}, {Mowla}, {Noirot}, {Oesch}, {Ono}, {Ortiz}, {Pan}, {Papovich}, {Pierel}, {Ricotti}, {Robbins}, {Schaerer}, {Schneider}, {Treu}, {Valentino}, {Windhorst}, {Bauer}, {Bromm}, {Egami}, {Gonz{\'a}lez-Otero}, {Kohno}, {Labbe}, {Matthee}, {Mun}, {Naidu}, \& {Tripodi}}]{Nakane2025}
{Nakane}, M., {Kokorev}, V., {Fujimoto}, S., {et~al.} 2025, \bibinfo{title}{{VENUS: A Strongly Lensed Clumpy Galaxy at $z\sim11-12$ behind the Galaxy Cluster MACS J0257.1-2325},} arXiv e-prints, arXiv:2511.14483, \dodoi{10.48550/arXiv.2511.14483}

\bibitem[{K.~V. {Nedkova} {et~al.}(2021){Nedkova}, {H{\"a}u{\ss}ler}, {Marchesini}, {Dimauro}, {Brammer}, {Eigenthaler}, {Feinstein}, {Ferguson}, {Huertas-Company}, {Johnston}, {Kado-Fong}, {Kartaltepe}, {Labb{\'e}}, {Lange-Vagle}, {Martis}, {McGrath}, {Muzzin}, {Oesch}, {Ordenes-Brice{\~n}o}, {Puzia}, {Shipley}, {Simmons}, {Skelton}, {Stefanon}, {van der Wel}, \& {Whitaker}}]{nedkova2021}
{Nedkova}, K.~V., {H{\"a}u{\ss}ler}, B., {Marchesini}, D., {et~al.} 2021, \bibinfo{title}{{Extending the evolution of the stellar mass-size relation at z {\ensuremath{\leq}} 2 to low stellar mass galaxies from HFF and CANDELS},} \mnras, 506, 928, \dodoi{10.1093/mnras/stab1744}

\bibitem[{K.~V. {Nedkova} {et~al.}(2024){Nedkova}, {Rafelski}, {Teplitz}, {Mehta}, {Degroot}, {Ravindranath}, {Alavi}, {Beckett}, {Grogin}, {H{\"a}u{\ss}ler}, {Koekemoer}, {Oyarz{\'u}n}, {Prichard}, {Revalski}, {Snyder}, {Sunnquist}, {Wang}, {Windhorst}, {Chartab}, {Conselice}, {Guo}, {Hathi}, {Hayes}, {Ji}, {Kim}, {Lucas}, {Mobasher}, {O'Connell}, {Sattari}, {Smith}, {Taamoli}, {Yung}, \& {The Uvcandels Team}}]{Nedkova2024}
{Nedkova}, K.~V., {Rafelski}, M., {Teplitz}, H.~I., {et~al.} 2024, \bibinfo{title}{{UVCANDELS: The Role of Dust on the Stellar Mass{\textendash}Size Relation of Disk Galaxies at 0.5 {\ensuremath{\leq}} z {\ensuremath{\leq}} 3.0},} \apj, 970, 188, \dodoi{10.3847/1538-4357/ad4ede}

\bibitem[{E.~J. {Nelson} {et~al.}(2016){Nelson}, {van Dokkum}, {Momcheva}, {Brammer}, {Wuyts}, {Franx}, {F{\"o}rster Schreiber}, {Whitaker}, \& {Skelton}}]{nelson2016b}
{Nelson}, E.~J., {van Dokkum}, P.~G., {Momcheva}, I.~G., {et~al.} 2016, \bibinfo{title}{{Spatially Resolved Dust Maps from Balmer Decrements in Galaxies at z \raisebox{-0.5ex}\textasciitilde 1.4},} \apjl, 817, L9, \dodoi{10.3847/2041-8205/817/1/L9}

\bibitem[{E.~J. {Nelson} {et~al.}(2021){Nelson}, {Tacchella}, {Diemer}, {Leja}, {Hernquist}, {Whitaker}, {Weinberger}, {Pillepich}, {Nelson}, {Terrazas}, {Nevin}, {Brammer}, {Burkhart}, {Cochrane}, {van Dokkum}, {Johnson}, {Marinacci}, {Mowla}, {Pakmor}, {Skelton}, {Speagle}, {Springel}, {Torrey}, {Vogelsberger}, \& {Wuyts}}]{Nelson2021}
{Nelson}, E.~J., {Tacchella}, S., {Diemer}, B., {et~al.} 2021, \bibinfo{title}{{Spatially resolved star formation and inside-out quenching in the TNG50 simulation and 3D-HST observations},} \mnras, 508, 219, \dodoi{10.1093/mnras/stab2131}

\bibitem[{A.~B. {Newman} {et~al.}(2012){Newman}, {Ellis}, {Bundy}, \& {Treu}}]{newman2012}
{Newman}, A.~B., {Ellis}, R.~S., {Bundy}, K., \& {Treu}, T. 2012, \bibinfo{title}{{Can Minor Merging Account for the Size Growth of Quiescent Galaxies? New Results from the CANDELS Survey},} \apj, 746, 162, \dodoi{10.1088/0004-637X/746/2/162}

\bibitem[{J.~B. {Oke} \& J.~E. {Gunn}(1983){Oke} \& {Gunn}}]{Oke1983}
{Oke}, J.~B., \& {Gunn}, J.~E. 1983, \bibinfo{title}{{Secondary standard stars for absolute spectrophotometry.},} \apj, 266, 713, \dodoi{10.1086/160817}

\bibitem[{Y. {Ono} {et~al.}(2025){Ono}, {Ouchi}, {Harikane}, {Yajima}, {Nakajima}, {Fujimoto}, {Nakane}, \& {Xu}}]{Ono2025}
{Ono}, Y., {Ouchi}, M., {Harikane}, Y., {et~al.} 2025, \bibinfo{title}{{Morphological Demographics of Galaxies at z {\ensuremath{\sim}} 10{\textendash}16: Log-normal Size Distribution and Exponential Profiles Consistent with the Disk Formation Scenario},} \apj, 991, 222, \dodoi{10.3847/1538-4357/adfc4d}

\bibitem[{K. Ormerod {et~al.}(2024)Ormerod, Conselice, Adams, Harvey, Austin, Trussler, Ferreira, Caruana, Lucatelli, Li, {et~al.}}]{ormerod2024}
Ormerod, K., Conselice, C., Adams, N., {et~al.} 2024, \bibinfo{title}{EPOCHS VI: the size and shape evolution of galaxies since z~ 8 with JWST Observations,} Monthly Notices of the Royal Astronomical Society, 527, 6110

\bibitem[{M.~E. {Orr} {et~al.}(2017){Orr}, {Hayward}, {Nelson}, {Hopkins}, {Faucher-Gigu{\`e}re}, {Kere{\v{s}}}, {Chan}, {Schmitz}, \& {Miller}}]{Orr2017}
{Orr}, M.~E., {Hayward}, C.~C., {Nelson}, E.~J., {et~al.} 2017, \bibinfo{title}{{Stacked Star Formation Rate Profiles of Bursty Galaxies Exhibit {\textquotedblleft}Coherent{\textquotedblright} Star Formation},} \apjl, 849, L2, \dodoi{10.3847/2041-8213/aa8f93}

\bibitem[{I. {Pasha} \& T.~B. {Miller}(2023){Pasha} \& {Miller}}]{pasha2023}
{Pasha}, I., \& {Miller}, T.~B. 2023, \bibinfo{title}{{pysersic: A Python package for determining galaxy structural properties via Bayesian inference, accelerated with jax},} The Journal of Open Source Software, 8, 5703, \dodoi{10.21105/joss.05703}

\bibitem[{F. Pedregosa {et~al.}(2011)Pedregosa, Varoquaux, Gramfort, Michel, Thirion, Grisel, Blondel, Prettenhofer, Weiss, Dubourg, Vanderplas, Passos, Cournapeau, Brucher, Perrot, \& Duchesnay}]{scikit-learn}
Pedregosa, F., Varoquaux, G., Gramfort, A., {et~al.} 2011, \bibinfo{title}{Scikit-learn: Machine Learning in {P}ython,} Journal of Machine Learning Research, 12, 2825

\bibitem[{C.~Y. {Peng} {et~al.}(2010){Peng}, {Ho}, {Impey}, \& {Rix}}]{peng2010}
{Peng}, C.~Y., {Ho}, L.~C., {Impey}, C.~D., \& {Rix}, H.-W. 2010, \bibinfo{title}{{Detailed Decomposition of Galaxy Images. II. Beyond Axisymmetric Models},} \aj, 139, 2097, \dodoi{10.1088/0004-6256/139/6/2097}

\bibitem[{D. Phan {et~al.}(2019)Phan, Pradhan, \& Jankowiak}]{phan2019}
Phan, D., Pradhan, N., \& Jankowiak, M. 2019, \bibinfo{title}{Composable Effects for Flexible and Accelerated Probabilistic Programming in NumPyro,} arXiv preprint arXiv:1912.11554

\bibitem[{A. {Pillepich} {et~al.}(2018){Pillepich}, {Nelson}, {Hernquist}, {Springel}, {Pakmor}, {Torrey}, {Weinberger}, {Genel}, {Naiman}, {Marinacci}, \& {Vogelsberger}}]{pillepich2018}
{Pillepich}, A., {Nelson}, D., {Hernquist}, L., {et~al.} 2018, \bibinfo{title}{{First results from the IllustrisTNG simulations: the stellar mass content of groups and clusters of galaxies},} \mnras, 475, 648, \dodoi{10.1093/mnras/stx3112}

\bibitem[{M. {Postman} \& M.~J. {Geller}(1984){Postman} \& {Geller}}]{Postman1984}
{Postman}, M., \& {Geller}, M.~J. 1984, \bibinfo{title}{{The morphology-density relation - The group connection.},} \apj, 281, 95, \dodoi{10.1086/162078}

\bibitem[{L. {Pozzetti} {et~al.}(2010){Pozzetti}, {Bolzonella}, {Zucca}, {Zamorani}, {Lilly}, {Renzini}, {Moresco}, {Mignoli}, {Cassata}, {Tasca}, {Lamareille}, {Maier}, {Meneux}, {Halliday}, {Oesch}, {Vergani}, {Caputi}, {Kova{\v{c}}}, {Cimatti}, {Cucciati}, {Iovino}, {Peng}, {Carollo}, {Contini}, {Kneib}, {Le F{\'e}vre}, {Mainieri}, {Scodeggio}, {Bardelli}, {Bongiorno}, {Coppa}, {de la Torre}, {de Ravel}, {Franzetti}, {Garilli}, {Kampczyk}, {Knobel}, {Le Borgne}, {Le Brun}, {Pell{\`o}}, {Perez Montero}, {Ricciardelli}, {Silverman}, {Tanaka}, {Tresse}, {Abbas}, {Bottini}, {Cappi}, {Guzzo}, {Koekemoer}, {Leauthaud}, {Maccagni}, {Marinoni}, {McCracken}, {Memeo}, {Porciani}, {Scaramella}, {Scarlata}, \& {Scoville}}]{Pozzetti2010}
{Pozzetti}, L., {Bolzonella}, M., {Zucca}, E., {et~al.} 2010, \bibinfo{title}{{zCOSMOS - 10k-bright spectroscopic sample. The bimodality in the galaxy stellar mass function: exploring its evolution with redshift},} \aap, 523, A13, \dodoi{10.1051/0004-6361/200913020}

\bibitem[{S.~H. {Price} {et~al.}(2024){Price}, {Bezanson}, {Labbe}, {Furtak}, {de Graaff}, {Greene}, {Kokorev}, {Setton}, {Suess}, {Brammer}, {Cutler}, {Leja}, {Pan}, {Wang}, {Weaver}, {Whitaker}, {Atek}, {Burgasser}, {Chemerynska}, {Dayal}, {Feldmann}, {F{\"o}rster Schreiber}, {Fudamoto}, {Fujimoto}, {Glazebrook}, {Goulding}, {Khullar}, {Kriek}, {Marchesini}, {Maseda}, {Miller}, {Muzzin}, {Nanayakkara}, {Nelson}, {Oesch}, {Shipley}, {Smit}, {Taylor}, {van Dokkum}, {Williams}, \& {Zitrin}}]{Price2024}
{Price}, S.~H., {Bezanson}, R., {Labbe}, I., {et~al.} 2024, \bibinfo{title}{{The UNCOVER Survey: First Release of Ultradeep JWST/NIRSpec PRISM spectra for \raisebox{-0.5ex}\textasciitilde700 galaxies from z\raisebox{-0.5ex}\textasciitilde0.3-13 in Abell 2744},} arXiv e-prints, arXiv:2408.03920, \dodoi{10.48550/arXiv.2408.03920}

\bibitem[{M.~J. {Rieke} {et~al.}(2023){Rieke}, {Kelly}, {Misselt}, {Stansberry}, {Boyer}, {Beatty}, {Egami}, {Florian}, {Greene}, {Hainline}, {Leisenring}, {Roellig}, {Schlawin}, {Sun}, {Tinnin}, {Williams}, {Willmer}, {Wilson}, {Clark}, {Rohrbach}, {Brooks}, {Canipe}, {Correnti}, {DiFelice}, {Gennaro}, {Girard}, {Hartig}, {Hilbert}, {Koekemoer}, {Nikolov}, {Pirzkal}, {Rest}, {Robberto}, {Sunnquist}, {Telfer}, {Wu}, {Ferry}, {Lewis}, {Baum}, {Beichman}, {Doyon}, {Dressler}, {Eisenstein}, {Ferrarese}, {Hodapp}, {Horner}, {Jaffe}, {Johnstone}, {Krist}, {Martin}, {McCarthy}, {Meyer}, {Rieke}, {Trauger}, \& {Young}}]{Rieke2023}
{Rieke}, M.~J., {Kelly}, D.~M., {Misselt}, K., {et~al.} 2023, \bibinfo{title}{{Performance of NIRCam on JWST in Flight},} \pasp, 135, 028001, \dodoi{10.1088/1538-3873/acac53}

\bibitem[{B.~E. {Robertson} {et~al.}(2023){Robertson}, {Tacchella}, {Johnson}, {Hainline}, {Whitler}, {Eisenstein}, {Endsley}, {Rieke}, {Stark}, {Alberts}, {Dressler}, {Egami}, {Hausen}, {Rieke}, {Shivaei}, {Williams}, {Willmer}, {Arribas}, {Bonaventura}, {Bunker}, {Cameron}, {Carniani}, {Charlot}, {Chevallard}, {Curti}, {Curtis-Lake}, {D'Eugenio}, {Jakobsen}, {Looser}, {L{\"u}tzgendorf}, {Maiolino}, {Maseda}, {Rawle}, {Rix}, {Smit}, {{\"U}bler}, {Willott}, {Witstok}, {Baum}, {Bhatawdekar}, {Boyett}, {Chen}, {de Graaff}, {Florian}, {Helton}, {Hviding}, {Ji}, {Kumari}, {Lyu}, {Nelson}, {Sandles}, {Saxena}, {Suess}, {Sun}, {Topping}, \& {Wallace}}]{Robertson2023}
{Robertson}, B.~E., {Tacchella}, S., {Johnson}, B.~D., {et~al.} 2023, \bibinfo{title}{{Identification and properties of intense star-forming galaxies at redshifts z > 10},} Nature Astronomy, 7, 611, \dodoi{10.1038/s41550-023-01921-1}

\bibitem[{E. {Rohr} {et~al.}(2022){Rohr}, {Feldmann}, {Bullock}, {{\c{C}}atmabacak}, {Boylan-Kolchin}, {Faucher-Gigu{\`e}re}, {Kere{\v{s}}}, {Liang}, {Moreno}, \& {Wetzel}}]{Rohr2022}
{Rohr}, E., {Feldmann}, R., {Bullock}, J.~S., {et~al.} 2022, \bibinfo{title}{{The galaxy-halo size relation of low-mass galaxies in FIRE},} \mnras, 510, 3967, \dodoi{10.1093/mnras/stab3625}

\bibitem[{S. Shen {et~al.}(2003)Shen, Mo, White, Blanton, Kauffmann, Voges, Brinkmann, \& Csabai}]{shen2003}
Shen, S., Mo, H.~J., White, S. D.~M., {et~al.} 2003, \bibinfo{title}{{The size distribution of galaxies in the Sloan Digital Sky Survey},} Monthly Notice of the Royal Astronomical Society, 343, 978

\bibitem[{M. {Shuntov} {et~al.}(2022){Shuntov}, {McCracken}, {Gavazzi}, {Laigle}, {Weaver}, {Davidzon}, {Ilbert}, {Kauffmann}, {Faisst}, {Dubois}, {Koekemoer}, {Moneti}, {Milvang-Jensen}, {Mobasher}, {Sanders}, \& {Toft}}]{Shuntov2022}
{Shuntov}, M., {McCracken}, H.~J., {Gavazzi}, R., {et~al.} 2022, \bibinfo{title}{{COSMOS2020: Cosmic evolution of the stellar-to-halo mass relation for central and satellite galaxies up to z {\ensuremath{\sim}} 5},} \aap, 664, A61, \dodoi{10.1051/0004-6361/202243136}

\bibitem[{C. {Simmonds} {et~al.}(2025){Simmonds}, {Tacchella}, {Curtis-Lake}, {D'Eugenio}, {Hainline}, {Johnson}, {Kravtsov}, {Pusk{\~A}{\textexclamdown}s}, {Robertson}, {Stoffers}, {Willott}, {Baker}, {Belokurov}, {Bhatawdekar}, {Bunker}, {Carniani}, {Chevallard}, {Curti}, {Duan}, {Helton}, {Ji}, {Looser}, {Maiolino}, {Maseda}, {Shivaei}, \& {Williams}}]{Simmonds2025}
{Simmonds}, C., {Tacchella}, S., {Curtis-Lake}, W. M.~E., {et~al.} 2025, \bibinfo{title}{{Bursting at the seams: the star-forming main sequence and its scatter at z=3-9 using NIRCam photometry from JADES},} \mnras, \dodoi{10.1093/mnras/staf1950}

\bibitem[{D. Simpson {et~al.}(2017)Simpson, Rue, Riebler, Martins, \& Sorbye}]{Simpson2017}
Simpson, D., Rue, H., Riebler, A., Martins, T.~G., \& Sorbye, S.~H. 2017, \bibinfo{title}{Penalising Model Component Complexity: A Principled, Practical Approach to Constructing Priors,} Statistical Science, 32, 1, \dodoi{10.1214/16-STS576}

\bibitem[{A. {Smercina} {et~al.}(2018){Smercina}, {Bell}, {Price}, {D'Souza}, {Slater}, {Bailin}, {Monachesi}, \& {Nidever}}]{Smercina2018}
{Smercina}, A., {Bell}, E.~F., {Price}, P.~A., {et~al.} 2018, \bibinfo{title}{{A Lonely Giant: The Sparse Satellite Population of M94 Challenges Galaxy Formation},} \apj, 863, 152, \dodoi{10.3847/1538-4357/aad2d6}

\bibitem[{R.~S. Somerville {et~al.}(2018)Somerville, Behroozi, Pandya, Dekel, Faber, Fontana, Koekemoer, Koo, P{\'{e}}rez-Gonz{\'{a}}lez, Primack, Santini, Taylor, \& van~der Wel}]{somerville2018}
Somerville, R.~S., Behroozi, P., Pandya, V., {et~al.} 2018, \bibinfo{title}{{The relationship between galaxy and dark matter halo size from z {\~{}} 3 to the present},} Monthly Notices of the Royal Astronomical Society, 473, 2714

\bibitem[{J. {Song} {et~al.}(2025){Song}, {Wang}, {Jia}, {Lyu}, {Chen}, {Wang}, {Li}, {Ding}, {Fang}, \& {Kong}}]{Song2025}
{Song}, J., {Wang}, E., {Jia}, C., {et~al.} 2025, \bibinfo{title}{{Transition from Outside-in to Inside-Out at $z\sim 2$: Evidence from Radial Profiles of Specific Star Formation Rate based on JWST/HST},} arXiv e-prints, arXiv:2512.01684, \dodoi{10.48550/arXiv.2512.01684}

\bibitem[{J. {Stern} {et~al.}(2021){Stern}, {Faucher-Gigu{\`e}re}, {Fielding}, {Quataert}, {Hafen}, {Gurvich}, {Ma}, {Byrne}, {El-Badry}, {Angl{\'e}s-Alc{\'a}zar}, {Chan}, {Feldmann}, {Kere{\v{s}}}, {Wetzel}, {Murray}, \& {Hopkins}}]{Stern2021}
{Stern}, J., {Faucher-Gigu{\`e}re}, C.-A., {Fielding}, D., {et~al.} 2021, \bibinfo{title}{{Virialization of the Inner CGM in the FIRE Simulations and Implications for Galaxy Disks, Star Formation, and Feedback},} \apj, 911, 88, \dodoi{10.3847/1538-4357/abd776}

\bibitem[{K.~A. {Suess} {et~al.}(2019){Suess}, {Kriek}, {Price}, \& {Barro}}]{suess2019}
{Suess}, K.~A., {Kriek}, M., {Price}, S.~H., \& {Barro}, G. 2019, \bibinfo{title}{{Half-mass Radii for ̃7000 Galaxies at 1.0 {\ensuremath{\leq}} z {\ensuremath{\leq}} 2.5: Most of the Evolution in the Mass-Size Relation Is Due to Color Gradients},} \apj, 877, 103, \dodoi{10.3847/1538-4357/ab1bda}

\bibitem[{K.~A. {Suess} {et~al.}(2021){Suess}, {Kriek}, {Price}, \& {Barro}}]{suess2021}
{Suess}, K.~A., {Kriek}, M., {Price}, S.~H., \& {Barro}, G. 2021, \bibinfo{title}{{Dissecting the Size-Mass and {\ensuremath{\Sigma}}$_{1}$-Mass Relations at 1.0 < z < 2.5: Galaxy Mass Profiles and Color Gradients as a Function of Spectral Shape},} \apj, 915, 87, \dodoi{10.3847/1538-4357/abf1e4}

\bibitem[{K.~A. {Suess} {et~al.}(2022){Suess}, {Bezanson}, {Nelson}, {Setton}, {Price}, {Dokkum}, {Brammer}, {Labb{\'e}}, {Leja}, {Miller}, {Robertson}, {Wel}, {Weaver}, \& {Whitaker}}]{suess2022}
{Suess}, K.~A., {Bezanson}, R., {Nelson}, E.~J., {et~al.} 2022, \bibinfo{title}{{Rest-frame Near-infrared Sizes of Galaxies at Cosmic Noon: Objects in JWST's Mirror Are Smaller than They Appeared},} \apjl, 937, L33, \dodoi{10.3847/2041-8213/ac8e06}

\bibitem[{K.~A. {Suess} {et~al.}(2024){Suess}, {Weaver}, {Price}, {Pan}, {Wang}, {Bezanson}, {Brammer}, {Cutler}, {Labbe}, {Leja}, {Williams}, {Whitaker}, {Dayal}, {de Graaff}, {Feldmann}, {Franx}, {Fudamoto}, {Fujimoto}, {Furtak}, {Goulding}, {Greene}, {Khullar}, {Kokorev}, {Kriek}, {Lorenz}, {Marchesini}, {Maseda}, {Matthee}, {Miller}, {Mitsuhashi}, {Mowla}, {Muzzin}, {Naidu}, {Nanayakkara}, {Nelson}, {Oesch}, {Setton}, {Shipley}, {Smit}, {Spilker}, {van Dokkum}, \& {Zitrin}}]{Suess2024}
{Suess}, K.~A., {Weaver}, J.~R., {Price}, S.~H., {et~al.} 2024, \bibinfo{title}{{Medium Bands, Mega Science: a JWST/NIRCam Medium-Band Imaging Survey of Abell 2744},} arXiv e-prints, arXiv:2404.13132, \dodoi{10.48550/arXiv.2404.13132}

\bibitem[{S. Tacchella {et~al.}(2018)Tacchella, Carollo, Schreiber, Renzini, Dekel, Genzel, Lang, Lilly, Mancini, Onodera, {et~al.}}]{tacchella2018}
Tacchella, S., Carollo, C.~M., Schreiber, N.~F., {et~al.} 2018, \bibinfo{title}{Dust Attenuation, Bulge Formation, and Inside-out Quenching of Star Formation in Star-forming Main Sequence Galaxies at z~ 2,} The Astrophysical Journal, 859, 56

\bibitem[{T. {Treu} {et~al.}(2023){Treu}, {Calabr{\`o}}, {Castellano}, {Leethochawalit}, {Merlin}, {Fontana}, {Yang}, {Morishita}, {Trenti}, {Dressler}, {Mason}, {Paris}, {Pentericci}, {Roberts-Borsani}, {Vulcani}, {Boyett}, {Bradac}, {Glazebrook}, {Jones}, {Marchesini}, {Mascia}, {Nanayakkara}, {Santini}, {Strait}, {Vanzella}, \& {Wang}}]{Treu2023}
{Treu}, T., {Calabr{\`o}}, A., {Castellano}, M., {et~al.} 2023, \bibinfo{title}{{Early Results From GLASS-JWST. XII. The Morphology of Galaxies at the Epoch of Reionization},} \apjl, 942, L28, \dodoi{10.3847/2041-8213/ac9283}

\bibitem[{I. Trujillo {et~al.}(2006)Trujillo, Forster~Schreiber, Rudnick, Barden, Franx, Rix, Caldwell, McIntosh, Toft, Haussler, Zirm, van Dokkum, Labbe, Moorwood, Rottgering, van~der Wel, van~der Werf, \& van Starkenburg}]{trujillo2006}
Trujillo, I., Forster~Schreiber, N.~M., Rudnick, G., {et~al.} 2006, \bibinfo{title}{{The Size Evolution of Galaxies since z ∼3: Combining SDSS, GEMS, and FIRES},} The Astrophysical Journal, 650, 18

\bibitem[{G. {van de Ven} \& A. {van der Wel}(2021){van de Ven} \& {van der Wel}}]{vandeven2021}
{van de Ven}, G., \& {van der Wel}, A. 2021, \bibinfo{title}{{Deprojecting S{\'e}rsic Profiles for Arbitrary Triaxial Shapes: Robust Measures of Intrinsic and Projected Galaxy Sizes},} \apj, 914, 45, \dodoi{10.3847/1538-4357/abf047}

\bibitem[{A. {van der Wel} {et~al.}(2012){van der Wel}, {Bell}, {H{\"a}ussler}, {McGrath}, {Chang}, {Guo}, {McIntosh}, {Rix}, {Barden}, {Cheung}, {Faber}, {Ferguson}, {Galametz}, {Grogin}, {Hartley}, {Kartaltepe}, {Kocevski}, {Koekemoer}, {Lotz}, {Mozena}, {Peth}, \& {Peng}}]{vanderwel2012}
{van der Wel}, A., {Bell}, E.~F., {H{\"a}ussler}, B., {et~al.} 2012, \bibinfo{title}{{Structural Parameters of Galaxies in CANDELS},} \apjs, 203, 24, \dodoi{10.1088/0067-0049/203/2/24}

\bibitem[{A. {van der Wel} {et~al.}(2014){van der Wel}, {Franx}, {van Dokkum}, {Skelton}, {Momcheva}, {Whitaker}, {Brammer}, {Bell}, {Rix}, {Wuyts}, {Ferguson}, {Holden}, {Barro}, {Koekemoer}, {Chang}, {McGrath}, {H{\"a}ussler}, {Dekel}, {Behroozi}, {Fumagalli}, {Leja}, {Lundgren}, {Maseda}, {Nelson}, {Wake}, {Patel}, {Labb{\'e}}, {Faber}, {Grogin}, \& {Kocevski}}]{vanderwel2014}
{van der Wel}, A., {Franx}, M., {van Dokkum}, P.~G., {et~al.} 2014, \bibinfo{title}{{3D-HST+CANDELS: The Evolution of the Galaxy Size-Mass Distribution since z = 3},} \apj, 788, 28, \dodoi{10.1088/0004-637X/788/1/28}

\bibitem[{A. {van der Wel} {et~al.}(2024){van der Wel}, {Martorano}, {H{\"a}u{\ss}ler}, {Nedkova}, {Miller}, {Brammer}, {van de Ven}, {Leja}, {Bezanson}, {Muzzin}, {Marchesini}, {de Graaff}, {Nelson}, {Kriek}, {Bell}, \& {Franx}}]{vanderwel2024}
{van der Wel}, A., {Martorano}, M., {H{\"a}u{\ss}ler}, B., {et~al.} 2024, \bibinfo{title}{{Stellar Half-mass Radii of 0.5 z < 2.3 Galaxies: Comparison with JWST/NIRCam Half-light Radii},} \apj, 960, 53, \dodoi{10.3847/1538-4357/ad02ee}

\bibitem[{P.~G. {van Dokkum} {et~al.}(2008){van Dokkum}, {Franx}, {Kriek}, {Holden}, {Illingworth}, {Magee}, {Bouwens}, {Marchesini}, {Quadri}, {Rudnick}, {Taylor}, \& {Toft}}]{vandokkum2008}
{van Dokkum}, P.~G., {Franx}, M., {Kriek}, M., {et~al.} 2008, \bibinfo{title}{{Confirmation of the Remarkable Compactness of Massive Quiescent Galaxies at z \raisebox{-0.5ex}\textasciitilde 2.3: Early-Type Galaxies Did not Form in a Simple Monolithic Collapse},} \apjl, 677, L5, \dodoi{10.1086/587874}

\bibitem[{P.~G. van Dokkum {et~al.}(2015)van Dokkum, Nelson, Franx, Oesch, Momcheva, Brammer, Schreiber, Skelton, Whitaker, Wel, Bezanson, Fumagalli, Illingworth, Kriek, Leja, \& Wuyts}]{vandokkum2015}
van Dokkum, P.~G., Nelson, E.~J., Franx, M., {et~al.} 2015, \bibinfo{title}{{Forming Compact Massive Galaxies},} Astrophysical Journal, 813

\bibitem[{A. Vehtari {et~al.}(2021)Vehtari, Gelman, Simpson, Carpenter, \& B{\"u}rkner}]{vehtari2021}
Vehtari, A., Gelman, A., Simpson, D., Carpenter, B., \& B{\"u}rkner, P.-C. 2021, \bibinfo{title}{Rank-normalization, folding, and localization: An improved R ̂ for assessing convergence of MCMC (with discussion),} Bayesian analysis, 16, 667

\bibitem[{B. {Wang} {et~al.}(2023{\natexlab{a}}){Wang}, {Leja}, {Bezanson}, {Johnson}, {Khullar}, {Labb{\'e}}, {Price}, {Weaver}, \& {Whitaker}}]{Wang2023}
{Wang}, B., {Leja}, J., {Bezanson}, R., {et~al.} 2023{\natexlab{a}}, \bibinfo{title}{{Inferring More from Less: Prospector as a Photometric Redshift Engine in the Era of JWST},} \apjl, 944, L58, \dodoi{10.3847/2041-8213/acba99}

\bibitem[{B. {Wang} {et~al.}(2023{\natexlab{b}}){Wang}, {Fujimoto}, {Labb{\'e}}, {Furtak}, {Miller}, {Setton}, {Zitrin}, {Atek}, {Bezanson}, {Brammer}, {Leja}, {Oesch}, {Price}, {Chemerynska}, {Cutler}, {Dayal}, {van Dokkum}, {Goulding}, {Greene}, {Fudamoto}, {Khullar}, {Kokorev}, {Marchesini}, {Pan}, {Weaver}, {Whitaker}, \& {Williams}}]{Wang2023b}
{Wang}, B., {Fujimoto}, S., {Labb{\'e}}, I., {et~al.} 2023{\natexlab{b}}, \bibinfo{title}{{UNCOVER: Illuminating the Early Universe-JWST/NIRSpec Confirmation of z > 12 Galaxies},} \apjl, 957, L34, \dodoi{10.3847/2041-8213/acfe07}

\bibitem[{B. {Wang} {et~al.}(2024{\natexlab{a}}){Wang}, {Leja}, {Labb{\'e}}, {Bezanson}, {Whitaker}, {Brammer}, {Furtak}, {Weaver}, {Price}, {Zitrin}, {Atek}, {Coe}, {Cutler}, {Dayal}, {van Dokkum}, {Feldmann}, {Marchesini}, {Franx}, {F{\"o}rster Schreiber}, {Fujimoto}, {Geha}, {Glazebrook}, {de Graaff}, {Greene}, {Juneau}, {Kassin}, {Kriek}, {Khullar}, {Maseda}, {Mowla}, {Muzzin}, {Nanayakkara}, {Nelson}, {Oesch}, {Pacifici}, {Pan}, {Papovich}, {Setton}, {Shapley}, {Smit}, {Stefanon}, {Suess}, {Taylor}, \& {Williams}}]{Wang2024}
{Wang}, B., {Leja}, J., {Labb{\'e}}, I., {et~al.} 2024{\natexlab{a}}, \bibinfo{title}{{The UNCOVER Survey: A First-look HST+JWST Catalog of Galaxy Redshifts and Stellar Population Properties Spanning 0.2 {\ensuremath{\lesssim}} z {\ensuremath{\lesssim}} 15},} \apjs, 270, 12, \dodoi{10.3847/1538-4365/ad0846}

\bibitem[{B. {Wang} {et~al.}(2024{\natexlab{b}}){Wang}, {Leja}, {de Graaff}, {Brammer}, {Weibel}, {van Dokkum}, {Baggen}, {Suess}, {Greene}, {Bezanson}, {Cleri}, {Hirschmann}, {Labb{\'e}}, {Matthee}, {McConachie}, {Naidu}, {Nelson}, {Oesch}, {Setton}, \& {Williams}}]{Wang2024b}
{Wang}, B., {Leja}, J., {de Graaff}, A., {et~al.} 2024{\natexlab{b}}, \bibinfo{title}{{RUBIES: Evolved Stellar Populations with Extended Formation Histories at z {\ensuremath{\sim}} 7{\textendash}8 in Candidate Massive Galaxies Identified with JWST/NIRSpec},} \apjl, 969, L13, \dodoi{10.3847/2041-8213/ad55f7}

\bibitem[{E. Ward {et~al.}(2024)Ward, de~la Vega, Mobasher, McGrath, Iyer, Calabro, Costantin, Dickinson, Holwerda, Hirschmann, {et~al.}}]{ward2024}
Ward, E., de~la Vega, A., Mobasher, B., {et~al.} 2024, \bibinfo{title}{Evolution of the Size--Mass Relation of Star-forming Galaxies Since z= 5.5 Revealed by CEERS,} The Astrophysical Journal, 962, 176

\bibitem[{J.~R. {Weaver} {et~al.}(2024){Weaver}, {Cutler}, {Pan}, {Whitaker}, {Labb{\'e}}, {Price}, {Bezanson}, {Brammer}, {Marchesini}, {Leja}, {Wang}, {Furtak}, {Zitrin}, {Atek}, {Chemerynska}, {Coe}, {Dayal}, {van Dokkum}, {Feldmann}, {F{\"o}rster Schreiber}, {Franx}, {Fujimoto}, {Fudamoto}, {Glazebrook}, {de Graaff}, {Greene}, {Juneau}, {Kassin}, {Kriek}, {Khullar}, {Maseda}, {Mowla}, {Muzzin}, {Nanayakkara}, {Nelson}, {Oesch}, {Pacifici}, {Papovich}, {Setton}, {Shapley}, {Shipley}, {Smit}, {Stefanon}, {Taylor}, {Weibel}, \& {Williams}}]{Weaver2024}
{Weaver}, J.~R., {Cutler}, S.~E., {Pan}, R., {et~al.} 2024, \bibinfo{title}{{The UNCOVER Survey: A First-look HST + JWST Catalog of 60,000 Galaxies near A2744 and beyond},} \apjs, 270, 7, \dodoi{10.3847/1538-4365/ad07e0}

\bibitem[{R.~H. {Wechsler} \& J.~L. {Tinker}(2018){Wechsler} \& {Tinker}}]{Wechsler2018}
{Wechsler}, R.~H., \& {Tinker}, J.~L. 2018, \bibinfo{title}{{The Connection Between Galaxies and Their Dark Matter Halos},} \araa, 56, 435, \dodoi{10.1146/annurev-astro-081817-051756}

\bibitem[{K.~E. Whitaker {et~al.}(2017)Whitaker, Bezanson, van Dokkum, Franx, van~der Wel, Brammer, Forster-Schreiber, Giavalisco, Labbe, Momcheva, Nelson, \& Skelton}]{whitaker2017}
Whitaker, K.~E., Bezanson, R., van Dokkum, P.~G., {et~al.} 2017, \bibinfo{title}{{Predicting Quiescence: The Dependence of Specific Star Formation Rate on Galaxy Size and Central Density at 0.5<z<2.5},} The Astrophysical Journal, 838

\bibitem[{L. {Yang} {et~al.}(2025){Yang}, {Kartaltepe}, {Franco}, {Ding}, {Achenbach}, {Arango-Toro}, {Casey}, {Drakos}, {Faisst}, {Gillman}, {Gozaliasl}, {Huertas-Company}, {Jin}, {Liu}, {Magdis}, {Massey}, {Silverman}, {Tanaka}, {Yu}, {Akins}, {Allen}, {Ilbert}, {Koekemoer}, {McCracken}, {Paquereau}, {Rhodes}, {Robertson}, {Shuntov}, \& {Toft}}]{Yang2025}
{Yang}, L., {Kartaltepe}, J.~S., {Franco}, M., {et~al.} 2025, \bibinfo{title}{{COSMOS-Web: Unraveling the Evolution of Galaxy Size and Related Properties at $2<z<10$},} arXiv e-prints, arXiv:2504.07185, \dodoi{10.48550/arXiv.2504.07185}

\bibitem[{L. {Zanisi} {et~al.}(2020){Zanisi}, {Shankar}, {Lapi}, {Menci}, {Bernardi}, {Duckworth}, {Huertas-Company}, {Grylls}, \& {Salucci}}]{Zanisi2020}
{Zanisi}, L., {Shankar}, F., {Lapi}, A., {et~al.} 2020, \bibinfo{title}{{Galaxy sizes and the galaxy-halo connection - I. The remarkable tightness of the size distributions},} \mnras, 492, 1671, \dodoi{10.1093/mnras/stz3516}

\bibitem[{Y. Zhang {et~al.}(Submitted)Zhang, Miller, Price, \& Suess}]{zhang2026}
Zhang, Y., Miller, T.~B., Price, S., \& Suess, K. Submitted, \bibinfo{title}{Everything Every Band All at Once I: A Global Morphology Catalog in Abell 2744 based on UNCOVER/MegaScience,}

\bibitem[{Y. {Zhang} {et~al.}(2026){Zhang}, {de Graaff}, {Setton}, {Price}, {Bezanson}, {del P. Lagos}, {Cutler}, {McConachie}, {Cleri}, {Cooper}, {Gottumukkala}, {Greene}, {Hirschmann}, {Khullar}, {Labbe}, {Leja}, {Maseda}, {Matthee}, {Miller}, {Nanayakkara}, {Suess}, {Wang}, {Whitaker}, \& {Williams}}]{Zhang2026_q}
{Zhang}, Y., {de Graaff}, A., {Setton}, D.~J., {et~al.} 2026, \bibinfo{title}{{RUBIES Spectroscopically Confirms the High Number Density of Quiescent Galaxies from 2 < z< 5},} \apj, 997, 252, \dodoi{10.3847/1538-4357/ae24e1}

\end{thebibliography}
\bibliographystyle{aasjournalv7}
\appendix

\section{Results of Symbolic Regression}

\begin{figure}
    \centering
    \includegraphics[width = 0.99\textwidth]{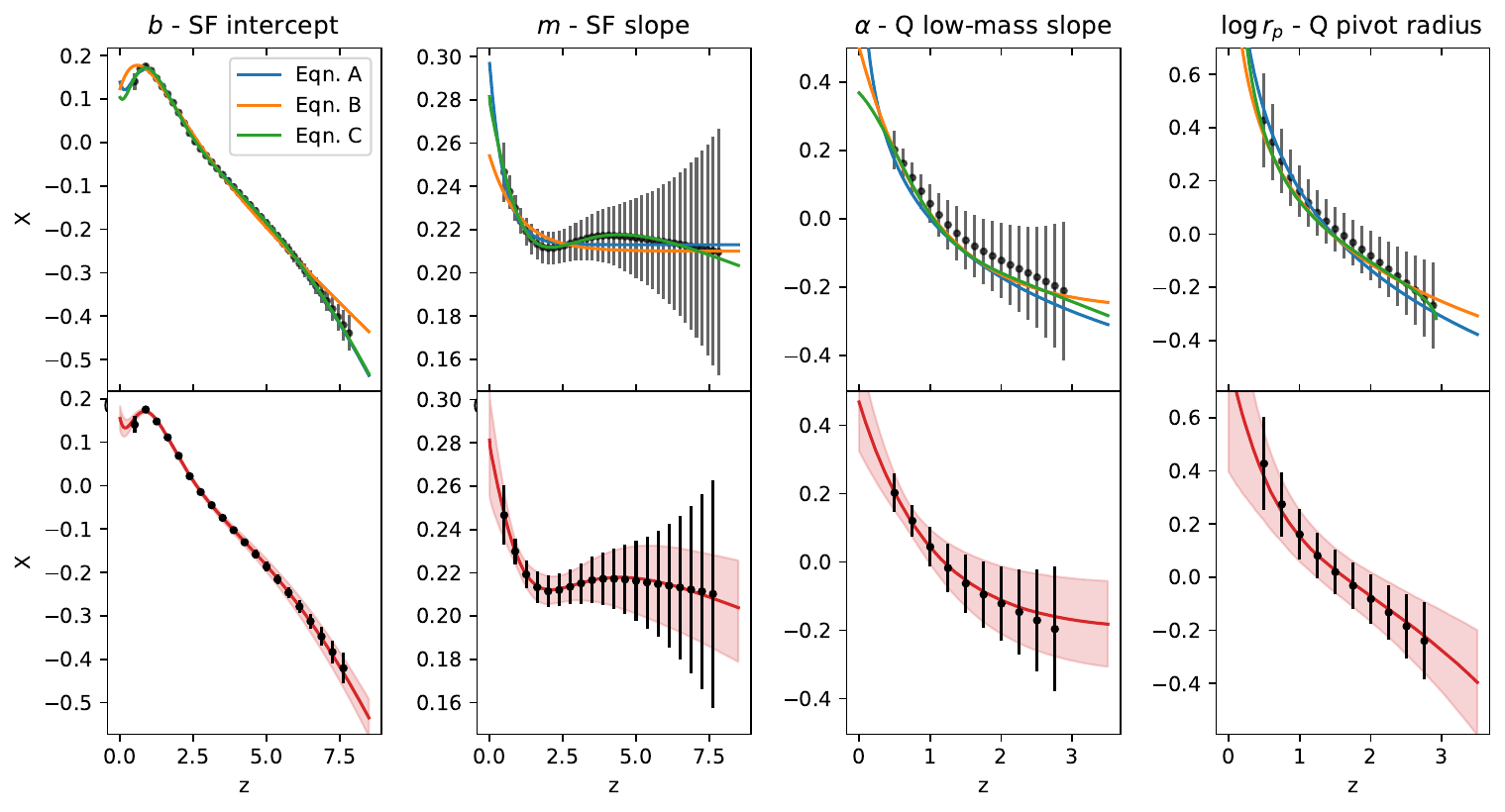}
    \caption{Visualizing the results of the symbolic regression procedure for the four parameters considered: $b$, $m$, $\alpha$, and $\log r_p$. The top three equations, ranked by their ``score'' in \texttt{pysr}, are shown compared to the evolution measure by the B-splines, shown as gray points in the top panel. These equations are shown in detail in Tab.~\ref{tab:sr_results}. For star-forming galaxies the additional point at $z=0$ from \citet{Asali2025} is shown as a square. In the bottom panels, for each selected equation, we show the median and 16th-84th percentile band corresponding to the uncertainty on the constants in Equations~\ref{eqn:b}-\ref{eqn:beta}. For each of the parameters this band matches well with the uncertainties measured using B-splines.}
    \label{fig:sr_results}
\end{figure}

Further details on the results of the symbolic regression fitting applied to the four size-mass parameters considered is displayed in Table~\ref{tab:sr_results}. For each we show the top three ( or four) equations ranked by ``Score'' which measure the increase in accuracy with respect to complexity. This is used in \texttt{pysr} to evaluate the ``best'' equations which match the data well but minimize complexity. The top three equations for each parameter are shown in Figure~\ref{fig:sr_results} compared to the evolution measured with B-splines. We also show the selected equation, and the uncertainty in the redshift evolution implied by the uncertainty in the constants as shown in Equations~\ref{eqn:b}-\ref{eqn:beta}.

When selecting an equation we considered the ``Score'', the accuracy and how well the parameters extrapolated beyond the redshift range. We also investigated how uncertainties on the constants, when translated to the redshift evolution, matched the uncertainties from measured using B-splines.

For the parameter $b$ we simply chose the equation with the highest Score as it is accurate and relatively simple. For $m$ we initially investigated the simpler equations, A and B, but noticed that uncertainties measured for the constants, when translated to the redshift evolution, were much lower than the uncertainties of the B-spline evolution. Thus we pick the more complex evolution which provides a better match, even if the uncertainties are still underestimated at $z>5$. For $\alpha$ we pick the second highest score equation, B, the equation A diverges as $z\rightarrow0$. Similarly for $\log(r_p)$, except in this case all three of the highest scored equations diverge. Therefore we move to the fourth equation, labeled D.

\begin{table}
    \centering
    \caption{Results of performing symbolic regression on the redshift evolution of the size-mass parameters}
    
    \subcaption*{$b$ - Star-forming galaxy Intercept}
    \begin{tabular}{c|c|c|c|c}
        Label & Complexity& Equation & Score & Loss\\
        \hline
        A$^\dagger$ & 11 & $(-0.0074 + \exp(-1.9*z) *(z^2 + 0.14)$ & 1.706 &$1.1 \times 10^{-5}$\\
        B & 10 &$z*(-0.070 + 0.28 * \exp(-z)) + 0.12$ & 0.438 & $5.8\times10^{-5}$\\
        C & 13 &$(-0.0074 + \exp(-1.9*z) *((z + 0.038)^2 + 0.10)$ & 0.287 & $5.9\times10^{-6}$\\
        \hline
    \end{tabular}
    
    \bigskip 
    \subcaption*{$m$ - Star-forming galaxy slope}
    
    \begin{tabular}{c|c|c|c|c}
        Label & Complexity& Equation & Score & Loss\\
        \hline
        A & 7 & $0.21 + 0.084\exp(-2z)$ & 1.02 &$7.1 \times 10^{-6}$\\
        B & 6 & $0.21 + 0.044\exp(-z)$ & 0.998 & $2.0\times10^{-5}$\\
        C$^\dagger$ & 14 & $\left(-0.042\,z^{3/2}\exp(-z) -0.013\,z^{1/2} + 0.079  \right)^{1/2} $ & 0.975 & $1.4\times10^{-7}$\\
        \hline
    \end{tabular}
    \bigskip 

    \subcaption*{$\alpha$ - Quiescent galaxy low-mass slope}
    
    \begin{tabular}{c|c|c|c|c}
        Label & Complexity& Equation & Score & Loss\\
        \hline
        A & 4 & $-0.57\, \log_{10}(z)$ & 3.37 & $3.5 \times 10^{-4}$\\
        B$^\dagger$ & 6 & $-0.27+ 0.77\exp(-z)$ & 2.507 & $2.0\times10^{-5}$\\
        C & 11 & $ -0.081\,z + 0.28\exp\left( -(z+0.18)^2 \right) $ & 1.18 & $1.4\times10^{-6}$\\
        \hline
    \end{tabular}
    \bigskip 

    \subcaption*{$\log_{10}(r_p)$ - Quiescent galaxy pivot radius}
    \begin{tabular}{c|c|c|c|c}
        Label & Complexity& Equation & Score & Loss\\
        \hline
        A & 4 & $ \log_{10}(1.47/z)$ & 2.50 & $1.1 \times 10^{-3}$\\
        B & 6 & $ 0.80\,\log_{10}(1.45/z)$ & 2.12 & $8.3\times10^{-5}$\\
        C & 8 & $ \left( -0.10 + 0.30/z \right)^{1/2} -0.32$ & 1.82 & $1.3\times10^{-5}$\\
        D$^{\dagger}$ & 10 & $-1.2\,z\exp(-z) -0.34z+0.21 $ & 1.51 & $1.5\times10^{-6}$\\
        \hline
    \end{tabular}
    \label{tab:sr_results}
    
\vspace{12pt}
\footnotesize{$^\dagger$ Represents the equation selected for each parameter. See text for further details on each choice.}\\
\end{table}

\label{app:sr}

\end{document}